\documentclass[num-refs]{wiley-article}

\usepackage[referable]{threeparttablex}
\usepackage{booktabs}
\usepackage{comment}
\usepackage{tikz}
\usetikzlibrary{bayesnet}
\usetikzlibrary{fit,positioning}
\usepackage{amsmath}
\usepackage{bbm}
\usepackage{bigints}
\usepackage{hyperref}
\usepackage{natbib}
\usepackage{xcolor, soul,color}
\usepackage{float}
\usepackage[outdir=./figures]{epstopdf}
\usepackage{subcaption}
\usepackage{caption}

\graphicspath{{./}{figures/}}
\papertype{Research Article}

\title{Uncovering Heterogeneity of Solar Flare Mechanism With Mixture Models}

\author[1\authfn{1}]{Bach Viet Do}
\author[1,3\authfn{1}]{Yang Chen}
\author[1\authfn{1}]{XuanLong Nguyen}
\author[2\authfn{2}]{Ward Manchester}

\affil[1]{Department of Statistics, University of Michigan, Ann Arbor, Michigan, MI 48109}
\affil[2]{Climate and Space Sciences and Engineering, University of Michigan, Ann Arbor, Michigan, MI 48109}
\affil[3]{Michigan Institute for Data Sciences, University of Michigan, Ann Arbor, Michigan, MI 48109}

\corraddress{Yang Chen, Department of Statistics, University of Michigan, Ann Arbor, Michigan, MI 48109}
\corremail{ychenang@umich.edu}

\fundinginfo{NSF DMS 2113397, NSF PHY 2027555}

\begin{document}

\begin{frontmatter}
\maketitle

\begin{abstract}
The physics of solar flares occurring on the Sun is highly complex and far from fully understood. However, observations show that solar eruptions are associated with the intense kilogauss fields of active regions, where free energies are stored with field-aligned electric currents. With the advent of high-quality data sources such as the Geostationary Operational Environmental Satellites (GOES) and Solar Dynamics Observatory (SDO)/Helioseismic and Magnetic Imager (HMI), recent works on solar flare forecasting have been focusing on data-driven methods. In particular, black box machine learning and deep learning models are increasingly adopted in which underlying data structures are not modeled explicitly. If the active regions indeed follow the same laws of physics, there should be similar patterns shared among them, reflected by the observations. Yet, these black box models currently used in the literature do not explicitly characterize the heterogeneous nature of the solar flare data, within and between active regions. In this paper, we propose two finite mixture models designed to capture the heterogeneous patterns of active regions and their associated solar flare events. With extensive numerical studies, we demonstrate the usefulness of our proposed method for both resolving the sample imbalance issue and modeling the heterogeneity for rare energetic solar flare events.

\end{abstract}
\end{frontmatter}

\section{Introduction} 
\label{sec:intro}

Solar flares originate from the explosions of magnetic energy caused by tangling, crossing, or reorganizing of magnetic field lines. Flares can last from minutes to hours and can disrupt space-earth radio communications increasing satellite drag when reaching certain thresholds. An example is the October 2003 superstorm event where the Sun unleashed powerful solar flares and coronal mass ejections that impacted the space environment of the Earth. In late October 28, 2003, the Sun produced the "Halloween Storms of 2003" dubbed by NASA \cite{HalloweenStorm2003}, whose impact on Earth caused airplanes to be rerouted, impacted satellite systems, and created power outages in Sweden. The Solar and Heliospheric Observatory (SOHO) was temporarily overwhelmed during the solar onslaught. \\

The energy release mechanism of solar flares is yet to be fully characterized. Observations have established that they are strongly associated with nonpotential magnetic fields, which store necessary free energy \cite{2019SW...17...10}. Most flares originate from localized intense kilogauss photospheric fields, which produce active regions (ARs). Accurate photospheric measurement of these fields has been greatly enhanced with the Helioseismic and Magnetic Imager (HMI) instrument on the Solar Dynamics Observatory (SDO) launched on February 2010 \cite{nasahmisdo}. HMI provides high-quality data in the form of high-cadence high-resolution vector magnetograms, which span the entire solar disk. These data are saved at a 12-minute cadence. The analysis and storage are subdivided into HMI Active Region Patches (HARPS), cutouts of the magnetograms. Time series of HARP data track the evolution of each AR from the time it appears until its disappearance, either by emergence/dispersion or rotating on/off the visible disk. From the 2D HARP data field, scalar quantities referred to as Space-weather HARP (or SHARP) are calculated, which includes 16 indices computed from the full 3-component vector magnetic field. These parameters are automatically calculated for HARPs and made available along with the HARP magnetogram data by the Joint Science Operations Center (JSOC) located at Stanford University  \cite{2014SolPhys...289...9}. \\

Machine learning (ML) algorithms have become increasingly common with Space Weather practitioners. At first, the line-of-sight (LOS) component of the photospheric magnetic field measured by the Michelson Doppler Imager (MDI) instrument (launched in 1995 as part of the Solar and Heliospheric Observatory) was used by several research groups to forecast solar flares using ML models \cite{2013SolPhys...283...1}, \cite{2018AAS...856...1}, \cite{2009SolPhys...254...1}, \cite{2009SolPhys...255...1}, \cite{2010IOP...10...8}. Later studies used SDO/HMI data, which provides the full vector magnetic field data with twice the spatial resolution and eight times the data cadence as MDI. Bobra \& Couvidat \cite{2015ApJ...798...135} employed the Support Vector Machine (SVM) trained with SHARP parameters for active region classification tasks; see also \cite{2014SolPhys...289...9}, \cite{2016ApJ...829...89}, \cite{LEKA201865} and \citep{2019SW...17...8}. Recently, deep learning models such as Long-Short Term Memory Networks (LSTM), Recurrent Neural Networks (RNN), and Convolutional Neural Networks (CNN) have also been adopted to exploit the correlated structure among the time series data; see \cite{2019SW...17...10}, \cite{2019ApJ...877...2}, \cite{2020SW...18...7}, \cite{2020ApJ...895...1} and \cite{2021arXiv...landa}. While these black box models have enjoyed predictive performance gains, their limitation is typically not being able to shed light on the underlying structures of the raw data, which can be utilized to gain new insights into the physics of solar flares. \\

\citet{2019SW...17...10} built a Long Short Term Memory (LSTM) neural network classifier with the HMI/SHARP patches' parameters from 1 May 2010 to 20 June 2018 as their covariate data. For the corresponding response variables, they took advantage of the data from the NOAA GOES flare list during the same time period (see \cite{1994SolPhys...154...2}). GOES flare data are provided both as a time series of soft X-ray intensities and a list of flare events, including start time, peak time, and peak X-ray intensity, recorded by space weather satellites. The GOES satellites are managed by National Oceanic and Atmospheric Administration (NOAA), and its spacecraft is located at a height of about 35,800 km, providing an uninterrupted view of the Sun. GOES’s main objective is collecting infrared radiation and solar reflection from the Earth’s surface \cite{1994SolPhys...154...2}. \\

The work in this paper closely follows the above data framework laid out by \citet{2019SW...17...10} with some differences. To make the task of binary classification manageable with LSTM, \citet{2019SW...17...10} considered only the B and M/X flares and excluded the prevalent C flare because their intensities straddle between the range of strong and weak flares, making the classification harder. In contrast, here we account for all data, including the C flares, as we wish to model flares' intensities as continuous values to closely resemble the observed data. Moreover, Chen et al. in \cite{2019SW...17...10} treated SDO/HMI stream data as time series where ARs are recorded from their initial appearance to disappearances. Here we consider each flare only at its peak time (at the highest intensity). As such, our data are not time series, and should be considered a collection of discrete events occurring at different time points. \\ 

In our work, we are interested in the shared properties of active regions. There have been works in the space weather literature that applied machine learning methods to classify active regions \citep{Colak2008AutomatedMC, Maloney2018class, nguyen2006class, smith2018}. The methods employed in these works include Support Vector Machines, Random Forest, K-Nearest Neighbor classification, and Neural
Networks, which do not explicitly take into account the rich statistical structure of the data. In addition, these black box models typically do not yield more insights about the underlying data structure. Most recently, \citet{Baeke2023} applied unsupervised learning methods such K-Means and Gaussian Mixture Model to cluster active regions. However, the authors cluster active regions on the data covariates. Our work in this paper clusters active regions based on the interaction between the response and the covariate of the data. We believe model-based clustering at this level would be more interesting and meaningful to space weather scientists. \\

It is scientifically reasonable to believe that solar flares across the Sun's active regions follow similar laws of physics, and so active regions' SHARP parameters should share some common data patterns. Nevertheless, to our knowledge, the heterogeneous nature of solar flare data has not been characterized or exploited in the space weather literature. Our contribution in this paper, which marks its difference from other works,  is to apply mixture models to detect and elucidate the heterogeneous patterns of active regions. The idea of mixture modeling is to describe a complicated data distribution as a weighted combination of simpler distributions \cite{titterington85, mclachlan200001}. They are especially useful in a setting where data naturally comes from a number of ``homogeneous'' subgroups within a population. For example, human height data can be considered a mixture of two subgroups, male and female. Mixture models have played a central role in machine learning, and statistics, with broad applications including bioinformatics, natural language and speech processing, and computer vision \cite{bishp2006}. A challenge of mixture modeling is the technical difficulty in parameter estimation. Finding the model's maximum likelihood estimates often involves solving a non-convex optimization problem \cite{bishp2006}. In practice, maximum likelihood estimation via the Expectation Maximization (EM) algorithm has been the workhorse for these models \cite{Dempster1977}. In the solar flare prediction problem, different active regions across the surface of the Sun seem to share certain common characteristics and are thus a good candidate for mixture modeling. We are proposing two types of mixture models. The first one is designed to characterize the heterogeneous pattern of active regions, as mentioned. The second one goes further and allows for the heterogeneity of individual flare events within an active region. As we will demonstrate, using mixture models for active regions does improve the predictive performance and confirms the validity of the empirical observation that active regions share similar patterns. The second proposed mixture model further improves the performance, albeit marginally, implying that heterogeneous patterns are not only restricted to active regions but also potentially extend to flare events within action regions. Since energetic solar flares are extremely rare events compared to low energy flares, which occur orders of magnitude more frequently, statistical inference for this type of data needs to address the data imbalance issue \cite{2015ApJ...798...135}. So another contribution of this paper is showing how to deal with the imbalance problem using the Expectation Maximization framework. \\

The paper is organized as follows. Section 2 describes the data preprocessing procedure. Section 3 proposes two types of mixture models designed to capture the heterogeneous properties of solar flare data. Section 4 provides the detailed data analysis results and interpretation. Section 5 concludes and briefly touches on future work.

\begin{table}[!ht]
  \centering
  \begin{threeparttable}
    \caption{Summary of notation used in the paper.}
    \label{tab:notation}
    \begin{tabular}{cc}
      \toprule
      \textbf{Notation} & \textbf{Description} \\
      \midrule
      $i$ & index of a flare event \\
      $r$ & index of an active region \\
      $k$ & index of a mixture component (linear mechanism) \\
      $K$ & the total number of mixture components (linear mechanisms) \\
      $n$ & the total number of flare events \\
      $n_r$ & the number of flare events in active region r \\
      $X_i$ & SHARP parameter covariates of flare event i \\
      $y_i$ & log intensity response of flare event i (in $\log_{10})$ \\
      $z^r$ & active region $r$'s mixture latent variable \\
      $z_i^r$ & active region $r$'s event $i$'s latent variable \\
      $\beta_k$ & the $k^{\text{th}}$ mixture component's linear regression coefficient \\
      $\sigma_k^2$ & the $k^{\text{th}}$ mixture component's linear regression variance\\
      $w_i$ & the weighted linear regression's weight for data point $i$ \\
      \bottomrule
    \end{tabular}
  \end{threeparttable}
\end{table}

\section{Data Preprocessing \& Feature Selections} \label{sec:prepro}

\subsection{Raw Data}

For the response variables, we take advantage of the recorded log intensities of flare events in the GOES data set \cite{1994SolPhys...154...2} ranging from 2 June 2010 to 29 December 2018. The flare events are recorded at their peak time (time at highest flare intensity). Although the theoretical distribution of the flare events should be a power law distribution, the observed distribution is different from the theoretical distribution because flares in lower energy levels are lost in the background and go undetected \cite{2020SW...18...7}. In this paper, we focus on the observed information. By scientific convention, solar flares belong to category B if their log intensities ($\log_{10}$) is within $(-\infty, -6)$, category C if $[-6, -5)$, category M $[-5, -4)$ and category X if $(-4, \infty)$. Figure \ref{fig:resy} shows the M/X flare events are far fewer than B/C. The data imbalance issue will be addressed in section \ref{subsec:imbalance}. \\

For covariates/features, we consider SHARP data \cite{nasahmisdo} from 860 HARPs during the same time period (2 June 2010 to 29 December 2018). There are approximately 7,000 HARPs, many occurring without flares. From these, to maintain the quality of the data, we down-select the HARPs to a group of 860 based on the criteria that (1) the longitude of the HARP should be within the range of $\pm 68^{\circ}$ from Sun central meridian, to avoid projection effects (see \cite{2019SW...17...10, 2015ApJ...798...135}) and (2) the missing SHARP parameters should be fewer than 5\% of all in the HARP, to make sure that the missing data is not significantly large to cause any bias in model training. \\

For this type of data, an important practical goal of any model is to forecast the future flare intensity given an observed value of SHARP parameters. As such, for each point in our dataset, we match the corresponding SHARP covariates with the GOES flare list (response variable) at the time point that is equal to the peak time subtracted by $\Delta t$ where $\Delta t$ is the prediction time window and can take values in $\{6, 12, 24, 36, 48\}$ hours. 

\begin{table}[!ht]
  \centering
  \scalebox{0.95}{
  \begin{threeparttable}
    \caption{Full list of SHARP parameters and their brief descriptions.}
    \label{tab:sharp}
    \begin{tabular}{cc}
      \toprule
      \textbf{Parameter} & \textbf{Description} \\
      \midrule
      \textbf{TOTUSJH} & Total unsigned current helicity \\
\textbf{TOTUSJZ} & Total unsigned vertical current \\
\textbf{SAVNCPP} & Sum of the modulus of the net current per polarity \\
\textbf{USFLUX} & Total unsigned flux \\
\textbf{ABSNJZH} & Absolute value of the net current helicity \\
\textbf{NACR} & The number of strong LOS magnetic field pixels in the patch \\
\textbf{MEANPOT} & Proxy for mean photospheric excess magnetic energy density \\
\textbf{TOTPOT} & Proxy for total photospheric magnetic free energy density \\
\textbf{SIZE ACR} & Deprojected area of active pixels \\
\textbf{SIZE} & Projected area of the image in microhemispheres \\
\textbf{MEANJZH} & Current helicity (Bz contribution) \\
\textbf{SHRGT45} & Fraction of area with shear $> 45^o$ \\
\textbf{MEANSHR} & Mean shear angle \\
\textbf{MEANJZD} & Vertical current density \\
\textbf{MEANALP} & Characteristic twist parameter, $\alpha$ \\
\textbf{MEANGBT} & Horizontal gradient of total field \\
\textbf{MEANGAM} & Mean angle of field from radial \\
\textbf{MEANGBZ} & Horizontal gradient of vertical field \\
\textbf{MEANGBH} & Horizontal gradient of horizontal field \\
\textbf{NPIX} & Number of pixels within the patch \\
     
      \bottomrule
    \end{tabular}
    \begin{tablenotes}
      \item[a]\label{tab:footnote} Note. SHARP = Space-Weather Helioseismic and Magnetic Imager-Active Region Patch; LOS = line of sight.
    \end{tablenotes}
  \end{threeparttable}}
\end{table}

\subsection{Feature Selection \& Preprocessing Procedure}

\begin{figure}[ht!]
    \centering

    \begin{subfigure}{0.48\textwidth}
        \centering
        \includegraphics[scale=0.2]{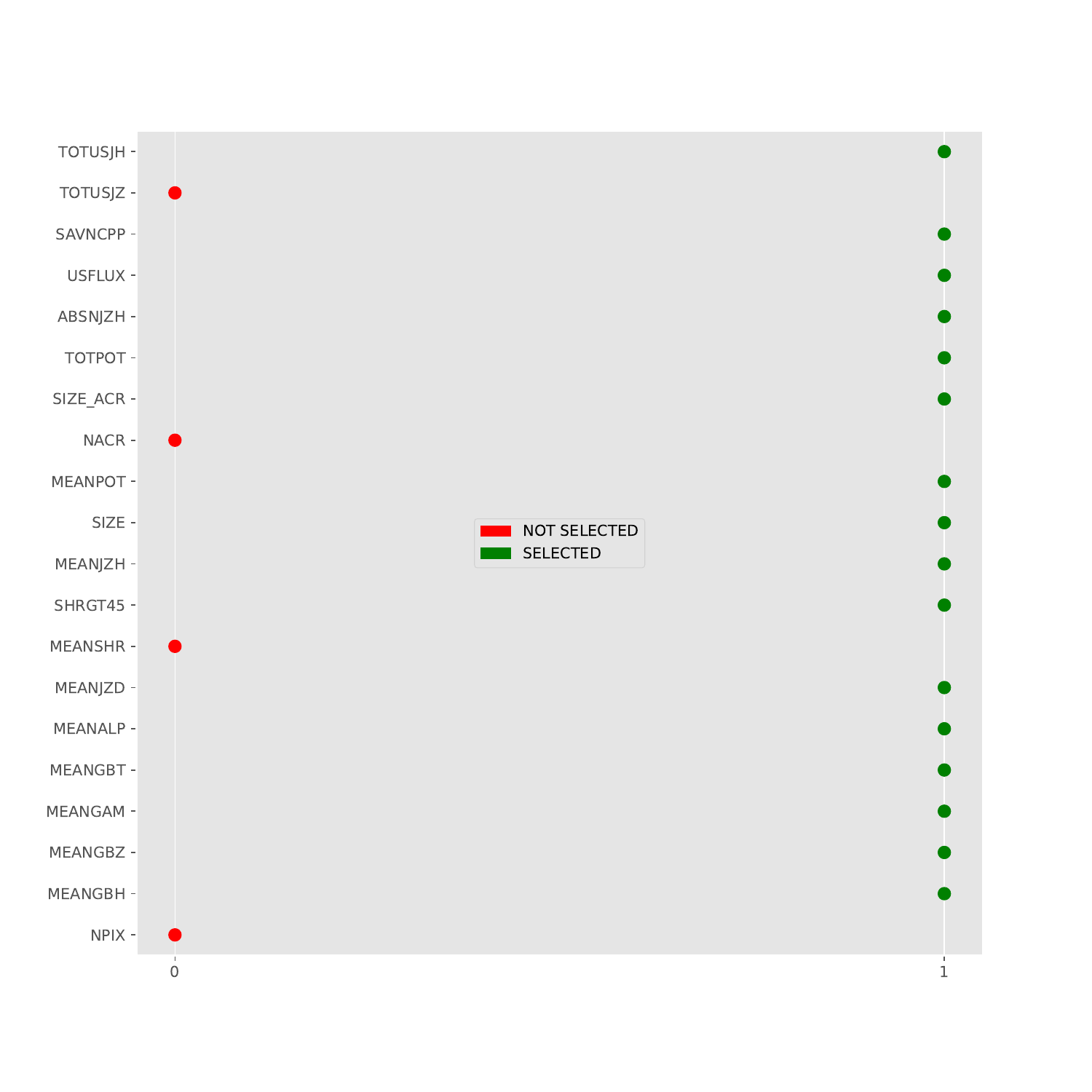}
        \caption{Only one of those extremely highly correlated features ($\ge 0.995$) is included.}
    \end{subfigure}
    \begin{subfigure}{0.48\textwidth}
        \centering
        \includegraphics[scale=0.47]{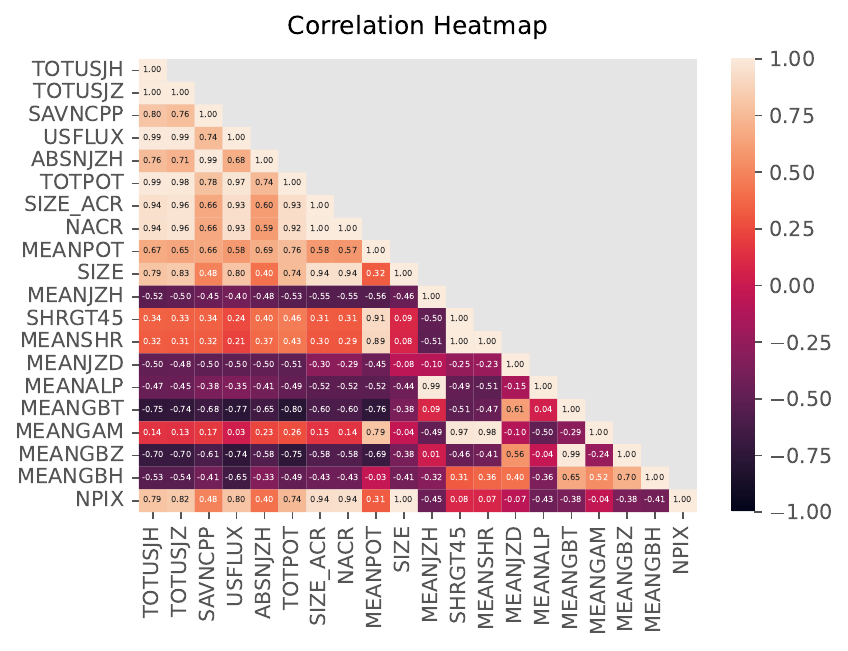}
        \caption{Data correlation of 20 SHARP features in table \ref{tab:sharp}.}
        \label{fig:datacorr}
    \end{subfigure}

    \begin{subfigure}{0.48\textwidth}
        \centering
        \includegraphics[scale=0.25]{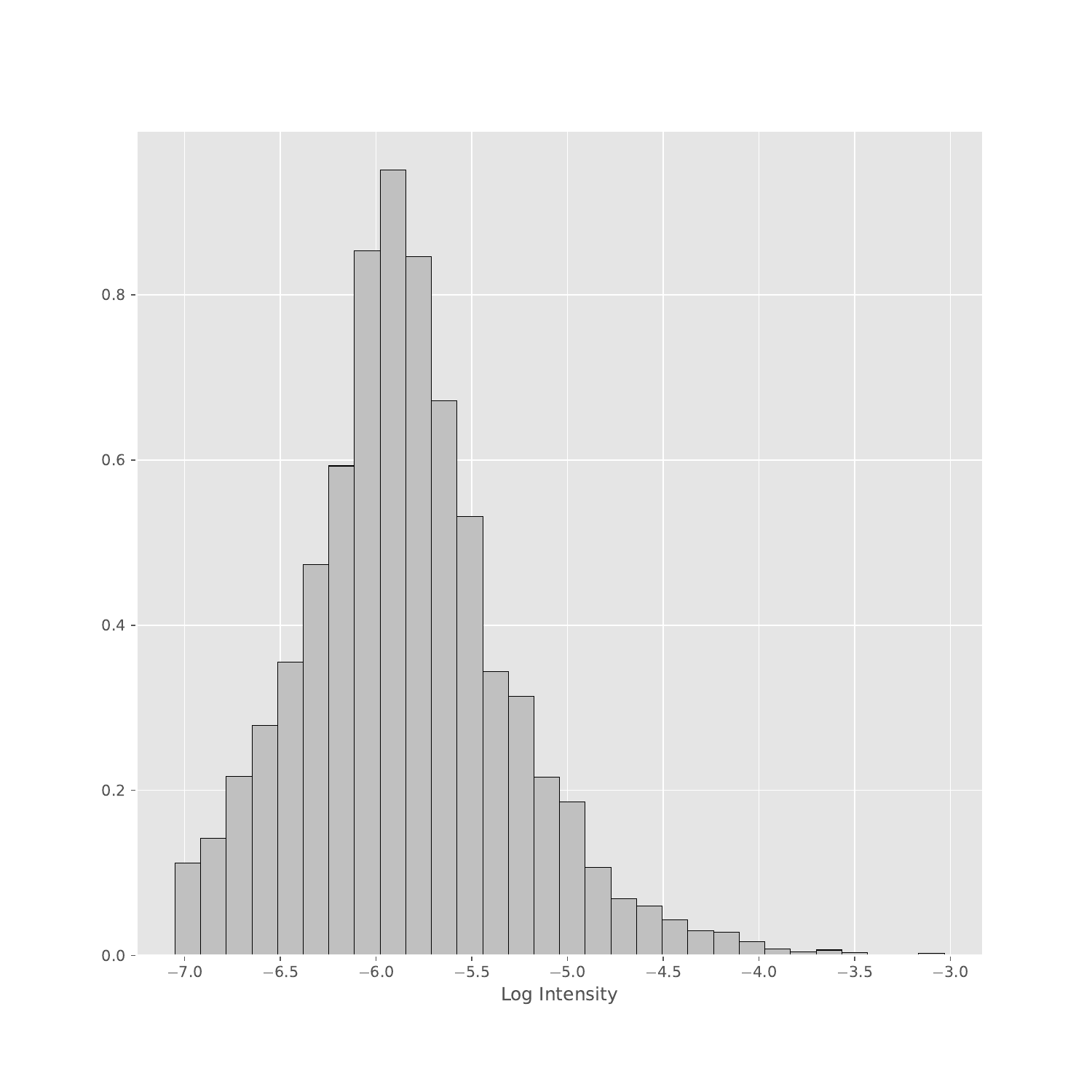}
        \caption{The histogram of log intensity responses (in $\log_{10}$).}
    \end{subfigure}
    \begin{subfigure}{0.48\textwidth}
        \centering
        \includegraphics[scale=0.25]{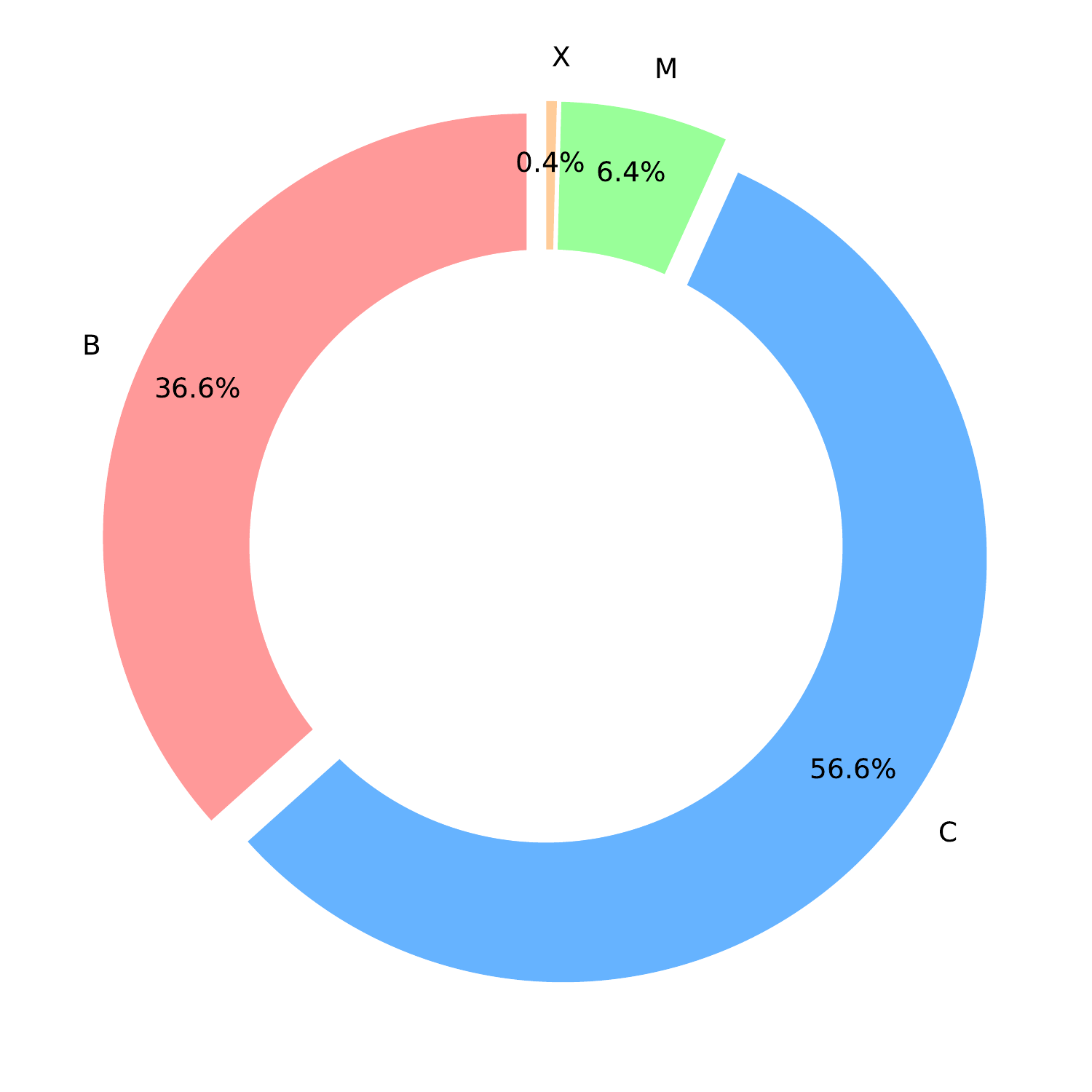}
        \caption{Categories of flares in the data set.}
        \label{fig:resy}
    \end{subfigure}

    \caption{Description of the HMI/SDO SHARP dataset.}
\end{figure}

The list of all covariates in the raw data is shown in Table \ref{tab:sharp}. These are the same features used in \cite{2019SW...17...10} and \cite{2015ApJ...798...135}. However, we only use a subset because our correlation analysis uncovered that some features are extremely highly correlated. Specifically, TOTUSJH/TOTUSJZ ($0.9959$), SIZE\_ACR/NACR ($0.9999$), SHRGT45/MEANSHR ($0.9969$) and SIZE/NPIX ($0.9999$) are extremely highly correlated. For each of these features, we keep one feature and leave out the other. After removing, we left with 16 covariates: TOTUSJH, SAVNCPP, USFLUX, ABSNJZH, TOTPOT, SIZE\_ACR, MEANPOT, SIZE, MEANJZH, SHRGT45, MEANJZD, MEANALP, MEANGBT, MEANGAM, MEANGBZ, MEANGBH. Their correlation matrix is displayed in Figure \ref{fig:datacorr}. \\

We will demonstrate shortly a key aspect of our models is modeling the flares' intensities by active regions (ARs). Consequently, we need to ensure similar sets of ARs in both training and testing. To facilitate this, we randomly split data into a train set and a test set, the latter containing 20\% of the total across all the ARs with two or more events. For those with only one event, we flip a biased coin with probability 0.2, assigning to the test set if the result was true, and to the train set otherwise. This random split scheme guarantees fair representation of each AR in both the train set and test set. Each feature in the train set was then standardized to have zero mean and standard deviation of one. We used these parameters to normalize the test set. Additionally, we further randomly divided the train set by ARs into a sub-train set and a validation set, with the latter equal to 20\% of the original set. We used this validation set for the model selection process that we will discuss in Section 4.

\section{Methodology} \label{sec:methodologies}

In this section, we describe our methodology for modeling the solar flare data described in the previous section. We begin by discussing a strategy to handle the data imbalance issue because space weather practitioners are mostly interested in the catastrophic flare events (M/X) of which occurrences are significantly fewer than the weak (B/C class) flares. After that, we apply mixture models to characterize the heterogeneity of the solar flare mechanisms. Specifically, we describe the mixture models MM-R and MM-H that characterizes the heterogeneous patterns among active regions, with MM-H an extension of MM-R.

\subsection{Approach to dealing with the data imbalance issue}
\label{subsec:imbalance}

The weighted likelihood and weighted maximum likelihood estimators (MLE) have been used in the literature for robust estimations, especially when outliers exist in the data \cite{1994Wiley...Field, 1993RSS...55...3, 1998ASS...93...442}. The idea is to down-weight the outlier data points so that they do not deteriorate the model's performance too much. A similar principle can be applied here to handle the imbalance problem, i.e., down-weighting the ``majority'' data points and/or up-weighting the ``minority'' points. 

To illustrate, we consider a simple example of a standard linear regression setting. Assume that the response $y_i$ given covariates $X_i$ follows the normal distribution: $y_i | X_i \sim N(X^T_i \beta, 1), i = 1, \ldots, n$ where $\beta$ is the linear regression coefficients. The least squares estimate for $\beta$ is given by $\hat{\beta} = \text{arg max}_{\beta} \sum_{i=1}^n -( y_i - X_i^T\beta)^2$. With highly imbalanced data like the solar flare data in section \ref{sec:prepro}, standard linear regression would not work well, and weighted linear regression may be employed instead. In that case, we need to calculate the weighted estimator $\tilde{\beta} = \text{arg max}_{\beta} \sum_{i=1}^n -w_i \cdot (y_i - X_i^T\beta)^2$ for known weights $w_i$ that may depend on the data. Moving beyond this simple example, the modeling we propose in the next sections is based on mixture models. Consider a simplified generative model as follows,
\begin{align*}
&y_i | X_i , z_i = k \sim \text{N}(y_i | X_i^T \beta_k, \sigma^2_k),\quad k = 1, \ldots, K,\quad i = 1,\ldots, n, 
\end{align*} 
where $K$ denotes the number of mixture components, and 
$z_i$ is a categorical variable with $P(z_i=k) = \pi_k$ for all $k$; $\sum_k \pi_k = 1$. Applying the weighted likelihood idea, we want to find the optimizer $\text{arg max}_{\theta} \sum_{i=1}^n w_i \cdot \log p(y_i | X_i, \theta)$ where $\theta := \{\beta_k, \sigma^2_k, \pi_k\}_1^K$. Recall that the original log likelihood is $l(\theta) := \sum_{i=1}^n \log p(y_i | X_i, \theta)$. For mixture models, it is not straightforward to directly maximize $l(\theta)$. The Expectation-Maximization (EM) algorithm is needed \cite{Dempster1977}. Typically, the EM algorithm works with the logarithm of the complete data likelihood, which is defined as $l(\pi, \beta, \sigma^2) := \sum_{k=1}^K \sum_{i=1}^n \mathbbm{1}(z_i = k) \cdot \left[ \log \pi_k + \log \text{N}(y_i | X_i^T \beta_k, \sigma_k^2) \right]$. In the E-step, the EM algorithm computes $\tau_{i,k} := p(z_i = k | X_i, y_i) := \cfrac{\pi_k \text{N}(y_i | X_i^T\beta_k, \sigma_k^2)}{\sum_{j=1}^K \pi_j \text{N}(y_i | X_i^T\beta_j, \sigma_j^2)}$, $k=1,\ldots, K$; then it maximizes the expected complete log likelihood, $\text{arg max}_{\beta, \sigma^2, \pi} Q(\beta, \sigma^2, \pi) := \text{arg max}_{\beta, \sigma^2, \pi} \sum_{i=1}^n \sum_{k=1}^K \tau_{i,k} \cdot \left[  \text{log} \pi_k + \log \text{N}(y_i | X^T_i \beta_k, \sigma_k^2) \right]$ in the M-step. That gives $\hat{\pi}_k = \cfrac{\sum_{i=1}^n \tau_{ik}}{n}$ and $\hat{\mu}_k = \cfrac{\sum_{i=1}^n \tau_{i,k} X_i}{\sum_{i=1}^n \tau_{i,k}}$ and $\hat{\sigma}^2_k = \cfrac{\sum_{i=1}^n \tau_{ik} \cdot (X_i - \hat{\mu}_k)^2}{\sum_{i=1}^n \tau_{ik}}$; $k=1,\ldots, K$. \\

To adapt the EM framework to find the MLE for the weighted likelihood, we note under the EM algorithm,  the log-likelihood is lower bounded by the expected complete log-likelihood, i.e., $l(\theta) \ge  Q(\theta)$, and by optimizing the lower bound $Q(\theta)$, EM algorithm yields the (local) maximum of the likelihood $l(\theta)$ \cite{bishp2006}. Under the weighted likelihood setting, we can also find a lower bound as follows:
\begin{align*}
\sum_{i=1}^n w_i \log p(y_i | X_i, \theta) &= \sum_{i=1}^n w_i \log \sum_{k=1}^K p(y_i, z_i = k | X_i, \theta) \\
&= \sum_{i=1}^n w_i \cdot \log \left[ \sum_{k=1}^K q(z_i = k) \cdot \cfrac{ p(y_i, z_i = k | X_i, \theta)}{q(z_i = k)} \right] \\
&\ge \sum_{i=1}^n w_i \cdot \sum_{k=1}^K q(z_i = k) \log\left(  \cfrac{ p(y_i, z_i = k | X_i, \theta)}{q(z_i = k)} \right).
\end{align*}

The last inequality is due to Jensen's inequality. Since logarithm is a concave function and $\sum_{k=1}^k q(z_i = k) =1$, we moved the log inside and left the $q$ outside. Here we can follow the usual EM procedure to find the optimal $q(z)$, E-step sets $q(z_i = k) = \mathbb{P}(z_i = k | X_i, y_i)$ then M-step maximizes the subsequent expression. Therefore, we observe that the weighted log likelihood is bounded below by the weighted expected complete log-likelihood. In other words, to find the weighted log likelihood estimator with EM framework, we can optimize the lower bound $\sum_{i=1}^n \sum_{k=1}^K w_i \cdot \tau_{i,k} \cdot \left[  \text{log} \pi_k + \log \text{N}(y_i | X^T_i \beta_k, \sigma_k^2) \right]$. Moreover, since $w_i(\cdot)$ is a known weight function, if we ensure that $0 < w_i \le C < \infty$ for all $i$ for some constant $C$ free of parameters, then by noting $\log \pi_k < 0$ we have

$\begin{aligned}
\sum_{i=1}^n \sum_{k=1}^K w_i \cdot \tau_{i,k} \cdot \left[  \text{log} \pi_k + \log \text{N}(y_i | X^T_i \beta_k, \sigma_k^2) \right] &= \sum_{i=1}^n \sum_{k=1}^K \tau_{i,k} \cdot \left[   w_i  \cdot \text{log}\pi_k + w_i \log \text{N}(y_i | X^T_i \beta_k, \sigma_k^2) \right] \\
&\ge \sum_{i=1}^n \sum_{k=1}^K \tau_{i,k} \cdot \left[   C \cdot \text{log}\pi_k + w_i \log \text{N}(y_i | X^T_i \beta_k, \sigma_k^2) \right]. \\
\end{aligned}$

Because $\sum_{k=1}^K \pi_k = 1$, it is easy to derive that 
\begin{align*}
    \text{argmax }_{\beta, \sigma^2, \pi} \sum_{i=1}^n \sum_{k=1}^K \tau_{i,k} \cdot \left[   C \cdot \text{log}\pi_k + w_i \log \text{N}(y_i | X^T_i \beta_k, \sigma_k^2) \right] \\
    = \text{argmax }_{\beta, \sigma^2, \pi} \sum_{i=1}^n \sum_{k=1}^K \tau_{i,k} \cdot \left[ \text{log}\pi_k + w_i \log \text{N}(y_i | X^T_i \beta_k, \sigma_k^2) \right].
\end{align*}
Therefore, the above expression is the lower bound of the weighted complete log likelihood and optimizing it in turn increases the data likelihood. The final expression is the lower bound to be optimized for our mixture models, which will be proposed in the subsequent sections. The justification of the algorithm comes from that the EM based inference algorithm will converge to a local maximum of the weighted likelihood function \cite{wu1983}.

\subsection{Dealing with Sample Imbalance Problem: weighting schemes} \label{sec:wi}

In this section, let $w_i$ be denoted by $w(y_i)$ to emphasize the fact that the weights only depend on $y_i$(s). By scientific convention the log intensity (in $\log_{10}$) $y_i \in (-\infty, -6)$ for flare category B, $y_i \in [-6, -5)$ for flare category C, $y_i \in [-5, -4)$ for flare category M and $y_i \in (-4, \infty)$ for flare category X. We propose the following scheme for $w(y_i)$,

\begin{align*} \label{eqn:wi} w(y_i) = \begin{cases} \cfrac{1}{\sum_{i=1}^n \mathbbm{1}(y_i \ge -4) / n } & \text{ if } y_i \ge -4 \\
\cfrac{1}{\sum_{i=1}^n \mathbbm{1}(-5 \le y_i < -4) / n}  & \text{ if } -5 \le y_i < -4 \\
\cfrac{1}{\sum_{i=1}^n \mathbbm{1}(-6 \le y_i < -5) / n} & \text{ if } -6 \le y_i < -5 \\
\cfrac{1}{\sum_{i=1}^n \mathbbm{1}(y_i < -6) / n} & \text{ if }  y_i < -6
\end{cases}.\end{align*} 

Note that as $n \to \infty$, by the strong law of large numbers,

\begin{align*} w(y) \overset{a.s.}{\to} \begin{cases} \cfrac{1}{\int_{-4}^\infty p_0(y)  \; dy} & \text{ if } y \ge -4 \\
\cfrac{1}{\int_{-5}^{-4} p_0(y) \; dy} & \text{ if } -5 \le y < -4 \\
\cfrac{1}{\int_{-6}^{-5} p_0(y) \; dy} & \text{ if } -6 \le y < -5 \\
\cfrac{1}{\int_{-\infty}^{-6} p_0(y) \; dy}& \text{ if }  y < -6
\end{cases},
\end{align*}
where $p_0(y)$ is the marginal distribution of y. The justification for choosing $w(y)$ this way is as follows. As the number of data points goes to infinity and assume that $(X,y)$ follows the "true" distribution $p_0(X, y)$, by the law of large numbers, the weighted score function becomes,

\begin{align*}
\cfrac{1}{n} \sum_{i=1}^n  w(y_i) \log p(y_i | X_i, \theta) &\overset{a.s.}{\to}\int w(y) \log p(y | x, \theta) p_0(x, y) dx dy \\
&= \int w(y) \cdot \left( \int \log p(y | x, \theta) \; p_0(x | y) dx \right) \cdot p_0(y) dy.
\end{align*}

Define $r(y, \theta) := \int \log p(y | x, \theta) \; p_0(x | y) dx$ and $I_B := (-\infty, -6), I_C := [-6, -5), I_M := [-5, -4), I_X := [-4, \infty)$. These intervals correspond to the scientific thresholds of flare category B, C, M, X, respectively. As the number of data points goes to infinity, the weighted log likelihood function can now be written as \\

$\begin{aligned}
\int w(y) \log p(y | x, \theta) p_0(x, y) dx dy &= \int w(y) r( y, \theta) p_0(y) dy \\
&= \int_{I_B} w(y)  r(y, \theta)  p_0(y) dy + \int_{I_C} w(y) r(y, \theta)  p_0(y) dy \\& + \int_{I_M} w(y)  r(y, \theta)  p_0(y) dy + \int_{I_X} w(y) r(y, \theta) p_0(y) dy. \\
\end{aligned}$ \\

As mentioned previously, M and X's number of data points are fewer compared to B and C ones (Figure \ref{fig:resy}), i.e., $p_0(y)$ places negligible masses on $I_M$ and $I_X$ compared to $I_B$ and $I_C$. As a consequence, the four terms in the last expression are of different scales. As such, setting $w(y)$ to the above proposed choice effectively ``normalizes'' and place these four components in the same scale and balance them out. In explicit,  \\

$\begin{aligned}
\int w(y) \log p(y | x, \theta) p_0(x, y) dx dy &= \cfrac{\int_{I_B} r(y, \beta)  p_0(y) dy }{\int_{I_B} p_0(y) dy} + \cfrac{\int_{I_C} r(y, \beta)  p_0(y) dy }{\int_{I_C} p_0(y) dy} \\ &+ \cfrac{\int_{I_M} r(y, \beta)  p_0(y) dy }{\int_{I_M} p_0(y) dy} + \cfrac{\int_{I_X} r(y, \beta)  p_0(y) dy }{\int_{I_X} p_0(y) dy}.
\end{aligned}$ \\

Finally, to tie this back to the last section, $w(y)$ is then being used as the weights in the weighted complete log likelihood. For instance, in the example of previous section, the weighted complete log likelihood is $l(\pi, \beta, \sigma^2) := \sum_{k=1}^K \sum_{i=1}^n \mathbbm{1}(z_i = k) \cdot w(y_i) \cdot \left[ \log \pi_k + \log \text{N}(y_i | X_i^T \beta_k, \sigma_k^2) \right]$.

\subsection{Mixture model over Active Regions (MM-R)} \label{sec:mm2R}

As described in section 2, $X_i$ and $y_i$ denote the SHARP parameter covariates and the corresponding log flare intensity response variable. We reiterate that $y_i$'s value is measured at time $\Delta t$ hours behind the $X_i$'s values since we want to model $y_i$ as the forecasted flare intensity at $\Delta t$ hours after observing SHARP values of $X_i$. Here we use mixture modeling to characterize both the heterogeneity and shared patterns among active regions. Let $K > 1$ be the number of mixture components, we model the heterogeneity to be shared across active regions (ARs) by the global parameters $\beta_k$ and $\sigma_k^2$ for $k = 1, 2, \ldots, K$. For $r = 1, \ldots, R$, each active region $r$ is equipped with a discrete latent variable $z^r$ taking values in $\{ 1, 2, \ldots, K \}$. If $z^r = k$, then all the flare events $\{X_i^r, y_i^r \}$ in the active region r follow the normal distribution $y_i^r \sim \text{N}\left( \cdot | \beta_k ^T \cdot X_i^r, \sigma_k^2 \right)$. Latent variable $z^r$ is introduced to capture the ``intrinsic'' categories of flaring mechanisms from SDO/HMI data . The values of $z^r$ do not necessarily correspond to the scientific B/C/M/X categories, but rather are the inferred ``clusters'' from the data. One important constraint under MM-R is that all the events under the same active region must have the same regression pattern parameterized by $\beta_{z^r}, \sigma^2_{z^r}$. 

Mathematically, the model is defined as follows. For AR $r = 1, \ldots, R$,
\begin{align*}
&z^r | \pi_{1:K} \sim \text{Cat}(\cdot | \pi) \\
&y_i^r | z^r = k \sim \text{N}\left(\cdot | \beta_k^T X_i^r, \sigma_k^2 \right), i = 1, \ldots, n_r.
\end{align*}

\begin{figure}[ht!]
    \centering
    
    \begin{tikzpicture}
    \tikzstyle{main}=[circle, minimum size = 10mm, thick, draw =black!80, node distance = 16mm]
    \tikzstyle{connect}=[-latex, thick]
    \tikzstyle{box}=[rectangle, draw=black!100]
    \node[latent] (z) [label=below:$z^r$] { };
    \node[latent] (pi) [left=of z, xshift=-0.5cm, label=above:$\pi$] {};
    \node[latent] (y) [right=of z,label=below:$y_i^r$] {};
    \node[obs] (x) [right=of y,label=below:$x_i^r$] { };
    \node[latent] (theta) [above=of y, yshift=+1.0cm, label=above:$\beta_k \text{ , } \sigma_k^2$] { };
    \edge {z} {y}
    \edge {x} {y}
    \edge {theta} {y}
    \edge {pi} {z}
    \node[rectangle, inner sep=0mm, fit= (y) (x),label=below right:$n_r$, xshift=8mm] {};
    \node[rectangle, inner sep=4.6mm,draw=black!100, fit= (y) (x)] {};
    \node[rectangle, inner sep=5.0mm, fit= (y) (x),label=below right:$R$, xshift=9mm] {};
    \node[rectangle, inner sep=9mm, draw=black!100, fit = (z) (y) (x)] {};
    \node[rectangle, inner sep=5.5mm, draw=black!100, fit= (theta)] {};
    \node[rectangle, inner sep=0mm, fit= (theta),label=below right:K, xshift=1.5mm, yshift=-1.5mm] {};
    \end{tikzpicture}
    
    \caption{Graphical Model for Mixture Models over ARs MM-R \label{fig:GM2R}.}
\end{figure}
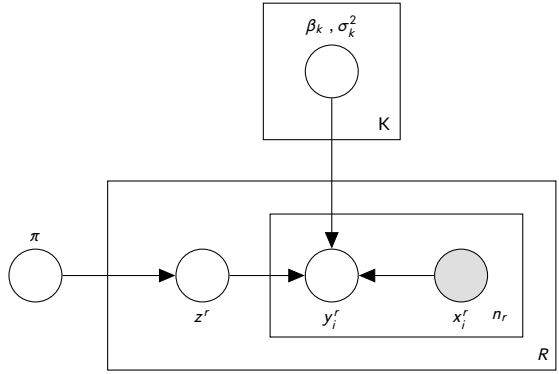

The model can also be represented as a probabilistic graphical model (see \cite{koller2009probabilistic}) as in Figure~\ref{fig:GM2R}. There are $2K + 1$ ``global'' parameters $\{ \beta_k, \sigma^2_k\}_{k=1}^K$ and $\pi$. Using the plate notation in this figure, there are $R$ conditionally i.i.d. latent variables $\{z^r \}_{r=1}^R$ for each of the unique active regions. Under each active region $r$, there are $n_r$ flare events $\{ (X_i^r, y_i^r) \}_{i=1}^{n_r}$.

Parameter estimation is achieved through an Expectation Maximization algorithm (see \cite{Dempster1977}) with the lower bound as a weighted complete log likelihood as defined in section \ref{sec:wi} and we optimize the weighted complete log likelihood to remediate the data imbalance problem. Under this model, rather than the standard expected complete log likehood, our Expectation Maximization (EM) algorithm optimizes the weighted Expected Complete Likelihood Optimization Problem, 
$${\text{argmax}_{\pi, \sigma^2, \beta} \sum_{k=1}^K \sum_{r=1}^R \mathbbm{E}[z^r = k | X^r, y^r] \left[ \log \pi_k - \sum_{i=1}^{n_r} \left( \cfrac{w_i}{2} \log \sigma_k^2 + \cfrac{w_i}{2\sigma_k^2} \cdot \left( y_i^r - \beta_k^T X_i^r\right)^2 \right) \right].}$$ 
The E-step computes:
    \begin{align*}
    &\tau_k^r = p(z^r = k | X^r, y^r) = \cfrac{\pi_k  \cdot \prod_{i=1}^{n_r} \text{N}\left(y_i^r | \beta_k^T X_i^r, \sigma_k^2 \right) }{\sum_{j=1}^K \pi_j \cdot \prod_{i=1}^{n_r} \text{N}\left(y_i^r | \beta_j^T X_i^r, \sigma_j^2 \right)}.
    \end{align*}
And the M-step gives:
    \begin{align*}
    &\hat{\pi}_k  =\cfrac{\sum_{r=1}^R \tau_k^r}{R}, \hat{\beta}_k = \left[\sum_{r=1}^R \tau_k^r \sum_{i=1}^{n_r} w_i X_i^r (X_i^r)^T \right]^{-1} \left[\sum_{r=1}^R \tau_k^r \sum_{i=1}^{n_r} w_i y_i^r X_i^r \right], \hat{\sigma}^2_k = \cfrac{\sum_{r=1}^R \tau_k^r \sum_{i=1}^{n_r} w_i \cdot (y_i^r - \hat{\beta}_k^T X_{i,r})^2}{ \sum_{r=1}^R n_r \cdot \tau_k^r}.
    \end{align*}
Next, for prediction, on the one hand, if the region of a new data point $\tilde{X}_i$ is not known, we estimate the log intensity $\hat{y}_i | \tilde{X}_i = \sum_{k=1}^K \pi_k  \cdot\tilde{X}_i^T \beta_k$. On the other hand, if its region is $r_i$ then $\hat{y}_i | r_i, \tilde{X}_i = \sum_{k=1}^K \tau_{r_i, k} \cdot \tilde{X}_i^T \beta_k$.

\subsection{Mixture Model over Flare Events (MM-H)} \label{sec:mm2H}

The mixture model MM-R in section \ref{sec:mm2R} requires all the flare events of an active region to follow the same regression pattern. This condition can be too restrictive. Note that in our dataset, each flare occurs at a different location and time within an active region and we can think of the entire data being a collection of events from many active regions. Now, it is reasonable to accommodate the possibility that the heterogeneous nature extends further into each individual flare event within an active region. To model this behavior, we can assign a latent variable $z_i^r$ to each data point $i$ but impose that $z_i^r \sim \text{Cat}(\cdot | \pi_i^r)$ where $\pi^r \in \mathbb{R}^K$ and $\sum_{k=1}^K \pi_k^r = 1$. The parameter $\pi_k^r$ captures a regional inclination for certain categories of flaring mechanism. If $\pi^r$(s) are extreme, e.g., taking value 1 in one coordinate and zeros elsewhere, the model MM-H is reduced to the last section mixture model MM-R. Note that this type of model is sometimes called mixed membership models in the statistical learning literature (see \cite{airoldi2014introduction}) where an active region is a group of members, where the members in this case are its flare events. As discussed previously, the weighted complete log likelihood is optimized to combat the unbalanced data.

Mathematically, the model is parameterized by $\beta_1, \ldots, \beta_K, \sigma_1^2, \ldots, \sigma_K^2, \pi^1, \ldots, \pi^R$ where $K$ is number of global paramters, $R$ number of active regions. For active region $r = 1, \ldots, R$,
\begin{align*}
    &z_i^r \sim \text{Cat}(\cdot | \pi^r), i = 1, \ldots, n_{r} \\
    &y_i^r | z_i^r = k, X_i^r \sim \text{N}\left(\cdot | \beta_k^T X_i^r, \sigma_k^2 \right).
    \end{align*}

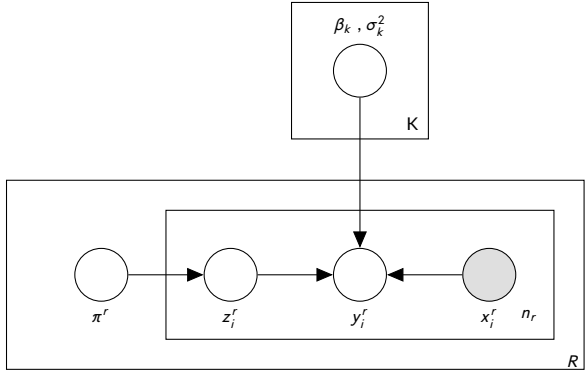
\begin{figure}[!ht]
    \centering
    \begin{tikzpicture}
    \tikzstyle{main}=[circle, minimum size = 10mm, thick, draw =black!80, node distance = 16mm]
    \tikzstyle{connect}=[-latex, thick]
    \tikzstyle{box}=[rectangle, draw=black!100]
    \node[latent] (pi) [label=below:$\pi^r$] { };
    \node[latent] (z) [right=of pi,label=below:$z_i^r$] {};
    \node[latent] (y) [right=of z,label=below:$y_i^r$] {};
    \node[obs] (x) [right=of y,label=below:$x_i^r$] { };
    \node[latent] (theta) [above=of y, yshift=+1.0cm, label=above:$\beta_k \text{ , } \sigma_k^2$] { };
    \edge {pi} {z}
    \edge {z} {y}
    \edge {x} {y}
    \edge {theta} {y}
    \node[rectangle, inner sep=0mm, fit= (z) (y) (x),label=below right:$n_r$, xshift=1.67cm] {};
    \node[rectangle, inner sep=5.0mm,draw=black!100, fit= (z) (y) (x)] {};
    \node[rectangle, inner sep=6.0mm, fit= (z) (y) (x),label=below right:$R$, xshift=1.68cm] {};
    \node[rectangle, inner sep=9mm, draw=black!100, fit = (pi) (z) (y) (x)] {};
    \node[rectangle, inner sep=5.5mm, draw=black!100, fit= (theta)] {};
    \node[rectangle, inner sep=0mm, fit= (theta),label=below right:K, xshift=1.5mm, yshift=-1.5mm] {};
    \end{tikzpicture}

    \caption{Graphical Model for Mixture Models over Flare Events MM-H \label{tab:GMMM2H}.}
\end{figure}

Figure \ref{tab:GMMM2H} is the probabilistic graphical model representation of model MM-H. There are $2K$ ``global'' parameters $\{\beta_k,\sigma_k^2\}_{k=1}^K$. With the plate notation, there are now $n = \sum_{r=1}^R n_r$ latent variables ${z_i^r}$ for each flare event $(X_i^r, y_i^r)$ of active region $r$. Under each active region $r$, the latent variable $z_i^r$ is a discrete random variable which takes values in $\{1, \ldots, K\}$ with probability weight $\pi^r$. Compared to the model described in section \ref{sec:mm2R}, MM-R only has $R$ latent variables $z^r$ with one ``global'' weight $\pi$. MM-H gives each active region the flexibility of having its own categorical weight $\pi^r$ over $K$ linear mechanisms.

Similarly to the previously discussed model, the likelihood is $p(y_i^r | x_i^r; \pi, \beta, \sigma^2) = \sum_{k=1}^K \pi_k^r \cdot N\left(\cdot | \beta_k^T x_i^r, \sigma_k^2 \right)$ and the weighted expected complete log likelihood optimization problem is equivalent to, 
\begin{equation*}
    \text{argmax}_{\beta, \pi, \sigma^2} \sum_{k=1}^K \sum_{r=1}^R \sum_{i=1}^{n_r} \mathbb{E}[z_i^r | y_i^r, X_i^r] \cdot \left( \log \pi_k^r - \cfrac{w_i}{2} \log \sigma_k^2 - \cfrac{w_i}{2\sigma_k^2} \cdot (y_i^r - \beta_k^T X_i^r)^2 \right).
\end{equation*}

Under these specifications, for parameter estimation, the iterative E-step computes 
\begin{equation*}
\tau_{i,k}^r = \mathbb{P}(z_i^r = k | y_i^r, X_i^r) = \cfrac{\pi^r_k  \cdot N\left(y_i^r | \beta_k^T X_i^r, \sigma_k^2 \right)}{\sum_{j=1}^K \pi^r_j \cdot N\left(y_i^r | \beta_j^T X_i^r, \sigma_j^2 \right)}, 
\end{equation*}
while the M-step performs the updates:
    \begin{align*}
    &{\hat{\pi}_k^r = \cfrac{\sum_{i=1}^{n_r} \tau_{i,k}^r}{n_r} }\\
    &\hat{\beta}_k = \left[\sum_{r=1}^R \sum_{i=1}^{n_r} \tau_{i,k}^r \cdot w_i \cdot X_i^r (X_i^r)^T \right]^{-1} \left[\sum_{r=1}^R \sum_{i=1}^{n_r} \tau_{i,k}^r \cdot y_i^r X_i^r \right] \\
    &\hat{\sigma}^2_k = \cfrac{\sum_{r=1}^R \sum_{i=1}^{n_r} \tau_{i,k}^r w_i \cdot (y_i^r - \hat{\beta}_k^T X_{i,r})^2}{ \sum_{r=1}^R \sum_{i=1}^{n_r} \tau_{i,k}^r}. \\
    \end{align*}
Finally, to perform prediction for a new data point $\tilde{X}_i$ of the region $r_i$, given $r_i, \tilde{X_i}$, we take $\hat{y}_i := \sum_{k=1}^K \pi^r_k \cdot \beta_k^T. \tilde{X_i}$.

\section{Model Selection \& Data Analysis Discussion}

The model selection for mixture models concerns the choice of $K$, the number of mixture components. In this section, we focus our investigation on the data set with the prediction time window $\Delta t = 6$ hours. Recall that with $\Delta t = 6$ hours, each response $y_i$ is matched with SHARP parameters $X_i$ 6 hours before in the data. Similar results can be obtained with other $\Delta t = 12, 24, 36, 48$ hours. For general regression problems, a common evaluation metric for model performance is the root mean squared error (rmse) = $\sqrt{\sum_{i=1}^n (y_i - X_i^T \hat{\beta} )^2}$. However, this metric can be misleading if the primary concern is the predictive performance of M/X future events because the data is imbalanced, with M/X events being rare. To assess the proposed models in a more balanced fashion, we can first discretize each of the continuous-valued $y_i$ into binary-valued $\tilde{y}_i \in \{0,1\}$ where $\tilde{y_i} = \mathbbm{1}(y_i > -5)$, then takes advantage of standard classifier metrics such as  precision, recall, f1-score (the harmonic mean of precision and recall) (see \cite{10.1007/978-3-540-31865-1_25}). If a model tries to improve the overall rmse performance by over-optimizing B/C events at the expense of M/X, the recall will suffer and lead to a low f1-score. The higher the f1 metrics (i.e., closer to 1), the better the performance. This approach also allows us to compare our models with other methods in the solar flare forecasting literature which are mostly black box machine learning classifiers. We also remind from section 2 that we split the original train set into a sub-train set and a validation set. 

The validation set is utilized to determine the optimal number of components ($K$) and the above discretization binary threshold. For the $6$-hour-dataset, the threshold $-5.0$ yields the best f1 performance in the validation. Next we explain with details on how to choose the best K.

\subsection{Model Selection}

\begin{figure}[H]
    \centering
    
    \begin{subfigure}[b]{0.48\textwidth}
         \centering
          \includegraphics[scale=0.2]{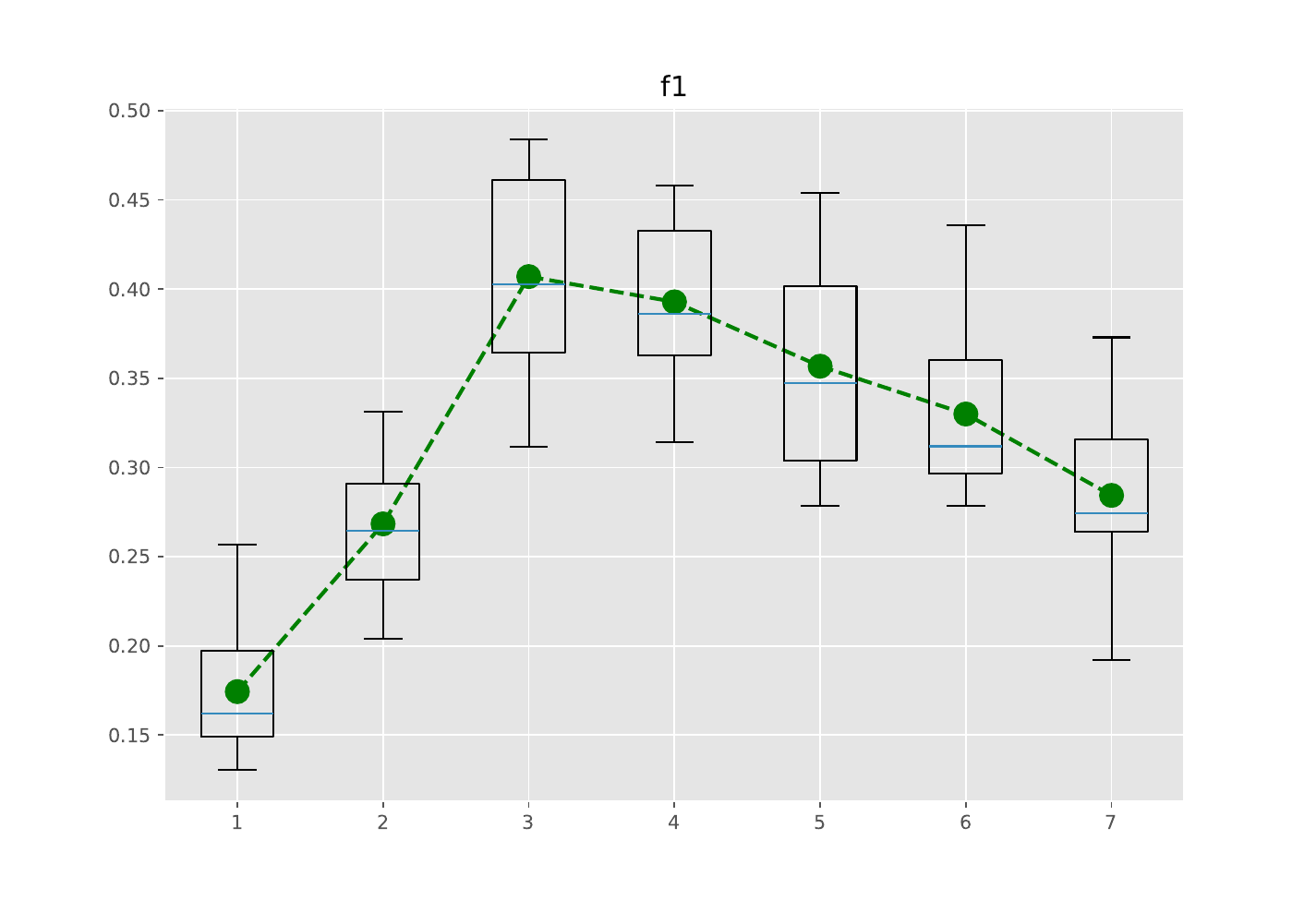}
         \caption{Validation set f1-score with different K for mixture model MM-R \ref{sec:mm2R} with prediction window $\Delta t = 6$ hours.}
         \label{fig:2RModelSelectiona}
     \end{subfigure}
     \hfill
    \begin{subfigure}[b]{0.48\textwidth}
         \centering
         \includegraphics[scale=0.2]{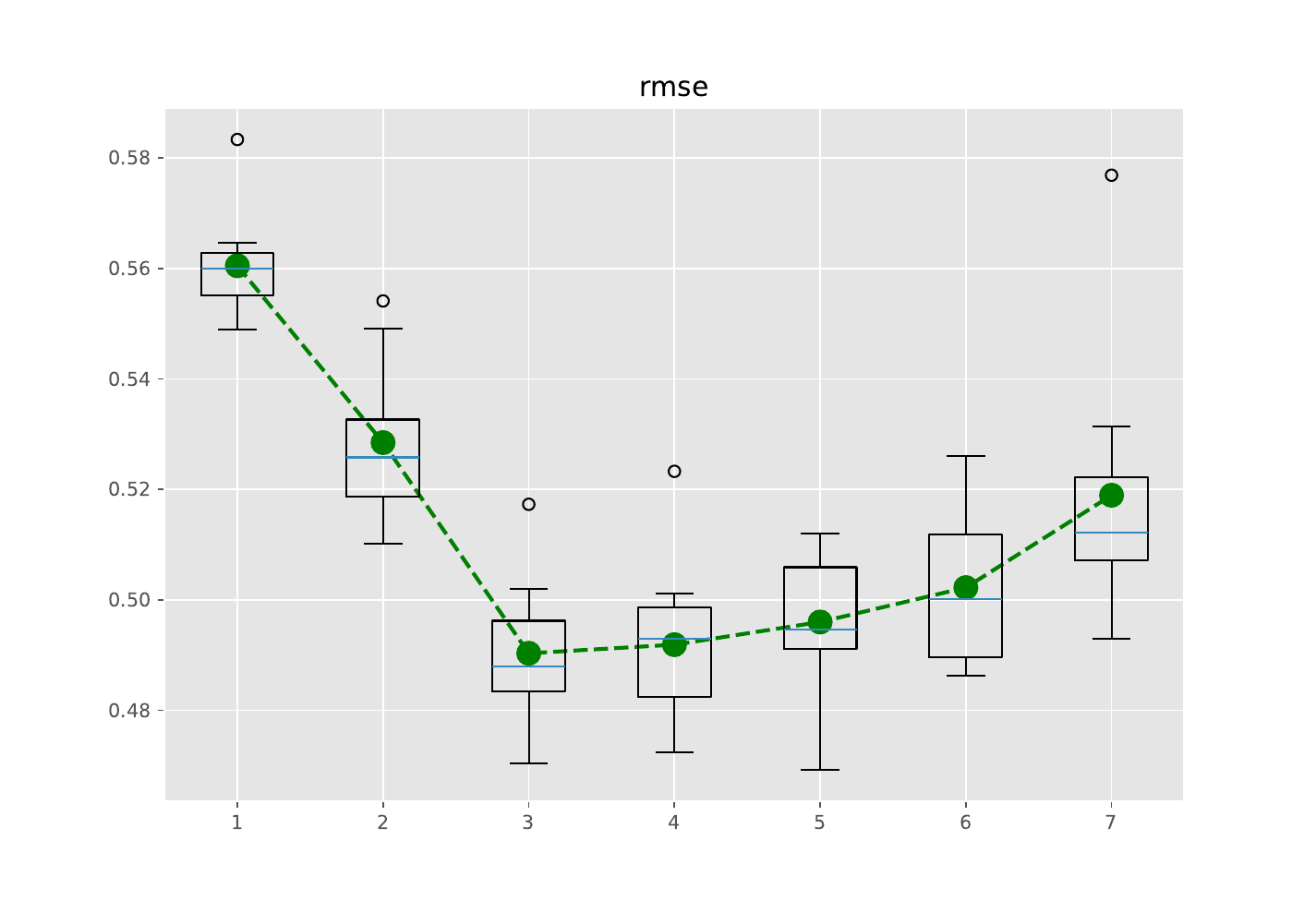}
         \caption{Validation set RMSE with different K for mixture model MM-R \ref{sec:mm2R} with prediction window $\Delta t = 6$ hours.}
        \label{fig:2RModelSelectionb}
     \end{subfigure}
    \begin{subfigure}[b]{0.48\textwidth}
         \centering
          \includegraphics[scale=0.2]{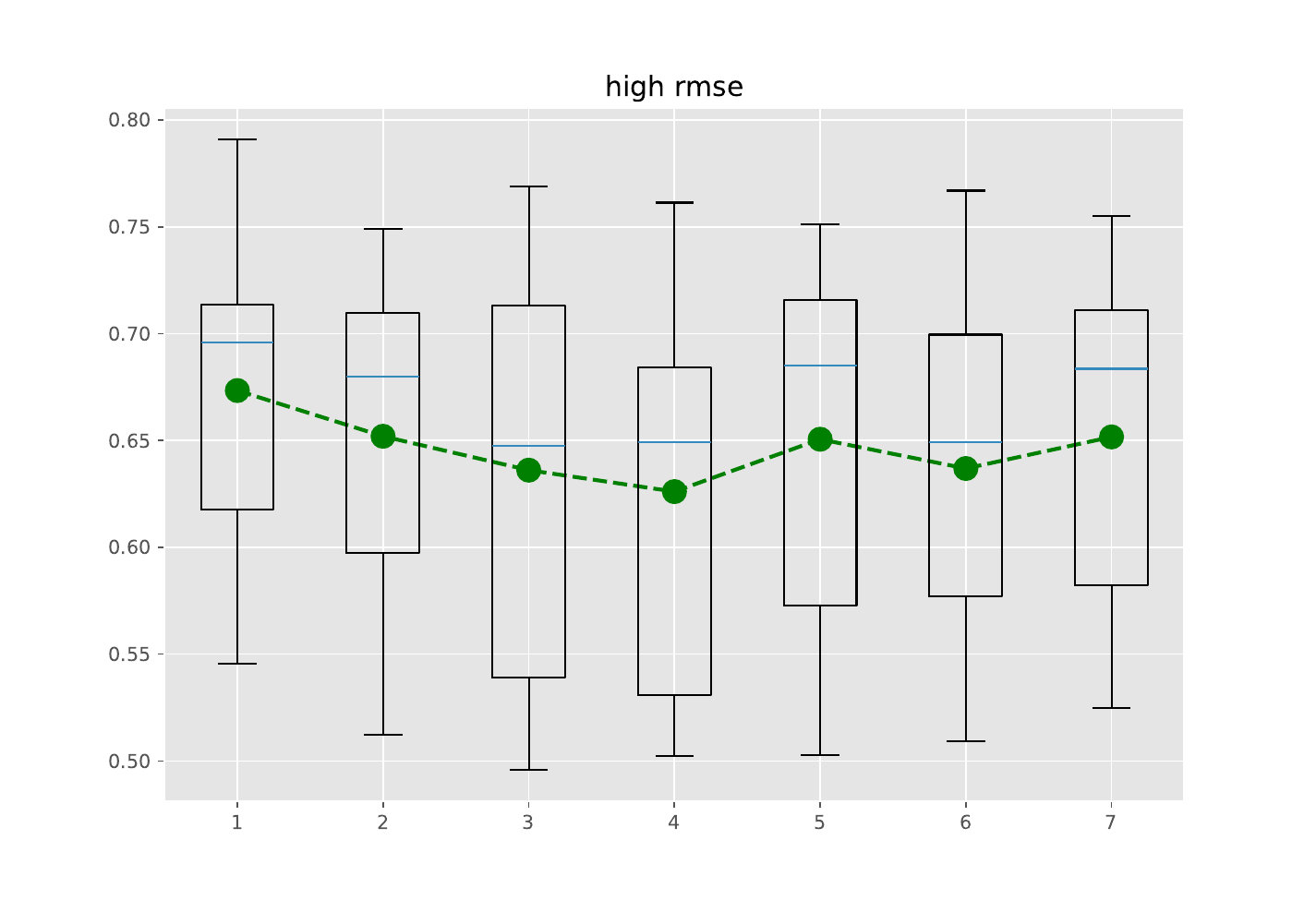}
         \caption{Validation set M/X Category RMSE with different K for mixture model MM-R \ref{sec:mm2R} with prediction window $\Delta t = 6$ hours.}
         \label{fig:2RModelSelectionc}
     \end{subfigure}
     \hfill
    \begin{subfigure}[b]{0.48\textwidth}
         \centering
         \includegraphics[scale=0.2]{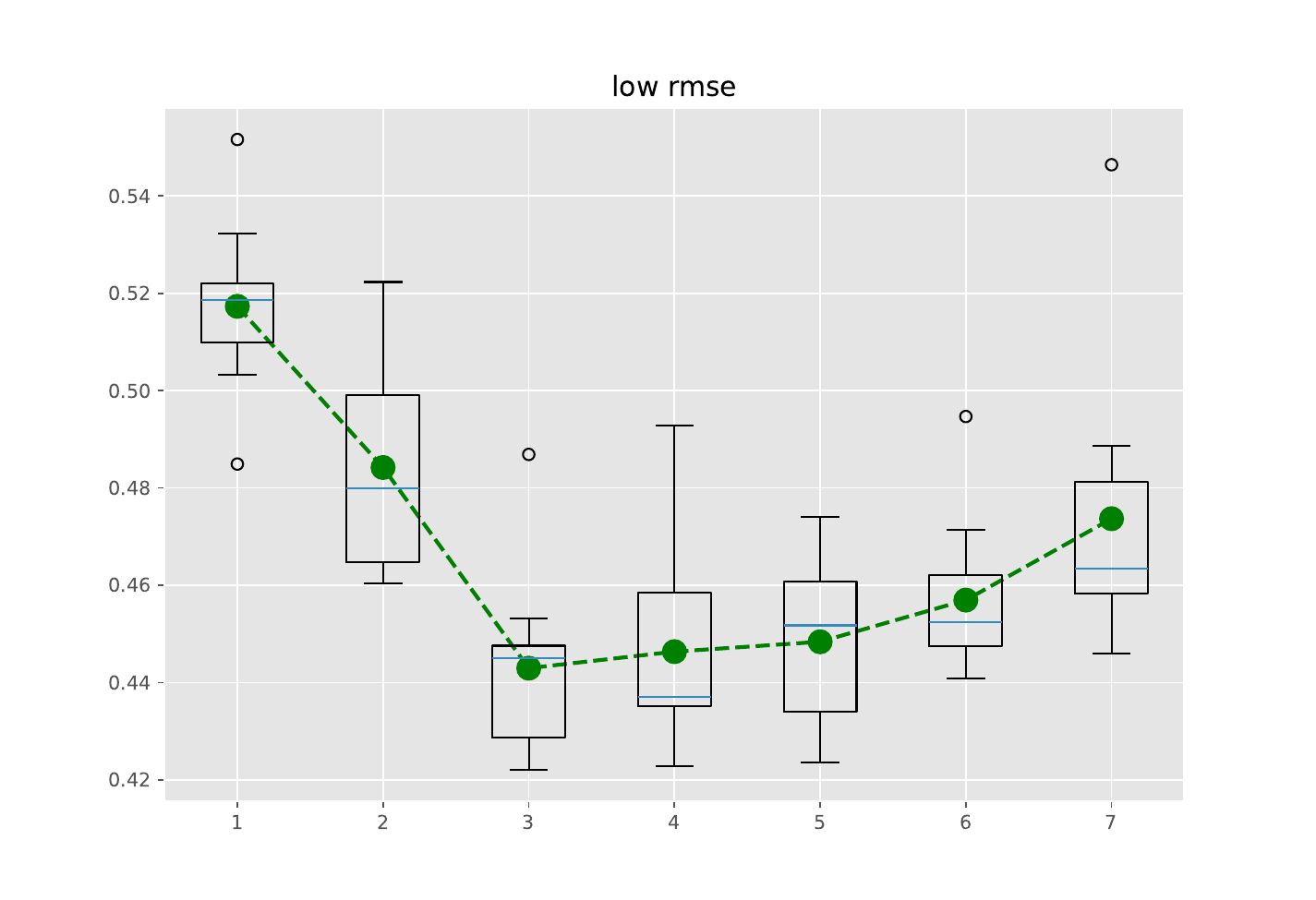}
         \caption{Validation set B/C Category RMSE with different K for mixture model MM-R \ref{sec:mm2R} with prediction window $\Delta t = 6$ hours.}
         \label{fig:2RModelSelectiond}
     \end{subfigure}
    \begin{subfigure}[b]{0.48\textwidth}
         \centering
          \includegraphics[scale=0.2]{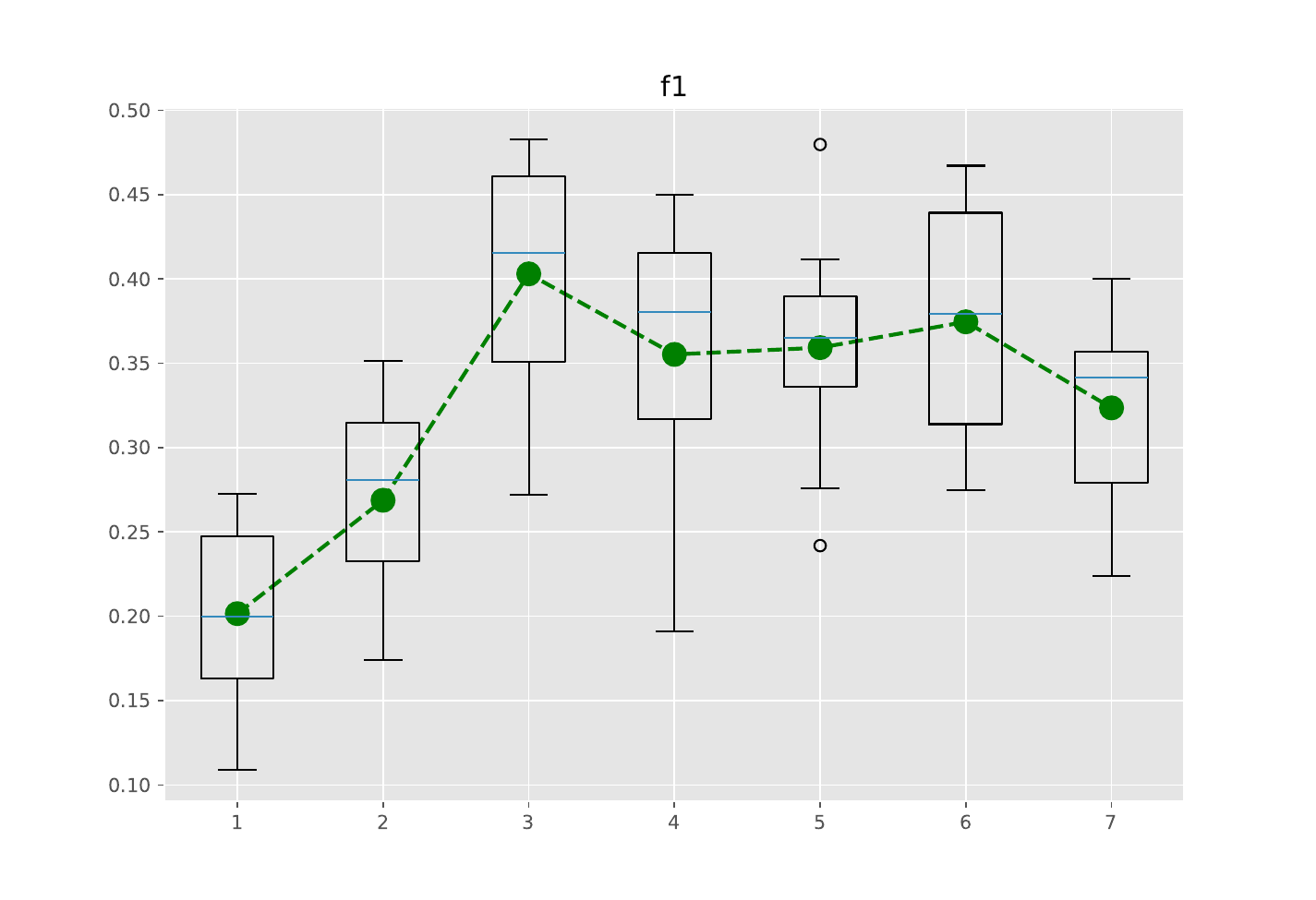}
         \caption{Validation set f1-score with different K for mixture model MM-H \ref{sec:mm2H} with prediction window $\Delta t = 6$ hours.}
         \label{fig:2RModelSelection2}
     \end{subfigure}
     \hfill
    \begin{subfigure}[b]{0.48\textwidth}
         \centering
         \includegraphics[scale=0.2]{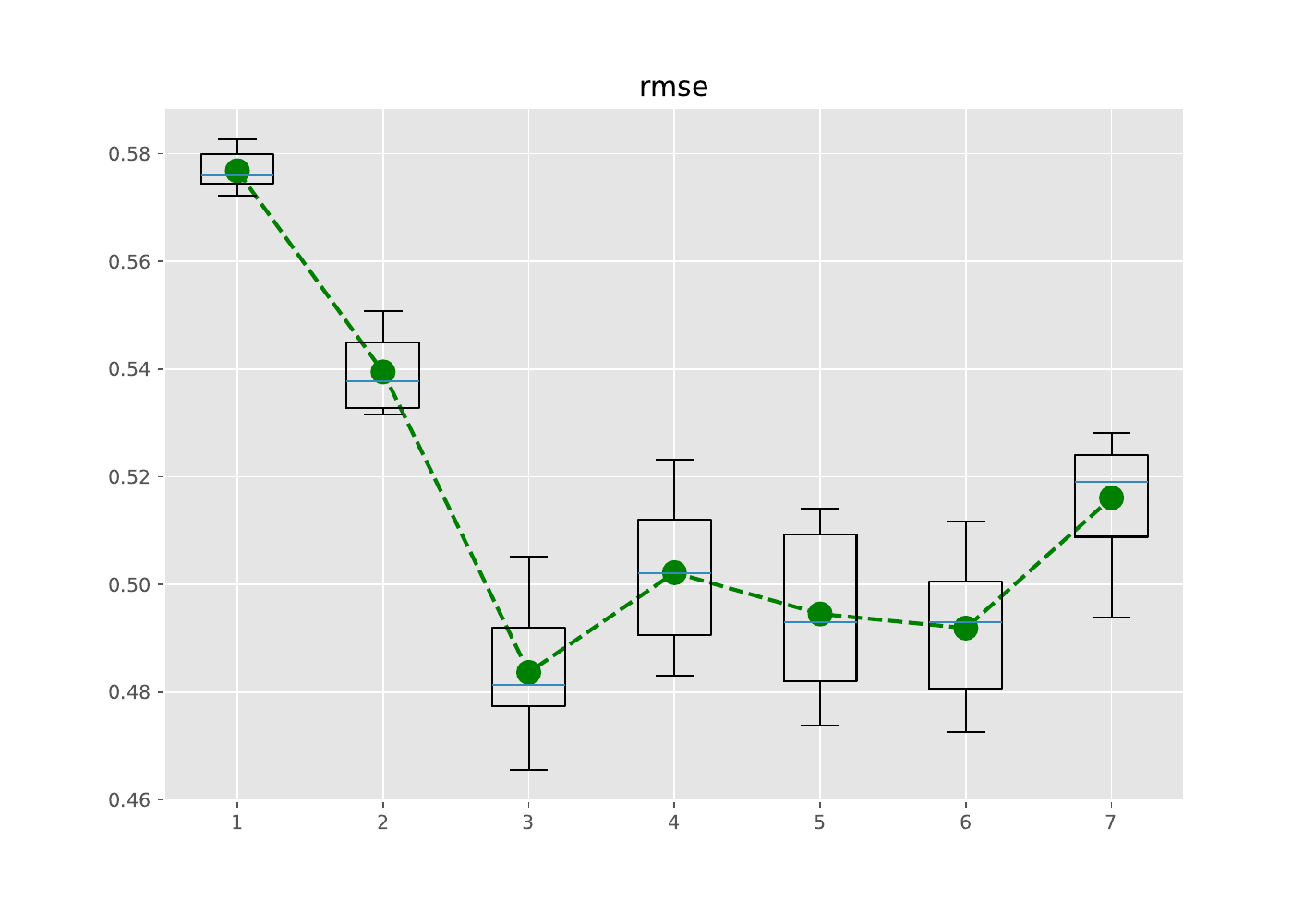}
         \caption{Validation set RMSE with different K for mixture model MM-H \ref{sec:mm2H} with prediction window $\Delta t = 6$ hours.}
    \end{subfigure}
    \begin{subfigure}[b]{0.48\textwidth}
         \centering
          \includegraphics[scale=0.2]{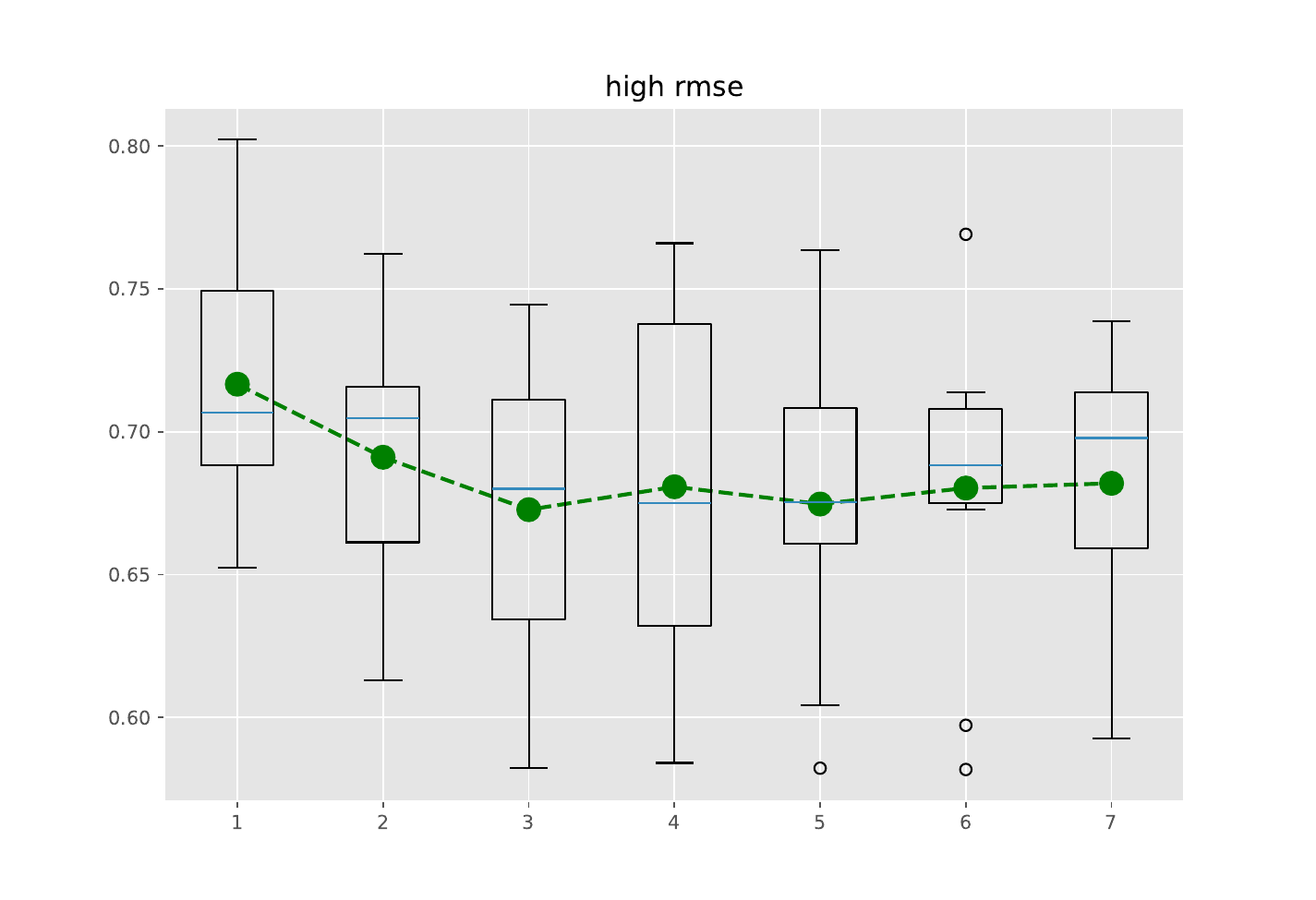}
         \caption{Validation Set M/X Category RMSE with different K for mixture model MM-H \ref{sec:mm2H} with prediction window $\Delta t = 6$ hours }
     \end{subfigure}
     \hfill
    \begin{subfigure}[b]{0.48\textwidth}
         \centering
         \includegraphics[scale=0.2]{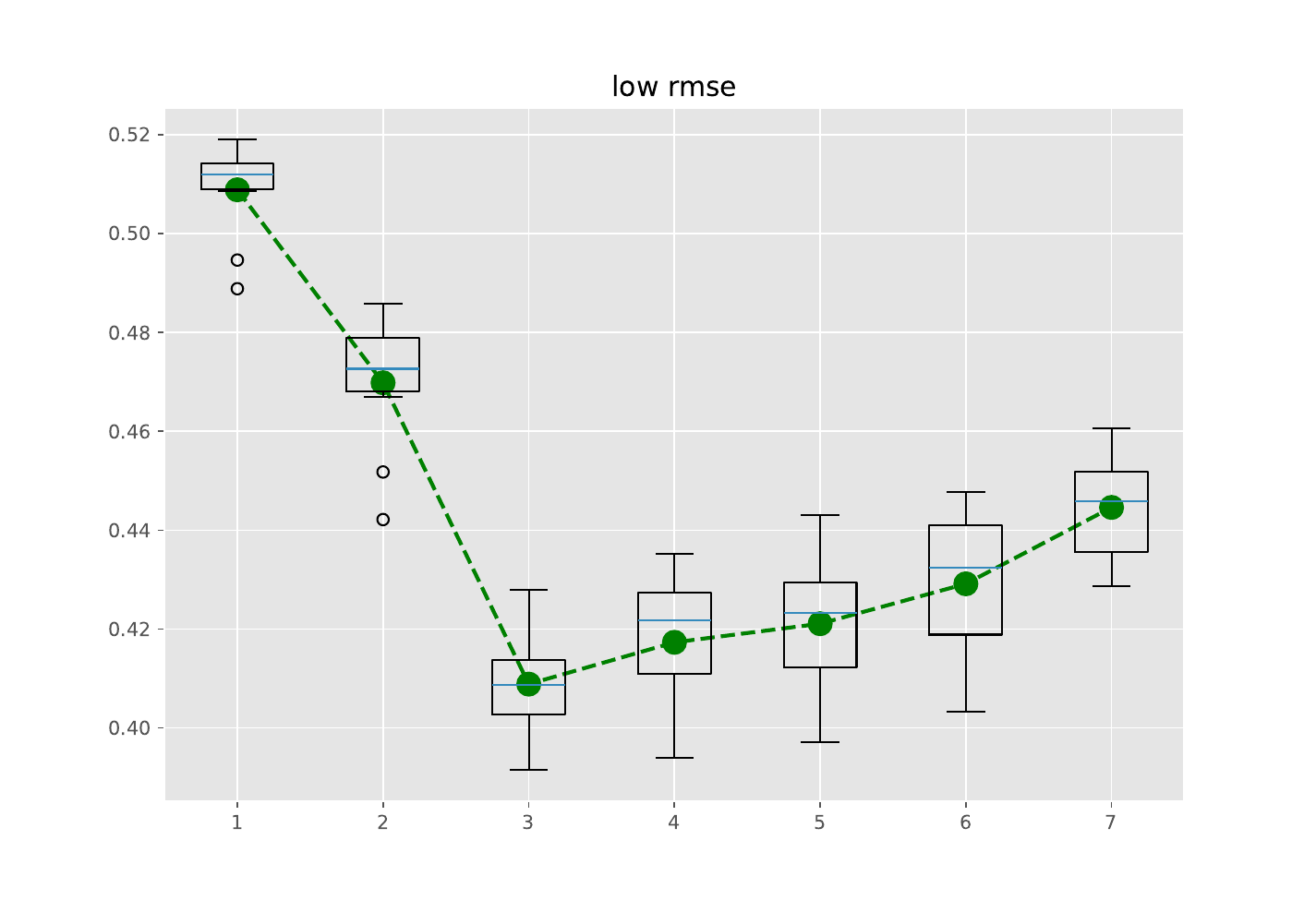}
         \caption{Validation set B/C Category RMSE with different K for mixture model MM-H \ref{sec:mm2H} with prediction window $\Delta t = 6$ hours.}
    \end{subfigure}
    \caption{Model selection's metrics for MM-R and MM-H. The x-axis is the number of mixture components to be selected.\label{selection:metric}}
\end{figure}

Note that when mixture component $K =1$, both models \ref{sec:mm2R} and \ref{sec:mm2H} reduce to a weighted linear regression model. To perform model selection, the test set is fixed and the original train set is randomly split into a sub-train set and a validation set for 100 repetitions. For each repetition, we train weighted linear regression (3.1), MM-R (\ref{sec:mm2R}), and MM-H (\ref{sec:mm2H}) on the sub-train set then apply them on the validation set to obtain the box plots in Figure \ref{selection:metric}. In the below figures, each box plot visualises the minimum, first quartile, median, third quartile and maximum of the collection of generated metrics over $100$ repetitions. The averages are also depicted as green dots. The numbers in the x-axis are the number of mixture components to be selected. For MM-R, setting $K = 3$ yields the best f1 score. Similar conclusion for $K$ can be seen from the break down of the rmse for B/C and M/X category. 

The B/C rmse is at the lowest for $K = 3, 4, 5$ and M/X rmse is at the lowest for $K = 3, 4$ as in Figures \ref{fig:2RModelSelectionc} and \ref{fig:2RModelSelectiond}. Taking all into consideration, we pick $K = 3$ for MM-R. Following a similar reasoning, we pick $K = 3$ for MM-H. Note that we combine the RMSE of M/X categories and B/C categories since the M/X represents strong flares and B/C represents weak flares. And we are most interested in early warnings of strong flares.

\subsection{Analysis of Mixture Model Fittings}

In this section, we demonstrate the clustering results of mixture model MM-R in section \ref{sec:mm2R} and MM-H in section \ref{sec:mm2H} after training on the observed data. We inspect clustering effects on the marginal space of response y, marginal space of covariate $X$ then we explore what clustering structure entails about the interactions between $y$ and $X$.  

\subsubsection{Response Space y}

\begin{figure}[ht!]
\centering
\begin{subfigure}{.48\textwidth}
  \centering
  \includegraphics[scale=0.28]{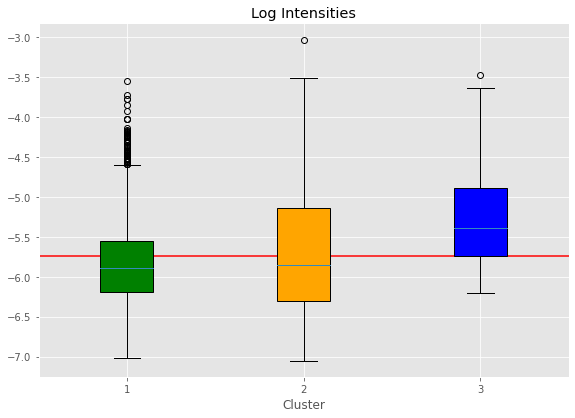}
  \caption{Mixture model over ARs MM-R (\ref{sec:mm2R})'s log intensity under each cluster.}
  \label{fig:2Rbpysigma2}
\end{subfigure}%
\begin{subfigure}{.48\textwidth}
  \centering
  \includegraphics[scale=0.45]{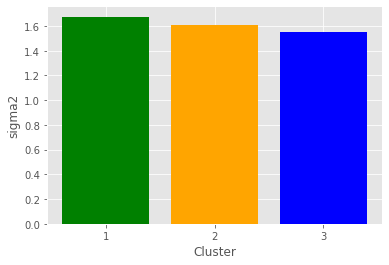}
  \caption{MM-R's estimated variance $\hat{\sigma}^2$ of each cluster.}
\end{subfigure}
\begin{subfigure}{.48\textwidth}
  \centering
  \includegraphics[scale=0.28]{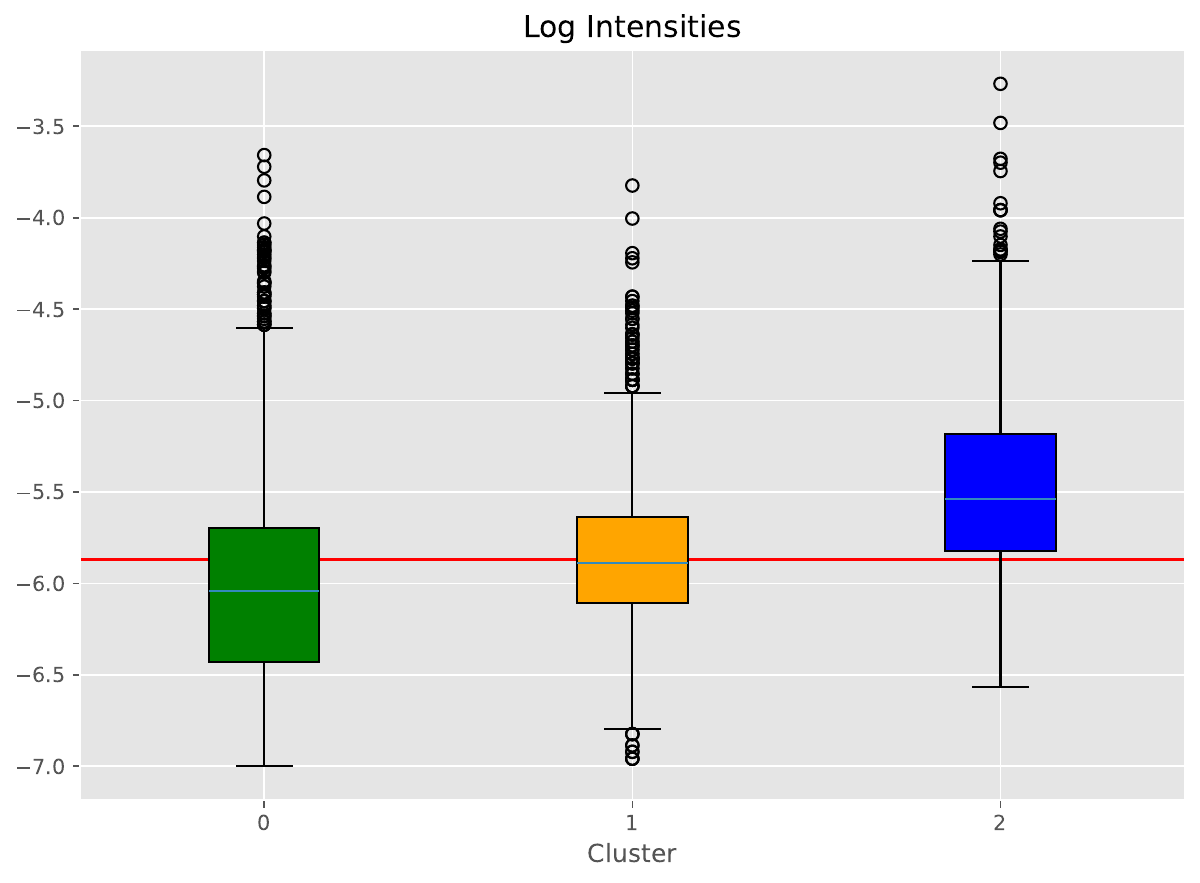}
  \caption{Mixture model over flare events MM-H (\ref{sec:mm2H})'s log intensity under each cluster.}
\end{subfigure}%
\begin{subfigure}{.48\textwidth}
  \centering
  \includegraphics[scale=0.45]{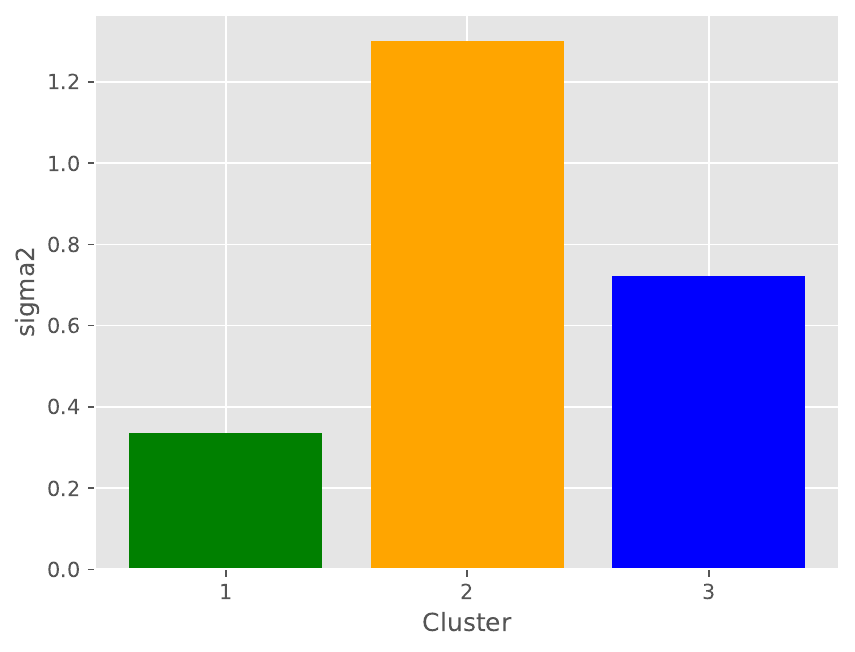}
  \caption{MM-H's estimated variance $\hat{\sigma}^2$ of each cluster.}
\end{subfigure}
\caption{MM-R and MM-H's log intensity and estimated variance of each cluster.}
\end{figure}

Examining each cluster's log intensities gives hint on the interpretation of the trained models' clustering structures. Specifically, by the above model selection procedure, we choose $K = 3$ for mixture model MM-R. The red line in Figure \ref{fig:2Rbpysigma2} is the average of the response y over the entire training dataset. The median of cluster 1 is below the line. On the other hand cluster 2's median is very close to it, and cluster 3's median is above. This suggests cluster 1 mostly consists of regions producing weak flares and cluster 3 contains those with strong flares while cluster 2 is in-between. A similar conclusion is made the mixture model over flare events MM-H with $K = 3$.

\subsubsection{Covariate space X}

\begin{figure}[!ht]
    \centering
    \begin{subfigure}{0.45\textwidth}
        \centering
        \includegraphics[scale=0.22]{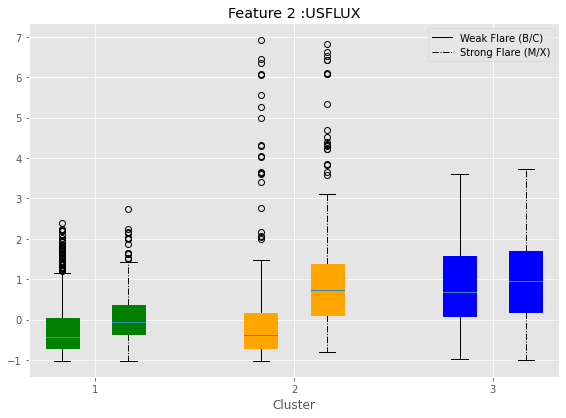}
        \caption{Total unsigned flux in Maxwells.}
        \label{fig:2Rf2}
    \end{subfigure}
    \begin{subfigure}{0.45\textwidth}
        \centering
        \includegraphics[scale=0.22]{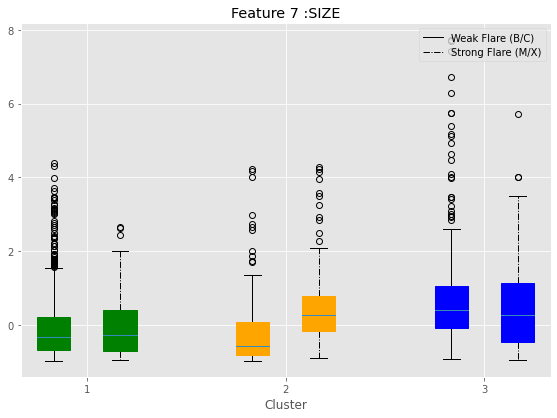}
        \caption{Projected area of patch on image in micro-hemisphere.}
        \label{fig:2Rf7}
    \end{subfigure}

    \begin{subfigure}{.45\textwidth}
        \centering
        \includegraphics[scale=0.22]{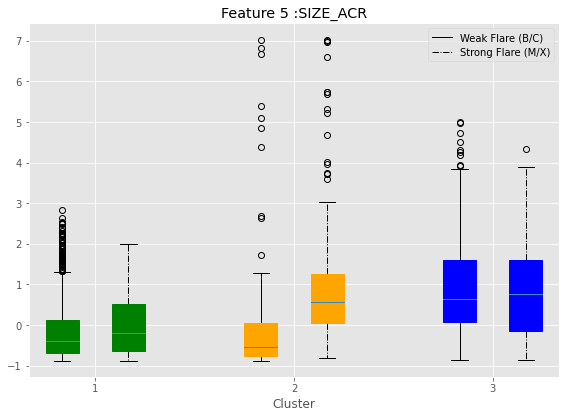}
        \caption{Projected area of active pixels on image in micro-hemisphere.}
        \label{fig:2Rf5}
    \end{subfigure}%
    \begin{subfigure}{0.45\textwidth}
        \centering
        \includegraphics[scale=0.22]{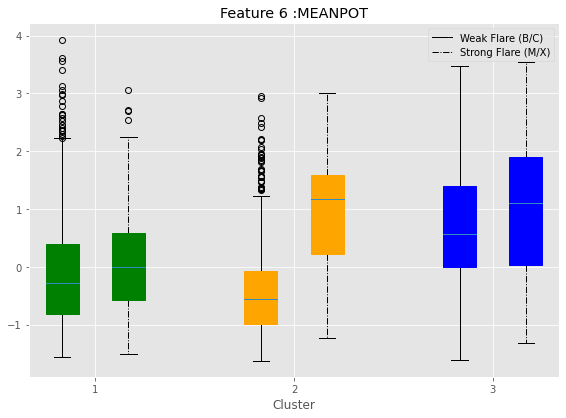}
        \caption{Mean photospheric excess magnetic energy density in ergs per cubic centimeter.}
        \label{fig:2Rf6}
    \end{subfigure}

    \begin{subfigure}{.45\textwidth}
        \centering
        \includegraphics[scale=0.22]{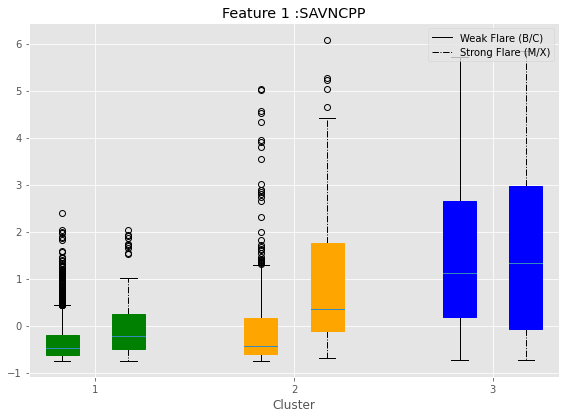}
        \caption{Sum of the Absolute Value of the Net Currents Per Polarity in Amperes.}
        \label{fig:2Rf1}
    \end{subfigure}%
    \begin{subfigure}{0.45\textwidth}
        \centering
        \includegraphics[scale=0.22]{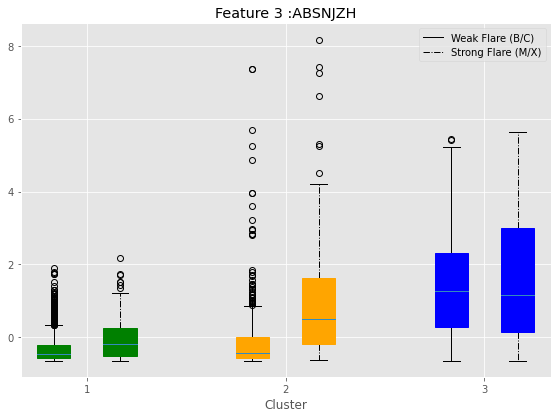}
    \caption{Absolute value of the net current helicity in G2/m.}
        \label{fig:2Rf3}
    \end{subfigure}

    \caption{Selected subset of covariate $X$ under each cluster of MM-R.}
\end{figure}

The same clustering interpretation can be extracted by inspecting covariate values for each cluster in Figures \ref{fig:2Rf2}, \ref{fig:2Rf7}, \ref{fig:2Rf5}, \ref{fig:2Rf6}  \ref{fig:2Rf1}, \ref{fig:2Rf3}. It has been scientifically observed that the flare intensities are connected to active region's magnetic properties. Using model MM-R, by inspecting the relevant magnetic covariate features we found that cluster 1 has lower numeric values under both weak (B/C) and weak strong (M/X) events compared to the others. In contrast, cluster 3 has the highest. This further corroborates the interpretation that cluster 1 with mainly weak flare events, cluster 3 populated with strong events, and cluster 2 in between. Similarly, we can reach the same conclusion for the model over flare events in MM-H; the corresponding plots are provided in the appendix.

\subsubsection{Interaction between covariate $X$ and response $y$}

Under the standard multivariate linear regression setting, the coefficient $\beta \in \mathbb{R}^d$ tells us how much the response y is expected to increase when an independent variable increases by one unit, holding all other independent variables constant. For heterogeneous regression responses, the interaction between covariates $X$ and $y$ is somewhat more delicate. In particular, for MM-R, there are three global coefficient parameters $\beta_1, \beta_2, \beta_3$. Now, how do we interpret them and talk about the interaction between y and  X?  

\begin{figure}[!ht]
    \centering
    \begin{subfigure}{.9\textwidth}
        \centering
        \includegraphics[scale=0.4]{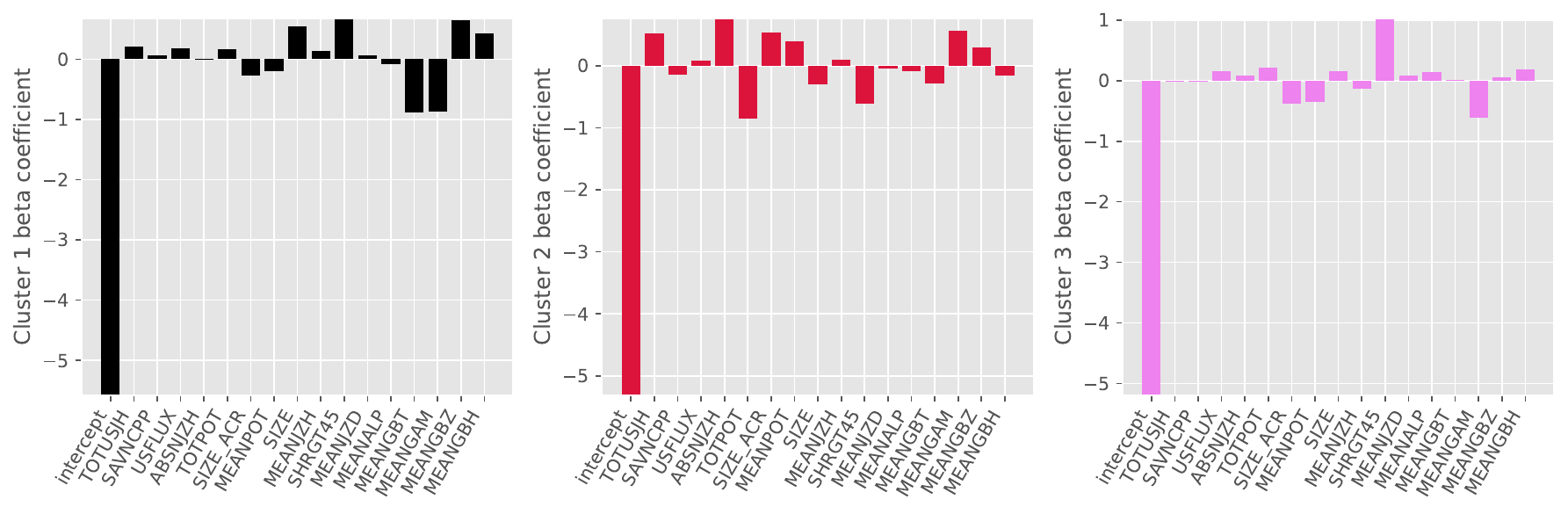}
        \caption{Mixture Model over ARs (\ref{sec:mm2R}) estimated $\{ \hat{\beta} \}_1^K$.}
        \label{fig:2Rbeta}
    \end{subfigure}%
    
    \begin{subfigure}{.9\textwidth}
        \centering
        \includegraphics[scale=0.4]{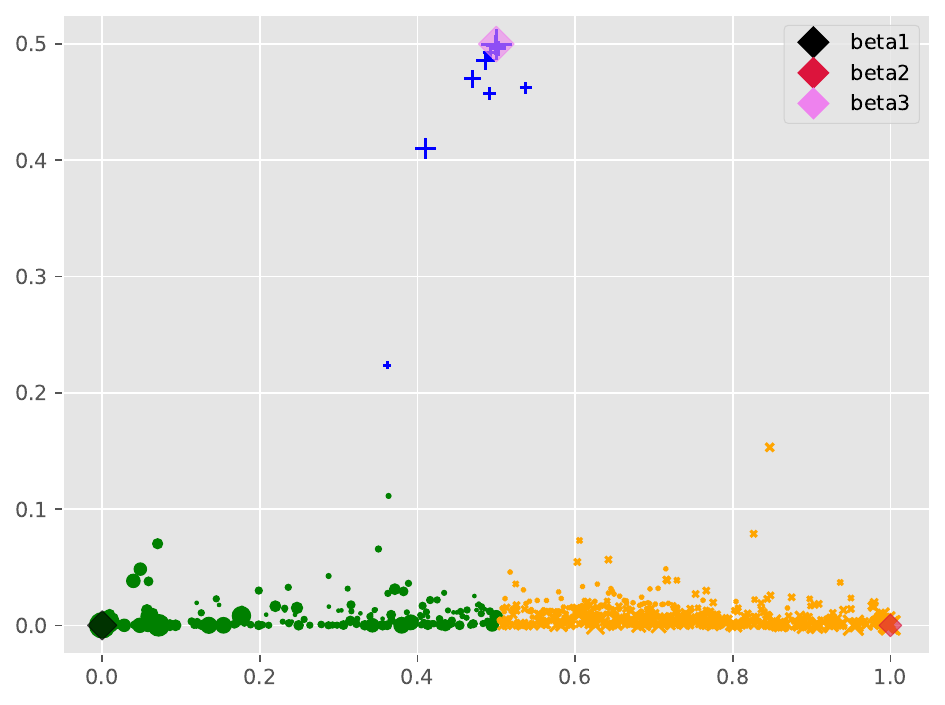}
        \caption{Mixture Model over ARs (\ref{sec:mm2R}) $\{ \hat{\beta} \}_1^K$'s mixing proportions for each active region r.  This plot shows the influence of each of the K linear mechanisms on ARs. The closer an AR to a vertex of the simplex, the stronger its influence.}
        \label{fig:2Rcluster}
    \end{subfigure}%

    \caption{Mixture model over ARs estimated $\{\hat{\beta}_{1}^K\}$ and each active region's mixing proportions of these values.}
\end{figure}

Recall that under MM-R, the predicted $\hat{y}$ for flare event $i$ in AR $r$ is given by

$\begin{aligned}
\hat{y}_i^r &= \sum_{k=1}^K \tau_k^r \beta_k^T X_i^r = \left[ \sum_{k=1}^K \tau_k^r \beta_k\right]^T \cdot X_i^r \\  
&=  \left[\sum_{k=1}^K p(z^r = k | X_1^r, \ldots, X_{n_r}^r, y_1^r, \ldots, y_{n_r}^r) \beta_k \right]^T \cdot X_i^r =: (\beta^r)^T X_i^r. \\
\end{aligned}$

Thus, under active region r, $\beta^r$ tells how much the response $y_i^r$ is expected to increase by increasing an independent variable by one unit, holding other variables fixed. Also, $\{\tau_k^r\}_1^K$ controls the influence of each of the $\beta_k$ on $\beta^r$, the interaction coefficient between y and X for region r. If we visualize $\beta_1, \beta_2$ and $\beta_3$ as three extreme points $e_1, e_2, e_3 = (0,0), (1,0), (0,1)$ in a 2D unit simplex and plot each AR r at the coordinate $\tilde{c}^r := \sum_{k=1}^3 \tau_k^r e_k$ (note that all flare events i under AR r have the same $\beta^r$), the position of AR r in the simplex indicates visually how much it is influenced by each of the global $\{ \beta_k \}_{k=1}^3$. Furthermore, we can inspect the mixture component assignments for all active regions under the trained model MM-R. Figure \ref{fig:2Rcluster} shows how the coefficient $\beta^r$ of each AR r under the influence of cluster 1 (green), 2 (yellow) and 3 (blue). Each AR is visualized as a colored dot in the unit simplex, and its size corresponds to the number of flare events recorded in that AR. In the figure \ref{fig:2Rcluster}, observe that active regions under cluster 1 (green) are mainly affected by $\hat{\beta}_1$ but $\hat{\beta}_2$ still plays some role while $\hat{\beta}_3$ has a negligible impact. By the same manner of reasoning, cluster 2 (yellow) is mainly influenced by $\hat{\beta}_2$, and $\hat{\beta}_1$ plays a relatively minor role. Cluster 3 (blue) is mostly affected by parameters $\hat{\beta}_3$. Figure \ref{fig:2Rbeta} shows the magnitudes of each features for $\hat{\beta}_1, \hat{\beta}_2$ and $\hat{\beta}_3$. Each has 17 bars. The first bar is the linear regression intercept coefficient, and the following bars are the coefficients of features TOTUSJH, SAVNCPP, USFLUX, ABSNJZH, TOTPOT, SIZE\_ACR, MEANPOT, SIZE, MEANJZH, SHRGT45, MEANJZD, MEANALP, MEANGBT, MEANGAM, MEANGBZ, MEANGBH. Their descriptions are in the table \ref{tab:sharp}. The same line of interpretation can be applied for MM-H of which similar plots are included in the appendix's figures \ref{fig:2Hbeta} and \ref{fig:2Hbetacluster}.

\subsection{Prediction Performance}

Note again that when the number of mixture component $K = 1$, models \ref{sec:mm2R} and \ref{sec:mm2H} are just the weighted linear regression model. To produce the figures in  tables \ref{tab:lr} \ref{tab:wlr}, \ref{tab:mm2R}, \ref{tab:mm2H}, we train standard linear regression, weighted linear regression, model MM-R (\ref{sec:mm2R}) and MM-H (\ref{sec:mm2H}) on the training data as described in the data section. We run MM-R and MM-H for 100 replications under each of which train and test are randomly split as described in section 2, then take the averages as final results. For each replication, we run our EM procedures 5 times and select the one with highest likelihood as the fitted model. We use the validate set to pick the best number of component K. Even though the models predict continuous responses and we can assess their root mean square errors (rmse), to help illustrate the models' performances against data imbalance of which rmse are not particularly helpful, we compute f1-score. Since this is a classification metric, continuous responses need to be converted into binary outcomes. We use the same validation set to pick the best binary splitting thresholds. The estimated binary responses are compared with the true labels of strong flare (M/X) and weak flares (B/C) in the data.

\begin{table}[!ht]
\centering
\begin{tabular}{|c||c|c|c|c|c|c||} 
 \hline
 \textit{metrics/Prediction window} & 6h & 12h & 24h & 36h & 48h \\
 \hline\hline
  \textbf{rmse} & 0.4543 & 0.4535 & 0.4569 & 0.471 & 0.46945 \\
  \textbf{accuracy} & 0.9315  & 0.9310  & 0.9336 & 0.9340  & 0.8972 \\
  \textbf{precision} & 0.28  & 0.200 & 0.16 &  0.15 & 0.3333\\
  \textbf{recall} & 0.051  & 0.0294 & 0.032 & 0.03  &  0.03508 \\
  \textbf{f1} &  0.086  & 0.0512 & 0.0533  & 0.061 & 0.06349 \\
 \hline
\end{tabular}
\caption{(Unweighted) Linear Regression's Test Performance.}
\label{tab:lr}
\end{table}

\begin{table}[!ht]
\centering
\begin{tabular}{|c||c|c|c|c|c|c||} 
 \hline
 \textit{metrics/Prediction window} & 6h & 12h & 24h & 36h & 48h \\
 \hline\hline
  \textbf{rmse} & 0.5873 & 0.5383 & 0.5572  &  0.5646 & 0.5798 \\
  \textbf{accuracy} & 0.8613  &  0.8990 & 0.896 & 0.8971 & 0.8972 \\
  \textbf{precision} & 0.3563  & 0.2158 & 0.2265 & 0.1840 & 0.2054 \\
  \textbf{recall} & 0.37825  & 0.2729 & 0.2622 & 0.2116 & 0.2212  \\
  \textbf{f1} &  0.3639  & 0.2407 & 0.2428 & 0.1967 & 0.212 \\
 \hline
\end{tabular}
\caption{Weighted Linear Regression's Test Performance.}
\label{tab:wlr}
\end{table}

\begin{table}[!ht]
\centering
\begin{tabular}{|c||c|c|c|c|c|c||} 
 \hline
 \textit{metrics/Prediction window} & 6h & 12h & 24h & 36h & 48h \\
 \hline\hline
  \textbf{rmse} & 0.4849 & 0.5140  & 0.5539  &  0.5419 & 0.5074  \\
  \textbf{accuracy} & 0.8487  & 0.8745  & 0.8729  & 0.8497  &  0.8574 \\
  \textbf{precision} & 0.3636 & 0.8594  & 0.2156  & 0.1959   & 0.2  \\
  \textbf{recall} & 0.4057 & 0.3839  & 0.3882 & 0.3372  & 0.2970  \\
  \textbf{f1} &  0.383  & 0.2935  &  0.2773  & 0.2479   & 0.2390  \\
 \hline
\end{tabular}
\caption{Mixture Model over ARs  (\ref{sec:mm2R})'s Test Performance. Note f1 score is computed based on \newline $K = 3, 3, 3, 3, 5$ and classification threshold = $-5.0, -5.0, -5.05, -5.1, -5.15$.}
\label{tab:mm2R}
\end{table}

\begin{table}[!ht]
\centering
\begin{tabular}{|c||c|c|c|c|c|c||} 
 \hline
 \textit{metrics/Prediction window} & 6h & 12h & 24h & 36h & 48h \\
 \hline\hline
  \textbf{rmse} & 0.4732 & 0.5058 & 0.5321 & 0.5338  & 0.5586  \\
  \textbf{accuracy} & 0.8433  & 0.8621  & 0.8663  & 0.84543 &  0.8485 \\
  \textbf{precision} & 0.3446  & 0.2486  & 0.2344 & 0.2 & 0.2 \\
  \textbf{recall} & 0.4693  & 0.4017 & 0.4 & 0.4875  & 0.3366 \\
  \textbf{f1} & 0.3933   & 0.3071 & 0.29 & 0.2708 & 0.2509  \\
 \hline
\end{tabular}
\caption{Mixture Model over Flare Events (\ref{sec:mm2H})'s Test Performance. Note f1 score is computed based on $K = 3, 3, 3, 6, 5$ and classification threshold = $-5.0, -5.0, -5.05, -5.05, -5.1$.}

\label{tab:mm2H}
\end{table}

The numbers in table \ref{tab:lr}, \ref{tab:wlr}, \ref{tab:mm2R}, \ref{tab:mm2H} show that Mixture Model MM-H (section \ref{sec:mm2H}) and MM-R perform similarly though the former is marginally better. MM-R and MM-H (section \ref{sec:mm2H}) significantly outperform the weighted linear regression. This result implies that adding more components improves the performance and thus supports the heterogeneity nature of the data. Model MM-H offers more flexibility by extending the heterogeneity pattern to individual flare events. But interestingly, this flexibility does not noticeably improve the defined metrics. This suggests the heterogeneity signal is most noticeable at the active region level. We also observe that performance degrades as the predicting time windows increase, which is expected. Finally, we remark that our predictive performance in the f1 metric ($0.383$) for 6-hour data is lower than in \citet{2019SW...17...10}. This is not surprising, as we assume linear relationships between covariates and responses and do not account for the temporal nature of solar flares, where future events may correlate with past ones. In contrast, Chen et al. (2019) constructed a sophisticated LSTM neural network to extract complex non-linear signals from the data. We describe potential future directions for improving model performance in the last section. However, the main contribution of this work is demonstrating how mixture models can cluster solar active regions based on the interaction mechanisms between their SHARP covariates and corresponding intensity responses, thus characterizing the heterogeneous nature of active regions.

\subsection{Case Studies}

Both models MM-R \ref{sec:mm2R} and MM-H \ref{sec:mm2H} perform similarly regarding f1 metrics even though MM-H is marginally better. As mentioned in the last section, this result seems to suggest flares from the same AR are intrinsically homogeneous in nature or homologous as known to the solar physics community \cite{Manchester2003,Sui2004,Liu2014,Romano2018} In contrast, individual ARs are most often heterogeneous. As such, we use MM-R for the case studies in this section. As stated in the introduction, the goal of this project is to characterize the heterogeneity among active regions (AR). The fitted models' cluster membership is particularly interesting because the model MM-R groups similar active regions together in each cluster. \\

Because there is randomness in the way the training and testing sets are created, mixture memberships might change in different replications of model fitting procedure. Thus to analyze the cluster assignment in a robust manner, we run the model fitting procedure for MM-R for $100$ repetitions. In each repetition, active regions are allocated to different mixture clusters. To standardize the meaning of clusters across repetitions, we assign the label `H', `L', `I' to mixture clusters of which the first quartile of its collections of log intensities is greater than $-5.75$, smaller than $-6.0$ and between $[-6, -5.75]$ respectively. By the design of the log intensity threshold for `H', `I' and `L', active regions allocated to 'H' labels should be reasonably active in terms of strong flare events and active regions to 'L' should be relatively quiet and 'I' in-the-between. \\

We observe that some active regions have the same labels for all $100$ repetitions. For example, ARs $11124$ and $11109$ are assigned to the label `L' in every one of the $100$ iterations. On the other hand, other active regions might be allocated to different labels over $100$ repetitions. For instance, AR $11967$ is assigned to cluster `H' $87$\% and `I' $13$\% of the $100$ repetitions or AR $12192$ $85$\% `H' and $15$\% `I'. The reason why an active regions may have different labels for different repetitions has to do with the randomness of data splitting. In particular, AR $11153$ has 28 flare events of which only one is M class and the rest are B and C ones. As such, if the M class event is included the training, it will skew the average log intensities higher than otherwise, and so affects its mixture cluster assignment. \\

The complete list of Active Region's membership is provided in the Supplementary. Here we mention the top-two ARs for each of the label `H', `I' and `L' in terms of the highest number of label assignment in $100$ repetitions.  Regarding 'L' label, ARs $11117$ and $11109$ are assigned to `L' $99$\% and $100$ \% of the times. For `H', ARs $11967$ and $12242$ are assigned to `H' $87$\% and $88$\% of the times. For `I', ARs $11261$ and $11087$ are allocated 'I' $62$\% and $59$\% of the $100$ repetitions. There are existing works in the space weather literature that corroborate that ARs $11967$ and $12242$ were known to produce strong flares \cite{Joshi2021, Duran2020} and \cite{Solovev2019, Maximov2019} respectively. The common trait of ARs with `I' label is that they have few strong flares among a majority of quiet flare events. In contrast, ARs with `L' labels have only quiet flare events. \\

\begin{figure}[ht!]
     \centering
      \begin{subfigure}[b]{0.9\textwidth}
         \centering
         \includegraphics[scale = 0.35]{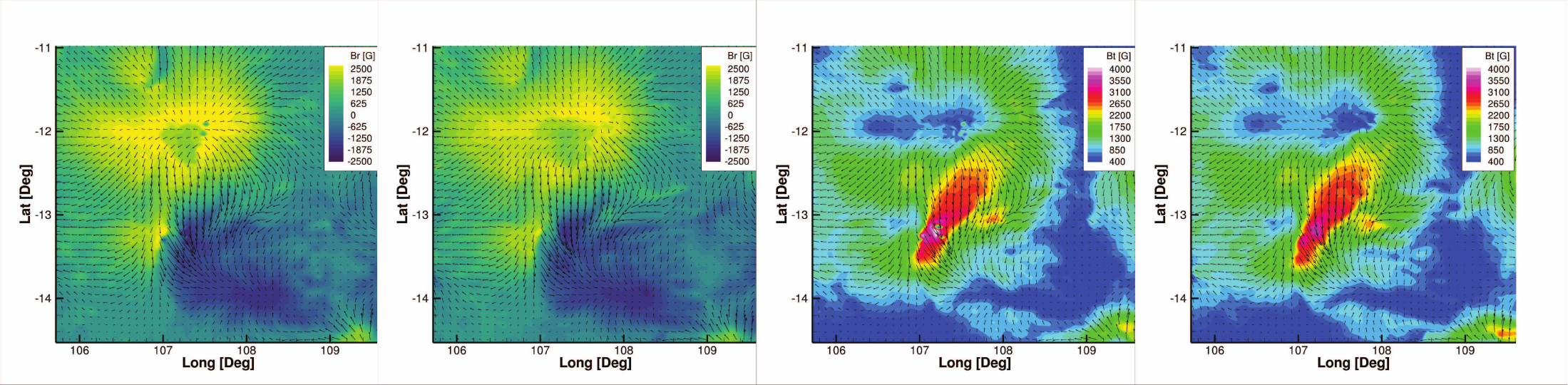}
        \caption{The left two images are radial components and right two ones are horizontal field components in AR 11967 (H label) at its strongest (2014.01.30\_16:11:00) and weakest flare events (2014.01.30\_13:36:00) during its existence. Arrows show the direction and relative magnitude of the horizontal magnetic field component.}
     \label{fig:11967weakhmi}
     \end{subfigure}
     \hfill
     \begin{subfigure}[b]{0.9\textwidth}
        \centering
        \includegraphics[scale = 0.18]{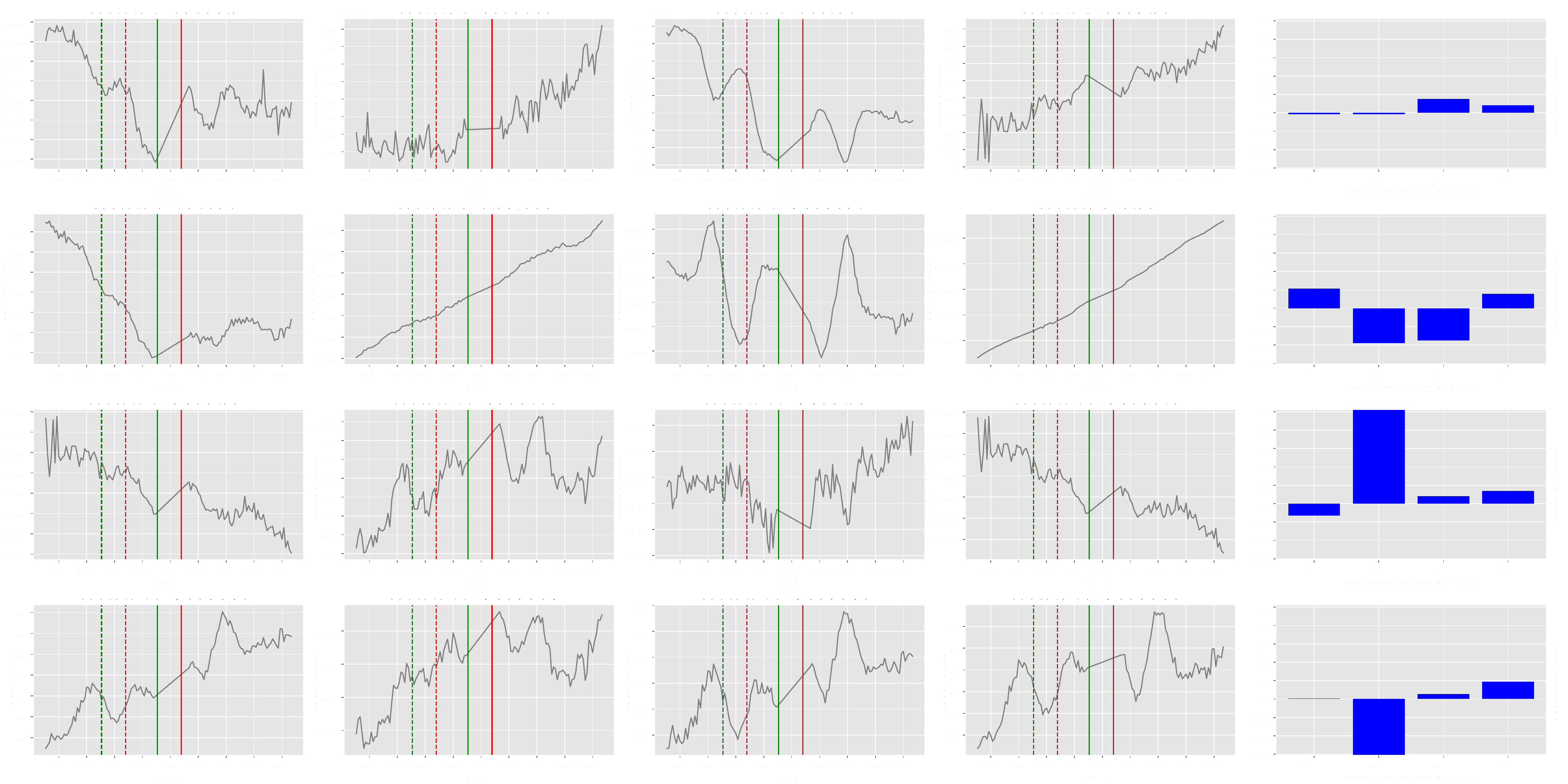}
        \caption{(Left) Evolution of SHARP parameters during the strongest (red) and weakest (green) flare events in AR 11967 (H label) and (Right) MM-R estimated $\hat{\beta}_k$ values. The dash lines are the times at their peak intensities. The solid lines are the 6-hour-before for covariates' regression. Green color is the weakest flare and red is the strongest one.}
        \label{fig:11967joinedts}
     \end{subfigure}
     \caption{Active Region 11967's SHARP parameters and HMI image during its strongest flare event and bar plots of regression coefficients on the right. AR 11967 existed from 2014/01/30 to 2014/02/08. The dash lines are the times at their peak intensities. The solid lines are the 6-hour-before for covariates' regression.}
     \label{fig:11967event}
\end{figure}

To complete our case study, we provide the HMI image and the temporal evolution of each SHARP parameters for the strongest and weakest flare events for AR $11967$ during its existence from $2014$, Jan $27$ to $2014$, Feb $09$ in Figures \ref{fig:11967event}. Refer to the appendix \ref{appendix:plots} for the same plots for all ARs $11117, 11261, 11967$ and $12242$. \\

All flaring active regions share basic features in common, which is a non-potential magnetic field forming a filament channel over a well-defined polarity inversion line \citep{Green2018}.  Beyond this basic feature, there are many possible avenues to eruption. Here, we summarize the observed magnetic structure and evolution of cluster members to determine if there are features or processes responsible for the heterogeneity of the the mixture models. The H cluster contains numerous X-class flares which received considerable attention in the published literature.   The conditions of AR 12242 leading to the X1.8 flare on 2014 December 20 is particularly well described.  Maximov et al. \cite{Maximov2019} describes the convergence of magnetic flux toward the polarity inversion line leading to a both a local and total maximum gradient of the magnetic field at the time of the flare. Solovev et al. \cite{Solovev2019} goes further to describe the formation of a magnetic flux rope by reconnection between the converging/colliding sunspots, which erupts to produce the flare.  This flaring process has been developed by numerous authors e.g. \cite{Chintzoglou2019,  Liu2020}.  In the case of AR 11967, a series of flares occurred at a sunspot light bridge, a region of extremely intense and highly sheared magnetic fields produced by flux emergence \cite{Duran2020, Kawabata2017}.  In both example, we find an intensification of the magnetic field. \\

For the I cluster, we again find well described events showing consistent patterns of evolution leading to the flares.  The strongest flare from AR 11087, a C2.7 event, occurred on 2010 July 13 10:51:00 UT is described in Joshi et al. \cite{Joshi2015}. The flare is characterized by the activation and partial eruption of an active region filament that produces a pair of flair ribbons. Another member of the I cluster, AR 11261, produced a series of flares including four M-class events, which arose from a complex system of four sunspots, one being in a delta configuration \cite{Thalmann2016}.  Ye et al. \cite{Ye2018} found that shearing of the photospheric magnetic field associated with flux emergence was a key driver of flares. These observations are consistent with Lorentz force driven shear flows powering solar eruptions \cite{Manchester2001, Manchester2003, Manchester2004, Manchester2007, Fang2010}. Similarly, Sarkar et al. \cite{Sarkar2019} found that the free energy necessary for flares from active region 11261 was provided by the shearing motion of moving magnetic features of opposite polarities near the polarity inversion line. The authors also found patterns in the time-evolution of the net Lorentz force associated with solar flares. \\

As stated earlier, the L cluster is dominated by weak flares, which poses two challenges.  First, the events are so low in energy they often occur without significant changes in the photospheric magnetic field and without clear precursors. Second, these low-energy events are far less documented in the literature. However, active region 11117 is an exception being described in detail in a series of papers by Jiang et al.\cite{Jiang2012, Jiang2016, Jiang2017}, which we recount here.  This region produced a series of small B-class flares observed on 2010 October 25 (a date in between our strong and weak flare events) culminating in a C2.3-class event.  As described in \cite{Jiang2012}, the coronal loops of active region 11117 (observed by AIA-171) remained largely unchanged but flare reconnection was observed at the location of a magnetic null derived from their nonlinear force-free field (NLFFF) model. This eruption event is consistent with topologically-driven reconnection models \cite{Titov2012, Liu2016}.  In this case, observations show shear and rotational motions at the photosphere providing a clear buildup of energy proceeding the flares, but no sudden changes precipitate a flare.  

\section{Conclusion \& Future work}

In this paper, our goal is the characterization of the heterogeneous patterns shared by different active regions on the surface of the Sun based on data driven approaches. We propose two types of mixture models: MM-R and MM-H. The first, MM-R, is designed to specify the heterogeneity across active regions. The second, MM-H, goes beyond to specify the heterogeneity for flaring patterns within an active region. As demonstrated, using mixture modeling improves the predictive performance on the solar flare prediction. The second model MM-H performs marginally better. Since the extension of heterogeneity to individual flare events within an active region in \ref{sec:mm2H} does not yield conclusive gains,  the mixture heterogeneous nature seems to be strongly due to active regions, while flaring events within active regions tend to be more homogeneous. Another contribution of this paper is showing how to deal with the imbalance problem using the Expectation Maximization framework.  \\

Significantly our work demonstrates the clustering results of mixture model MM-R. We observed three clusters, namely `H', `I', `L'. As our mixture model is designed to do clustering at the level of interaction between covariates and responses, it implies three distinct linear mechanism for three clusters. Moreover, the `H' group of ARs produces significantly more strong flares, while ARs in group `I' have few strong flares and a majority of weak flare events. In contrast, ARs with `L' labels have only weak flare events. \\

An equally important result is based on the fact that MM-H is marginally better than MM-R demonstrates that flares from the same AR are intrinsically homogeneous. This result is fully consistent with what was already know about homologous flares, that the magnetic configuration remains similar between successive flares and is reformed between events flare. Such flares are readily explained by reconnection of coronal magnetic fields resulting in flare ribbons in the chromosphere \cite{Sui2004, Liu2014, Janvier2023}, which is now considered the standard model in solar physics.  The energized field is in the form of a sheared core or filament channel, which persists or reforms by shearing motions after subsequent flares \cite{Manchester2003,Romano2018}.  \\

The mixture model also discerns heterogeneity between active regions in three distinct clusters.  The H cluster is representative of the most energetic events.  These flares follow sudden intensification of magnetic fields and their gradients, which can follow from emergence of intense magnetic fields or large-scale collisions of opposite polarities. This evolution comes with clear and distinct signatures of the covariates.  The I-class events follow a buildup of energy from shear and rotational flows associated with lower levels of emerging magnetic fields.   At lower energies, we find a buildup of energy that eventually activates flares from a topological feature. The I and L classes show more similarity with a shear and rotational flows producing an energy buildup.  However, while the I class is associated with flux emergence the least energetic L class events follow an energy buildup with little emergence occurring over the time-scale of the flare events. In this sense, we find a clear pattern related to the relative disruption the photospheric magnetic field driving the flare events. \\

In this work, we assume independent linear relationship between $X_i$ and $y_i$ for each data point in a mixture cluster. This does not take into account the fact that flares occur through time and there may be temporal correlation between a future event and past ones. The next step is to adopt more sophisticated regression methods such as Gaussian Process Regression. Moreover, we can also apply a more powerful approach to determine the number of mixture components. Existing methodologies like Dirichlet Process or Hierarchical Dirichlet Process seems to be a promising direction to pursue.

\newpage


\bibliography{main.bib}

\newpage

\appendix

\section{EM Algorithm Derivation}

For mixture model MM-R defined in section \ref{sec:mm2R}, we can write the complete likelihood as

\begin{align*}
p(y_1, y_2, \ldots y_n, z^1, \ldots, z^R | X_1, X_2, \ldots, X_n)) &= \prod_{r=1}^R p(z^r) \prod_{i=1}^{n_r} p(y_i^r |z^r, X_i^r) \\
&= \prod_{r=1}^R \prod_{k=1}^K \left[\pi_k \prod_{i=1}^{n_r} p(y_i^r |z^r = k, X_i^r) \right]^{\mathbbm{1}(z^r = k)}. \\ 
\end{align*}

So the complete log likelihood can be written as

\begin{align*}
l(\beta, z) &= \log p(y_1, y_2, \ldots y_n, z^1, \ldots, z^R | X_1, X_2, \ldots, X_n)) \\
&= \sum_{r=1}^R \sum_{k=1}^K \mathbbm{1}(z^r = k) \cdot \left[ \log \pi_k +  \sum_{i=1}^{n_r} \log p(y_i^r | z_i^r = k , X_i^r)  \right].
\end{align*}

Since $y_i^r | X_i^r, z_i^r = k \sim \text{N}(\cdot | \beta_k^T X_i^r, \sigma^2_k)$, by denoting $\tau^r_k := \mathbbm{E}[z^r = k | X^r, y^r] $, the expected complete log likelihood is 

\begin{align*}
el(\beta, z)  &= \sum_{k=1}^K \sum_{r=1}^R \mathbbm{E}[z^r = k | X^r, y^r] \cdot \left[ \log \pi_k - \sum_{i=1}^{n_r} \left( \cfrac{1}{2} \log \sigma_k^2 - \cfrac{1}{2\sigma_k^2} \cdot \left( y_i^r - \beta_k^T X_i^r\right)^2 \right) \right] \\
&= \sum_{k=1}^K \sum_{r=1}^R \tau_k^r \cdot \left[ \log \pi_k - \sum_{i=1}^{n_r} \left( \cfrac{1}{2} \log \sigma_k^2 - \cfrac{1}{2\sigma_k^2} \cdot \left( y_i^r - \beta_k^T X_i^r\right)^2 \right) \right]. \\
\end{align*}

Now, as explained in previous sections, to combat the data unbalanced issue, we optimize a weighted version of the complete log likelihood

\[ \text{arg max}_{\beta, \pi, \sigma^2} \sum_{k=1}^K \sum_{r=1}^R \tau_k^r \cdot \left[ \log \pi_k - \sum_{i=1}^{n_r} \left( \cfrac{w_i}{2} \log \sigma_k^2 - \cfrac{w_i}{2\sigma_k^2} \cdot \left( y_i^r - \beta_k^T X_i^r\right)^2 \right) \right].  \]

For the M-step, taking derivatives w.r.t each parameters and setting to zeros, it is straightforward to see that

\begin{align*}
    &\hat{\pi}_k = \cfrac{\sum_{r=1}^R \tau_k^r}{R} \\
    &\hat{\beta}_k = \left[\sum_{r=1}^R \tau_k^r \sum_{i=1}^{n_r} w_i X_i^r (X_i^r)^T \right]^{-1} \left[\sum_{r=1}^R \tau_k^r \sum_{i=1}^{n_r} w_i y_i^r X_i^r \right] \\
    &\hat{\sigma}^2_k = \cfrac{\sum_{r=1}^R \tau_k^r \sum_{i=1}^{n_r} w_i \cdot (y_i^r - \hat{\beta}_k^T X_i^r)^2}{ \sum_{r=1}^R n_r \cdot \tau_k^r}.
\end{align*}

For the E-step, 
\begin{align*}
\tau^r_k &= \mathbb{E}(z^r = k | X^r, y ^r) \\
&= \cfrac{\pi_k p(y^r| X^r, z^r = k)}{ \sum_{j=1}^K \pi_j p(y^r| X^r, z^r = j)} \\
&= \cfrac{\pi_k \prod_{i=1}^{n_r} p(y^r_i| X_i^r, z^r = k)}{ \sum_{j=1}^K \pi_j \prod_{i=1}^{n_r} p(y^r_i| X_i^r, z^r = j)} \\
&= \cfrac{\pi_k \cdot \prod_{i=1}^{n_r} \text{N}\left(y_i^r | \beta_k^T X_i^r, \sigma_k^2 \right) }{\sum_{j=1}^K \pi_j  \cdot \prod_{i=1}^{n_r} \text{N}\left(y_i^r | \beta_j^T X_i^r, \sigma_j^2 \right)}.
\end{align*}

Similarly, with model MM-H define in section \ref{sec:mm2R}, the expected complete log likelihood is

\begin{align*}
el(\beta, z)  &= \sum_{k=1}^K \sum_{r=1}^R  \sum_{i=1}^{n_r} \tau_{i,k}^r\cdot \left[ \log \pi_k^r -  \left( \cfrac{1}{2} \log \sigma_k^2 - \cfrac{1}{2\sigma_k^2} \cdot \left( y_i^r - \beta_k^T X_i^r\right)^2 \right) \right].
\end{align*}

where $\tau_{i,k}^r = \mathbb{E}(z_i^r = k | y_i^r, X_i^r)$. Again to deal with the data imbalance, we work with the weighted optimization instead

\[ \text{arg max}_{\beta, \pi, \sigma^2} \sum_{k=1}^K \sum_{r=1}^R \sum_{i=1}^{n_r} \tau_{i,k}^r \cdot \left( \log \pi_k^r - \cfrac{w_i}{2} \log \sigma_k^2 - \cfrac{w_i}{2\sigma_k^2} \cdot (y_i^r - \beta_k^T X_i^r)^2 \right). \]

For M-step, it is easy to derive

\begin{align*}
    &\hat{\pi}_k^r = \cfrac{\sum_{i=1}^{n_r} \tau_{i,k}^r}{n_r} \\
    &\hat{\beta}_k = \left[\sum_{r=1}^R \sum_{i=1}^{n_r} \tau_{i,k}^r \cdot w_i \cdot X_i^r (X_i^r)^T \right]^{-1} \left[\sum_{r=1}^R \sum_{i=1}^{n_r} \tau_{i,k}^r \cdot y_i^r X_i^r \right] \\
    &\hat{\sigma}^2_k = \cfrac{\sum_{r=1}^R \sum_{i=1}^{n_r} \tau_{i,k}^r w_i \cdot (y_i^r - \hat{\beta}_k^T X_{i,r})^2}{ \sum_{r=1}^R \sum_{i=1}^{n_r} \tau_{i,k}^r}.
\end{align*}

For E-step, 

\begin{align*}
\tau_{i,k}^r &= \mathbb{E}(z_i^r = k | X^r_i, y ^r_i) \\
&= \cfrac{\pi_k^r p(y^r_i| X_i^r, z_i^r = k)}{ \sum_{j=1}^K \pi_j p(y^r_i| X_i^r, z_i^r = j)} \\
&= \cfrac{\pi_k \cdot \text{N}\left(y_i^r | \beta_k^T X_i^r, \sigma_k^2 \right) }{\sum_{j=1}^K \pi_j \cdot \text{N}\left(y_i^r | \beta_j^T X_i^r, \sigma_j^2 \right)}.
\end{align*}

\newpage

\section{Case Study Plot \& Additional Plots}
\label{appendix:plots}


\begin{figure}[H]
     \centering
      \begin{subfigure}[b]{0.9\textwidth}
         \centering
         \includegraphics[scale = 0.33]{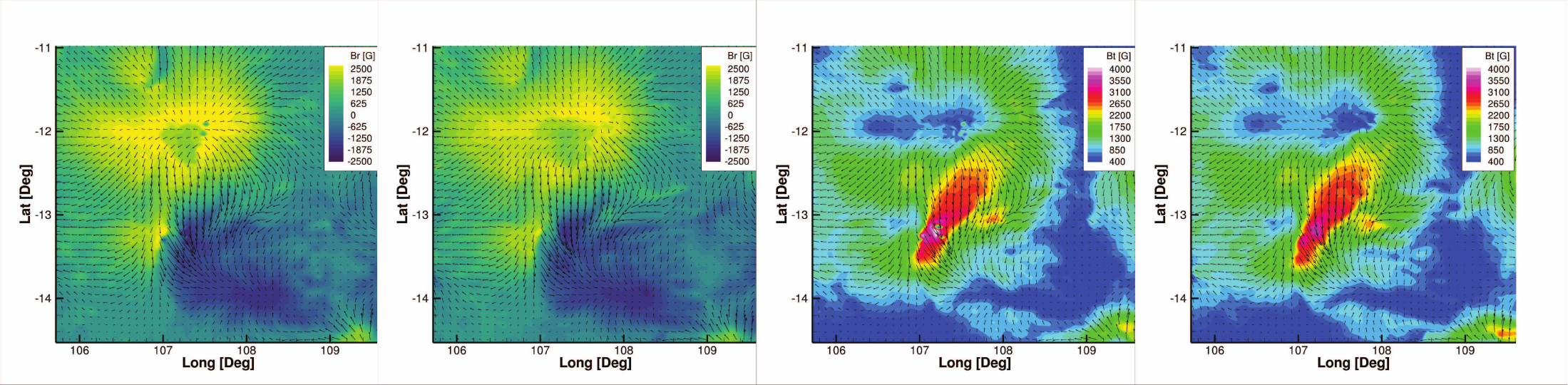}
        \caption{The left two images are radial components and right two ones are horizontal filed in AR 11967 (H label) at its strongest (2014.01.30\_16:11:00) and weakest flare events (2014.01.30\_13:36:00) during its existence.}
     \label{fig:appendix:11967weakhmi}
     \end{subfigure}
     \hfill
     \begin{subfigure}[b]{0.9\textwidth}
        \centering
        \includegraphics[scale = 0.16]{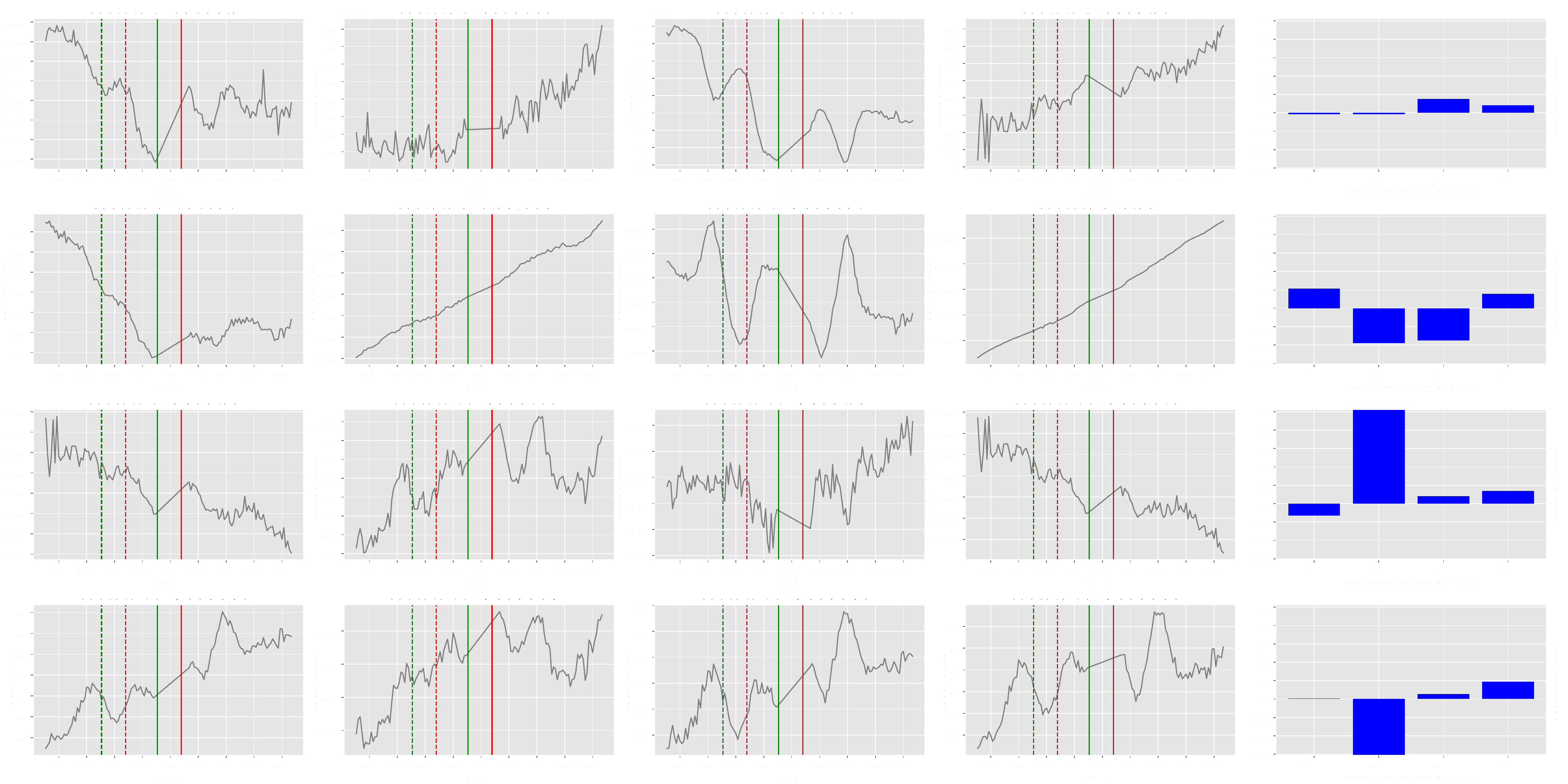}
        \caption{(Left) Evolution of SHARP parameters during the strongest (red) and weakest (green) flare events in AR 11967 (H label) and (Right) MM-R estimated $\hat{\beta}_k$ values. The dash lines are the times at their peak intensities. The solid lines are the 6-hour-before for covariates' regression. Green color is the weakest flare and red is the strongest one.}
        \label{fig:appendix:11967joinedts}
     \end{subfigure}
     \caption{Active Region 11967's SHARP parameters and HMI images. AR 11967 existed from 2014/01/30 to 2014/02/08.}
     \label{fig:appendix:11967event}
\end{figure}

\newpage


\begin{figure}[H]
     \centering
     \begin{subfigure}[b]{0.9\textwidth}
         \centering
        \includegraphics[scale = 0.33]{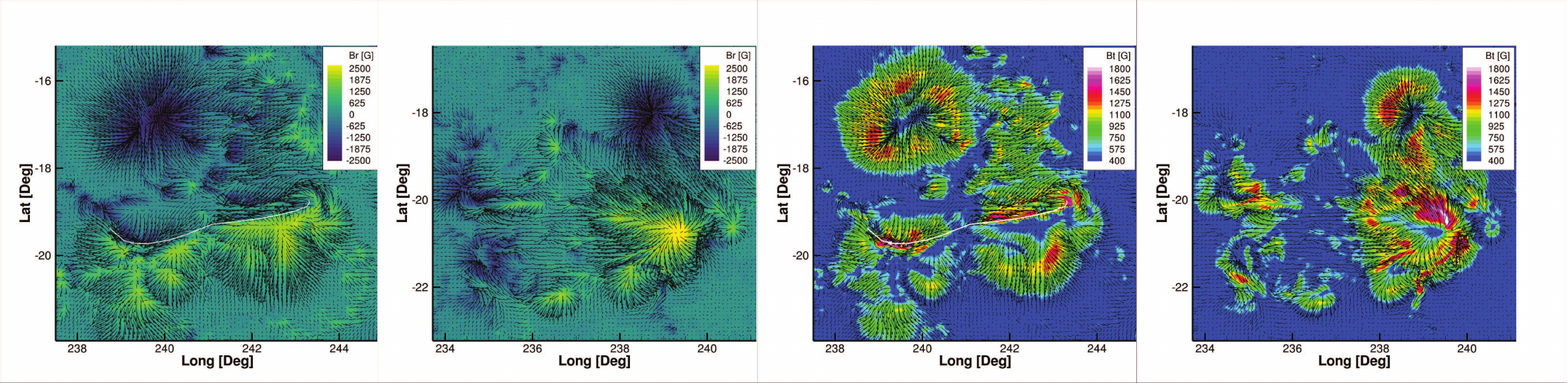}
     \end{subfigure}
     \hfill
     
     \caption{The left two images are radial components and right two ones are horizontal filed in AR 12242 (H label) at its strongest (2014.12.20\_00:28:00) and weakest flare events (2014.12.16\_00:05:00) during its existence. AR 12242 existed from 2014/12/14 to 2014/12/21.}
\end{figure}

\begin{figure}[!ht]
     \centering
     \begin{subfigure}[b]{0.9\textwidth}
         \centering
        \includegraphics[scale = 0.15]{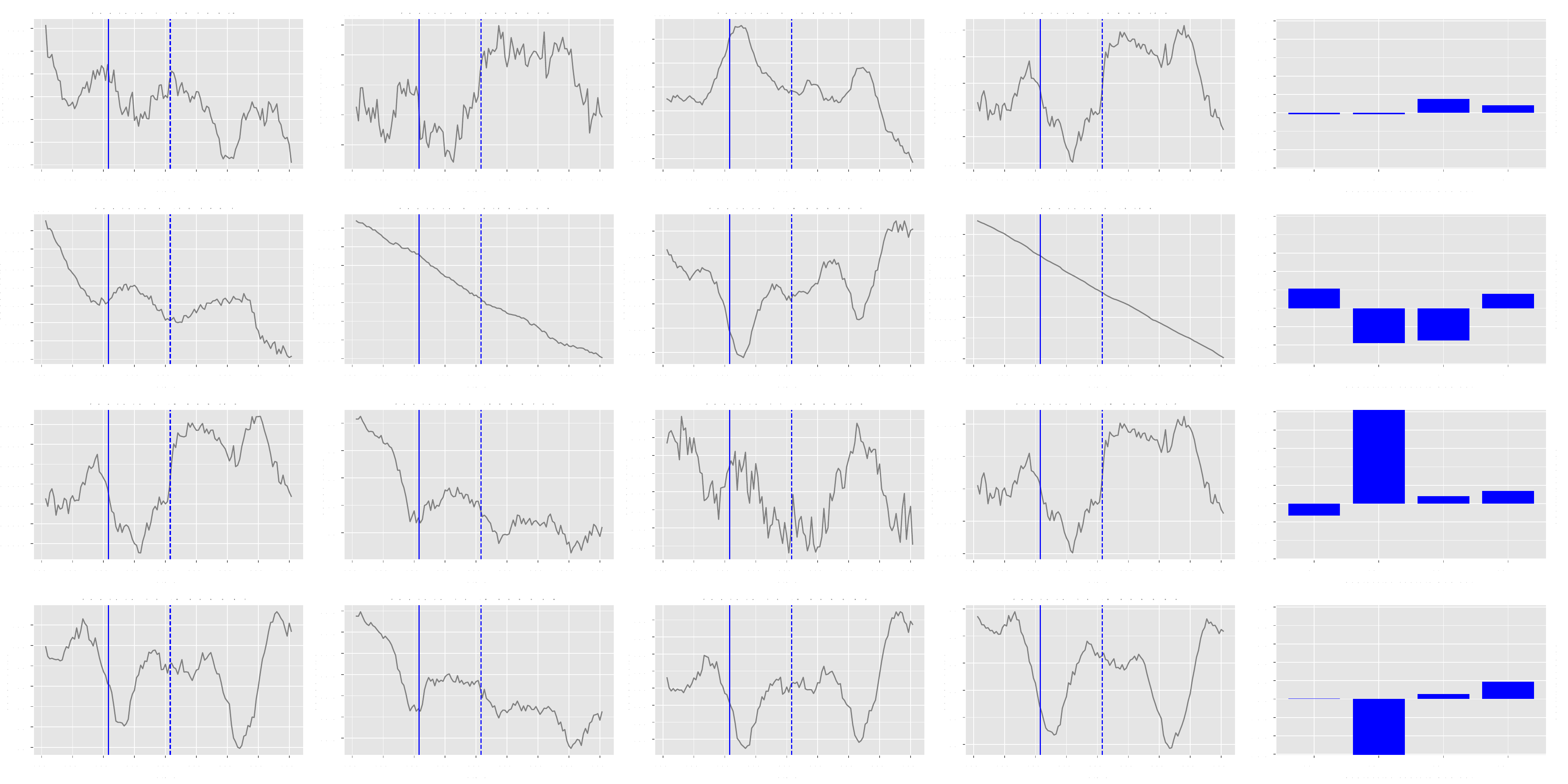}
        \caption{(Left) Evolution of SHARP parameters during the strongest flare event in AR 12242 (H label) and (Right) MM-R estimated $\hat{\beta}_k$ values.}
     \end{subfigure}
     \hfill
     \begin{subfigure}[b]{0.9\textwidth}
         \centering
        \includegraphics[scale = 0.15]{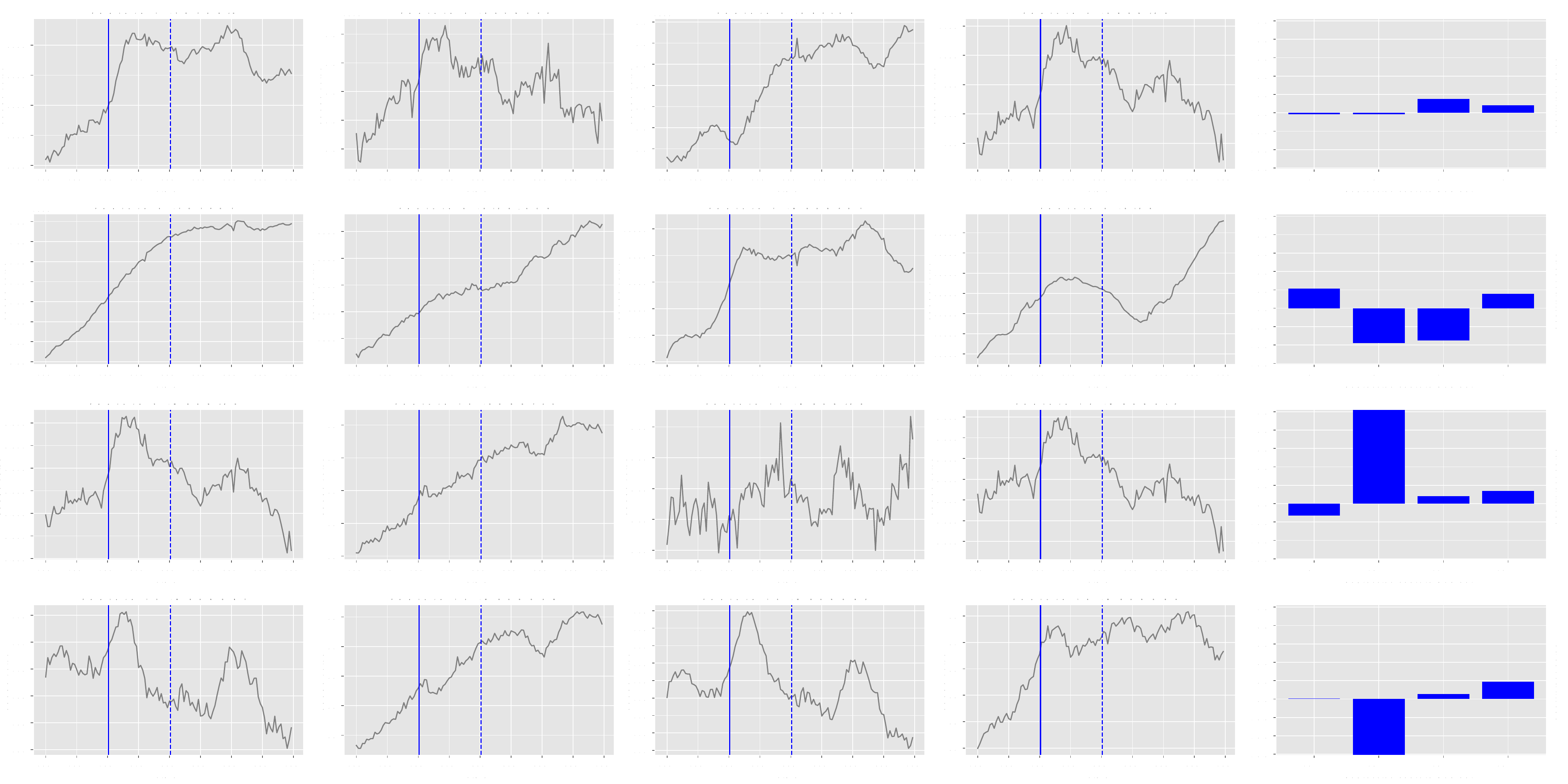}
        \caption{Evolution of SHARP parameters during the weakest flare event in AR 12242 (H label) and (Right) MM-R estimated $\hat{\beta}_k$ values.}
     \end{subfigure}
     \caption{Active Region 12242's SHARP parameters. AR 12242 existed from 2014/12/14 to 2014/12/21.}
\end{figure}



\begin{figure}[H]
     \centering
     \begin{subfigure}[b]{0.9\textwidth}
         \centering
        \includegraphics[scale = 0.33]{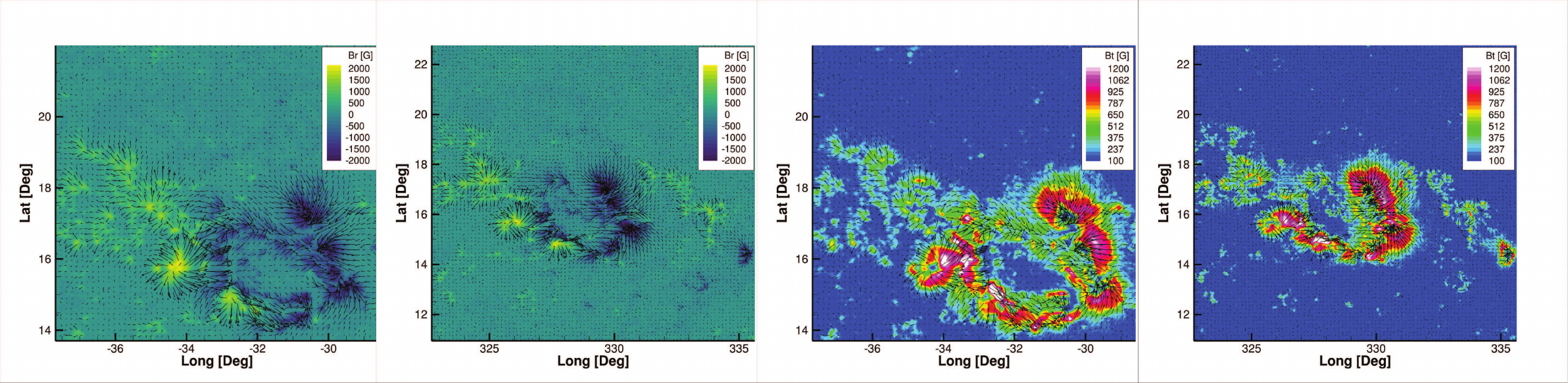}
     \end{subfigure}
     \hfill
     
     \caption{The left two images are radial components and right two ones are horizontal filed in AR 11261 (I label) at its strongest (2011.07.30\_02:09:00 and weakest flare events (	2011.07.30\_19:41:00 during its existence. AR 11261 existed from 2011/7/28 to 2011/8/5.}
\end{figure}

\begin{figure}[!ht]
     \centering
     \begin{subfigure}[b]{0.9\textwidth}
         \centering
        \includegraphics[scale = 0.15]{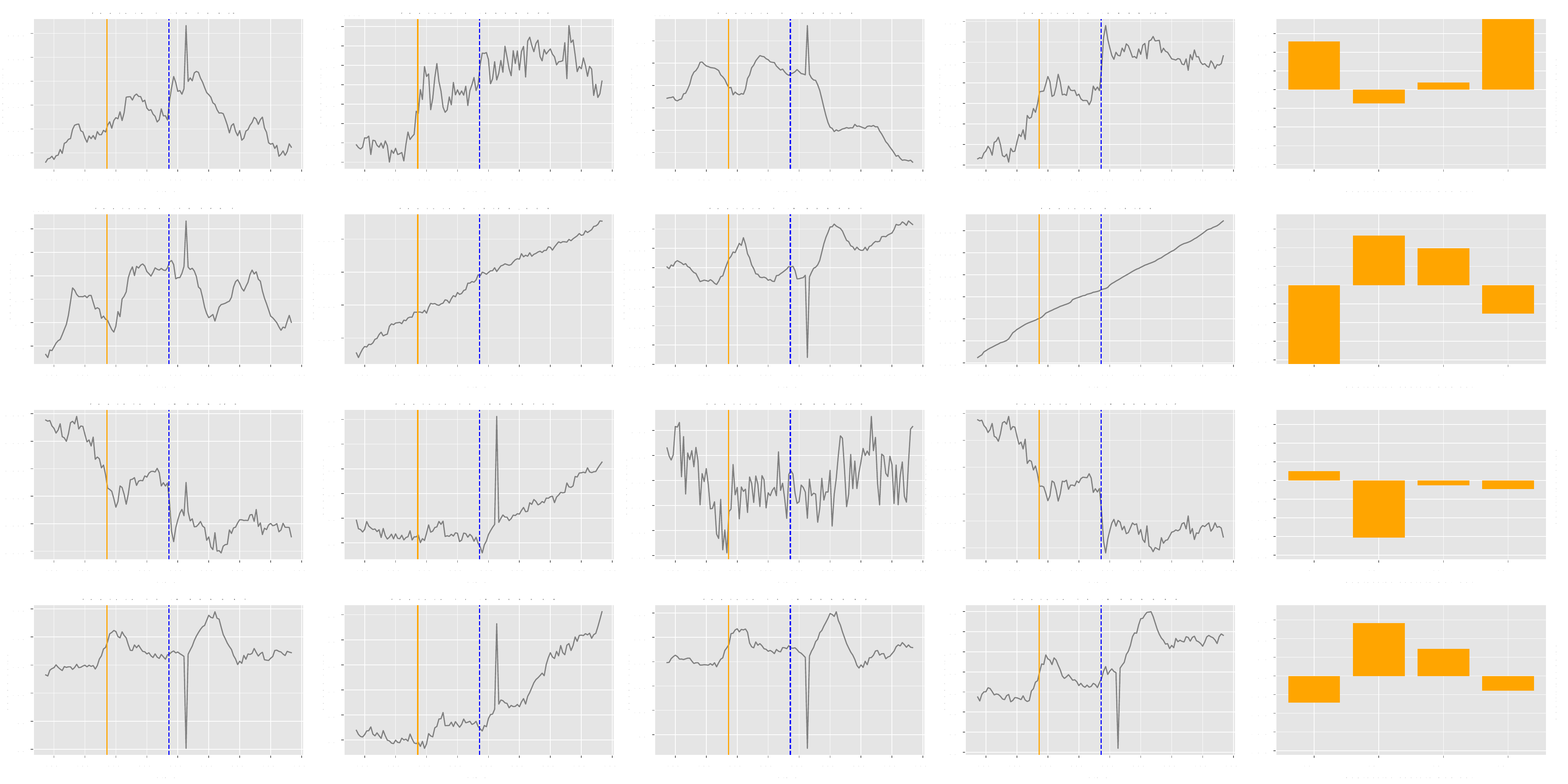}
        \caption{(Left) Evolution of SHARP parameters during the strongest flare event in AR 11261 (I label) and (Right) MM-R estimated $\hat{\beta}_k$ values.}
     \end{subfigure}
     \hfill
     \begin{subfigure}[b]{0.9\textwidth}
         \centering
        \includegraphics[scale = 0.15]{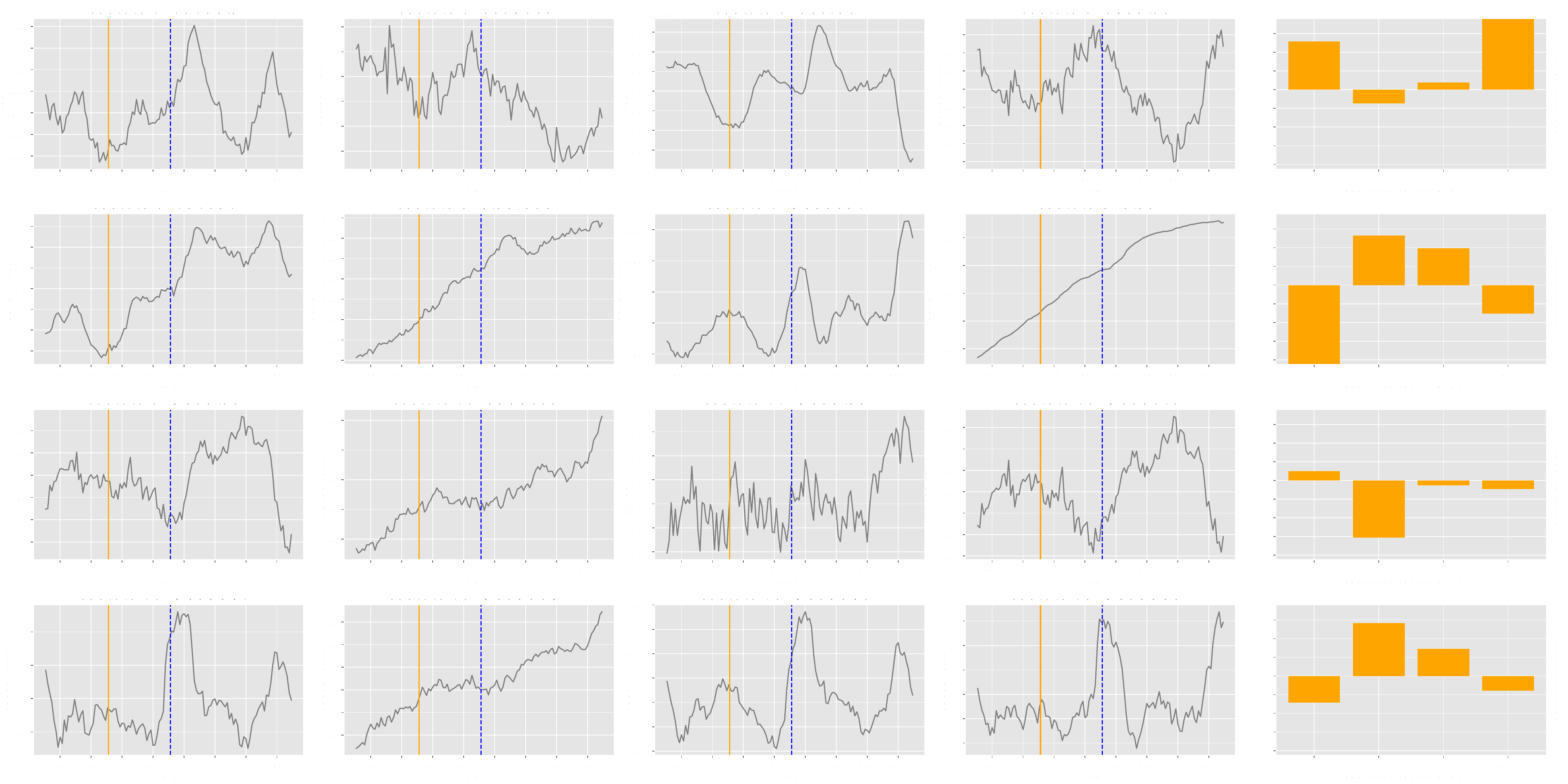}
        \caption{Evolution of SHARP parameters during the weakest flare event in AR 11261 (I label) and (Right) MM-R estimated $\hat{\beta}_k$ values.}
     \end{subfigure}
     \caption{Active Region 11261's SHARP parameters.  AR 11261 existed from 2011/7/28 to 2011/8/5}
\end{figure}



\begin{figure}[H]
     \centering
     \begin{subfigure}[b]{0.9\textwidth}
         \centering
        \includegraphics[scale = 0.33]{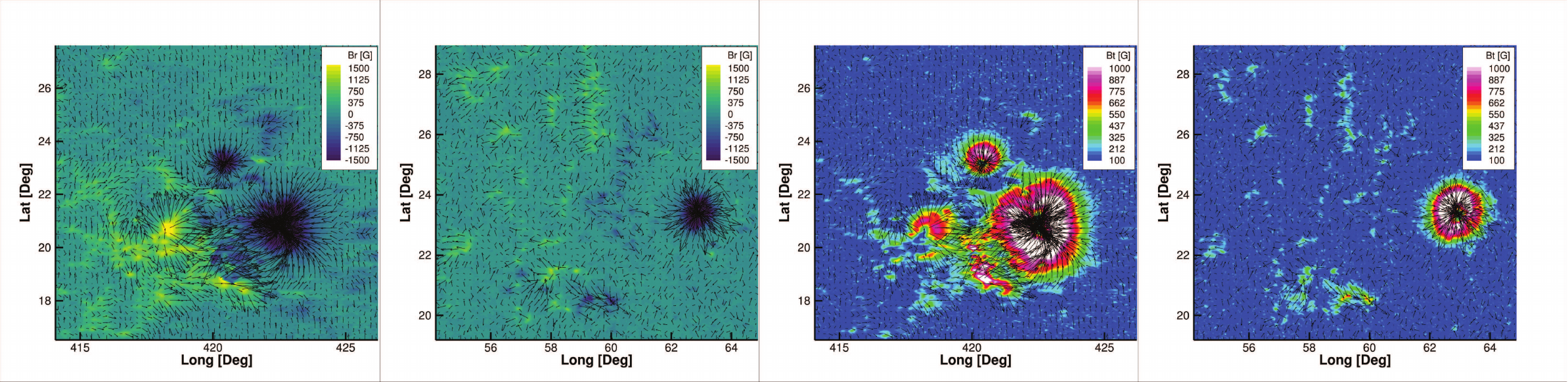}
     \end{subfigure}
     \hfill

     \caption{The left two images are radial components and right two ones are horizontal filed in AR 11117 (L label) at its strongest (2010.10.31\_04:31:00) and weakest flare events (2010.10.22\_07:59:00) during its existence. AR 11117 existed from 2010/10/21 to 2010/10/31.}
\end{figure}

\begin{figure}[!ht]
     \centering
     \begin{subfigure}[b]{0.9\textwidth}
         \centering
        \includegraphics[scale = 0.15]{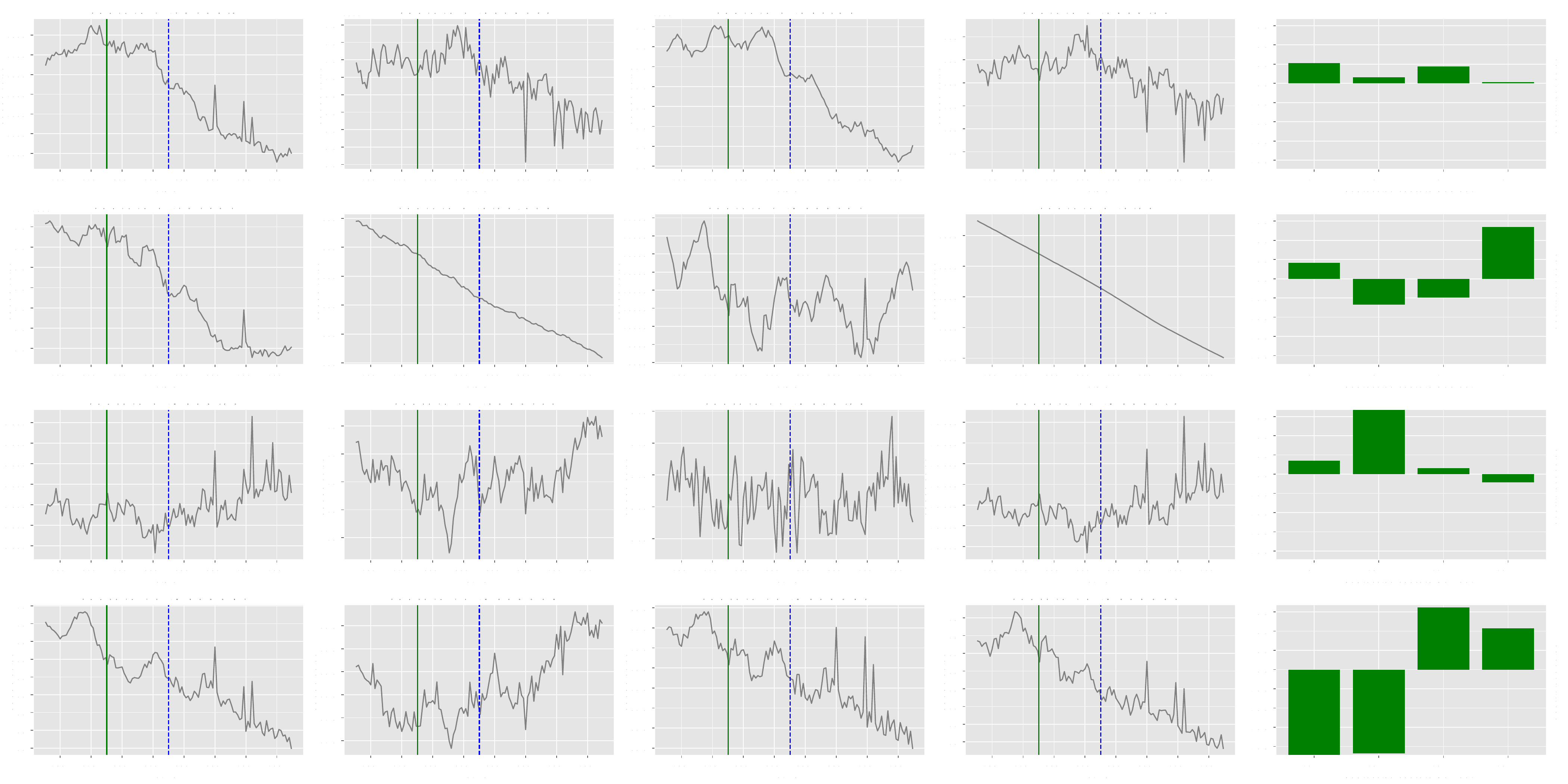}
        \caption{(Left) Evolution of SHARP parameters during the strongest flare event in AR 11117 (L label) and (Right) MM-R estimated $\hat{\beta}_k$ values.}
     \end{subfigure}
     \hfill
     \begin{subfigure}[b]{0.9\textwidth}
         \centering
        \includegraphics[scale = 0.15]{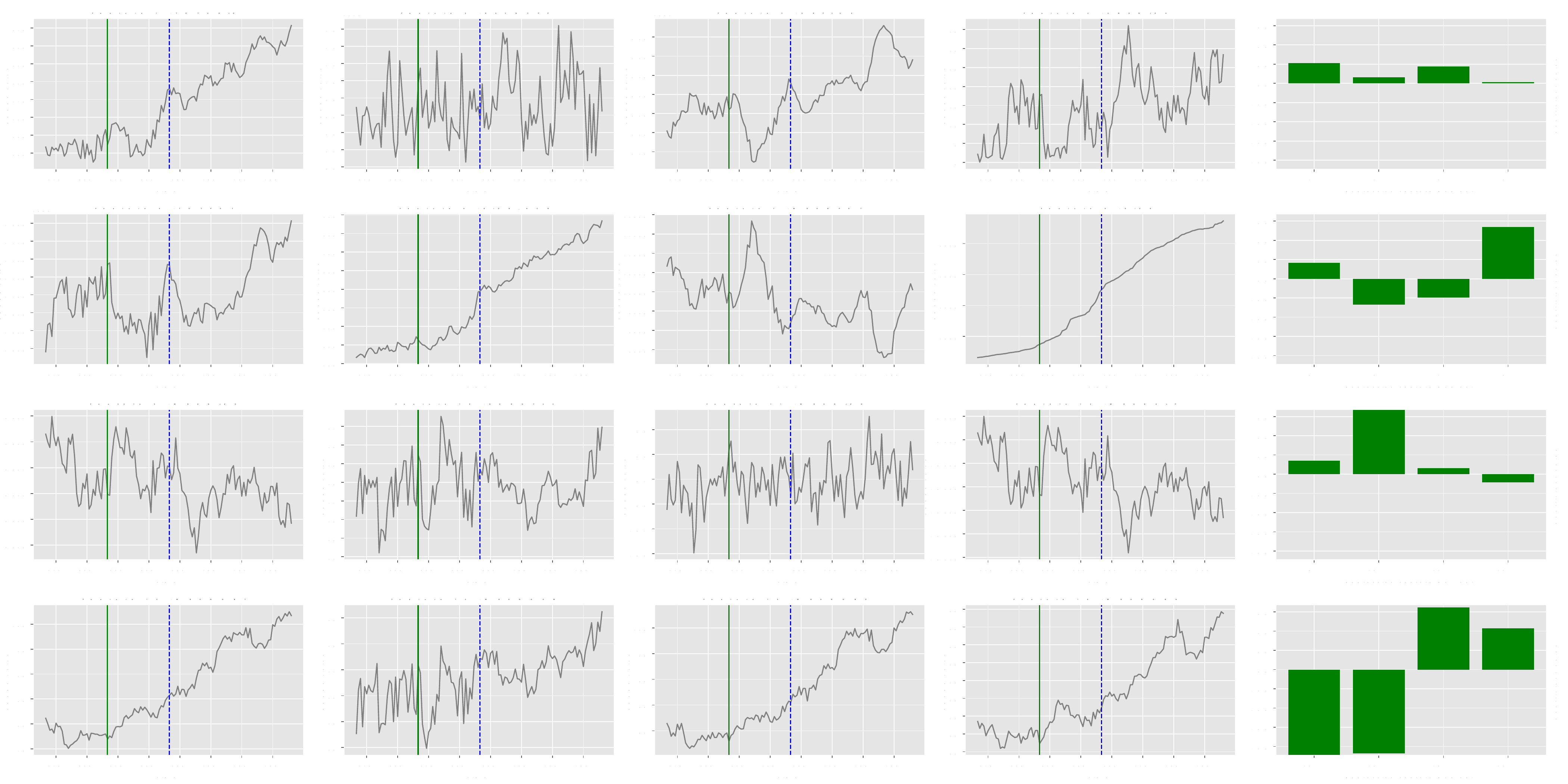}
        \caption{Evolution of SHARP parameters during the weakest flare event in AR 11117 (L label) and (Right) MM-R estimated $\hat{\beta}_k$ values.}
     \end{subfigure}
     \caption{Active Region 11117's SHARP parameters. AR 11117 existed from 2010/10/21 to 2010/10/31.}
\end{figure}

\newpage


\begin{figure}[H]
    \centering
    
    \begin{subfigure}{.9 \textwidth}
        \centering
        \includegraphics[scale=0.4]{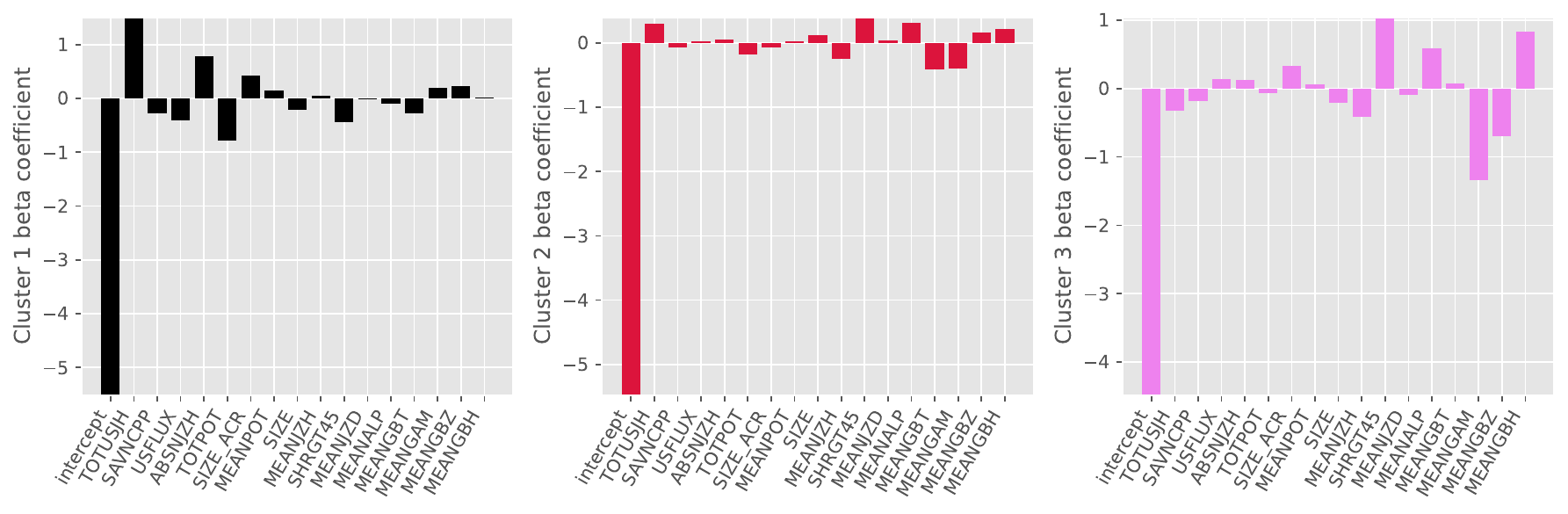}
        \caption{Mixture Model over flare events (\ref{sec:mm2H}) estimated $\{\hat{\beta} \}_1^K$.}
        \label{fig:2Hbeta}
    \end{subfigure}%
    
    \begin{subfigure}{.9 \textwidth}
        \centering
        \includegraphics[scale=0.45]{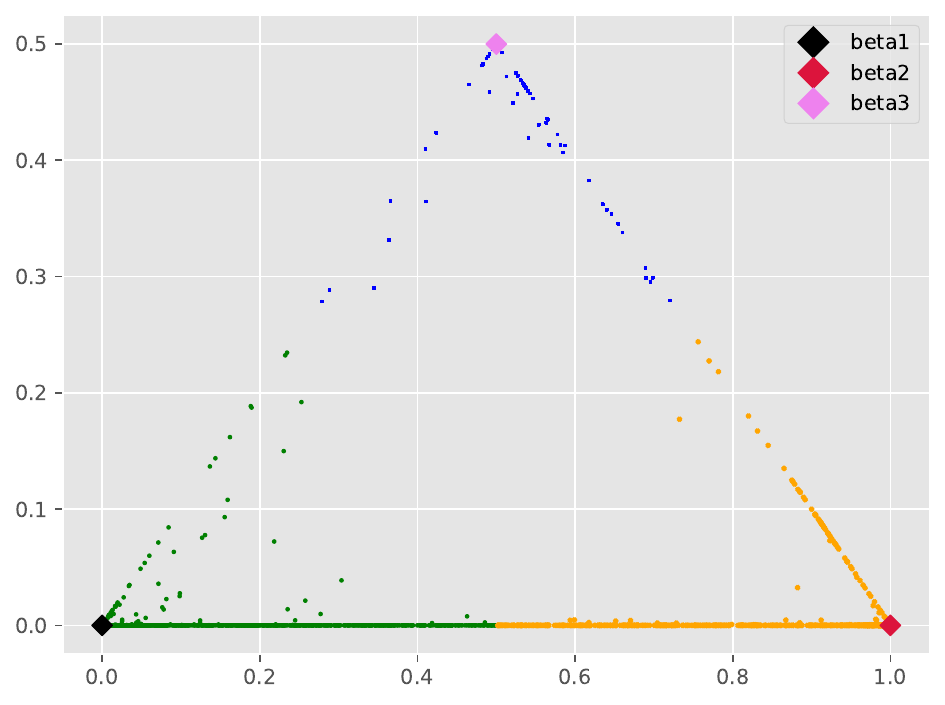}
        \caption{Mixture Model over flare events (\ref{sec:mm2H}) mixing proportion of estimated $\{\hat{\beta} \}_1^K$ for each flare event. This plot shows the influence of each of the K linear mechanisms on ARs' flare events. The closer an event to a vertex of the simplex, the stronger its influence.}
        \label{fig:2Hbetacluster}
    \end{subfigure}%
    \caption{MM-H 's estimated $\{\hat{\beta} \}_1^K$ and the mixing proportion w.r.t these $\hat{\beta}$ for each flare event.}
\end{figure}

\begin{figure}[!ht]
    \centering
    \begin{subfigure}{.45\textwidth}
        \centering
        \includegraphics[scale=0.32]{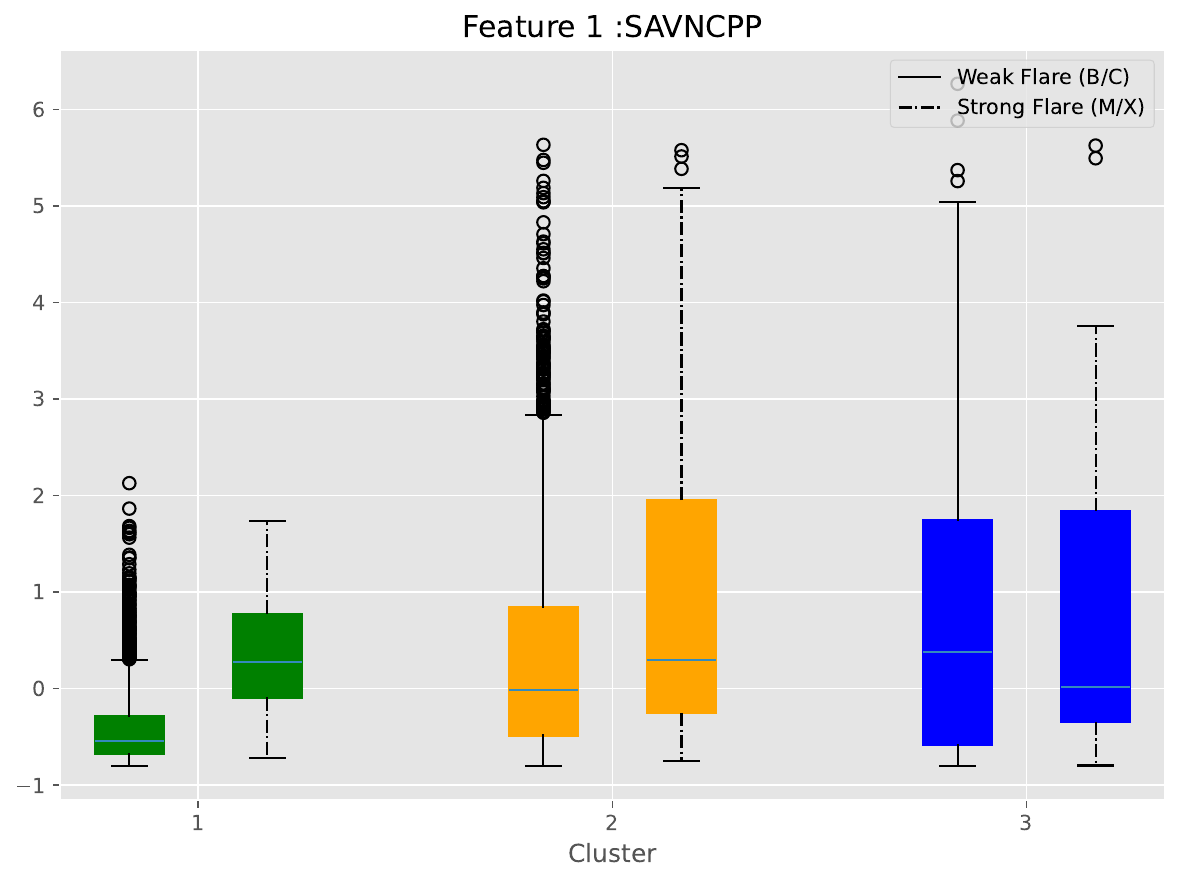}
        \caption{Mixture Model over flare events MM-H (\ref{sec:mm2H}) Sum of the Absolute Value of the Net Currents Per Polarity in Amperes.}
        \label{fig:2Hf1}
    \end{subfigure}%
    \begin{subfigure}{0.45\textwidth}
        \centering
        \includegraphics[scale=0.32]{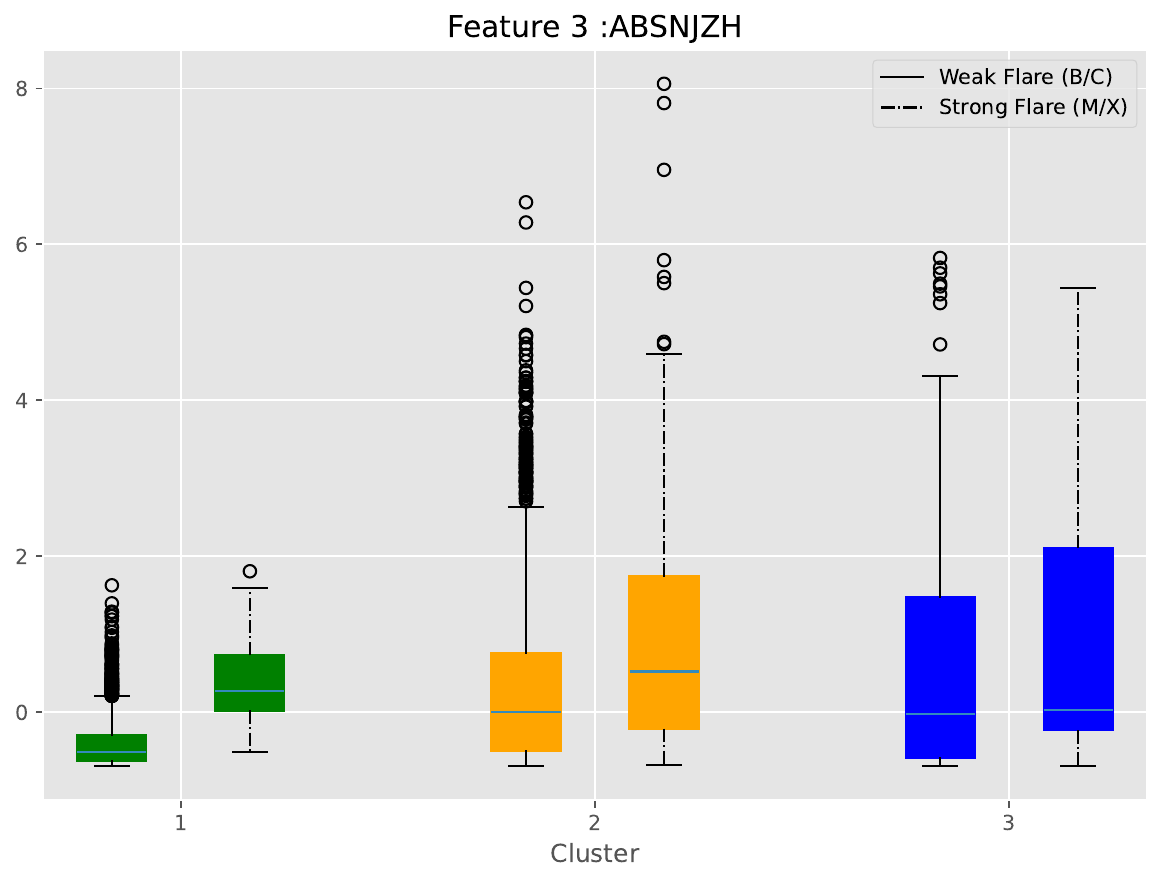}
    \caption{Absolute value of the net current helicity in G2/m.}
        \label{fig:2Hf3}
    \end{subfigure}
\end{figure}

\begin{figure}[!ht]
    \centering
    
    \begin{subfigure}{.45\textwidth}
        \centering
        \includegraphics[scale=0.32]{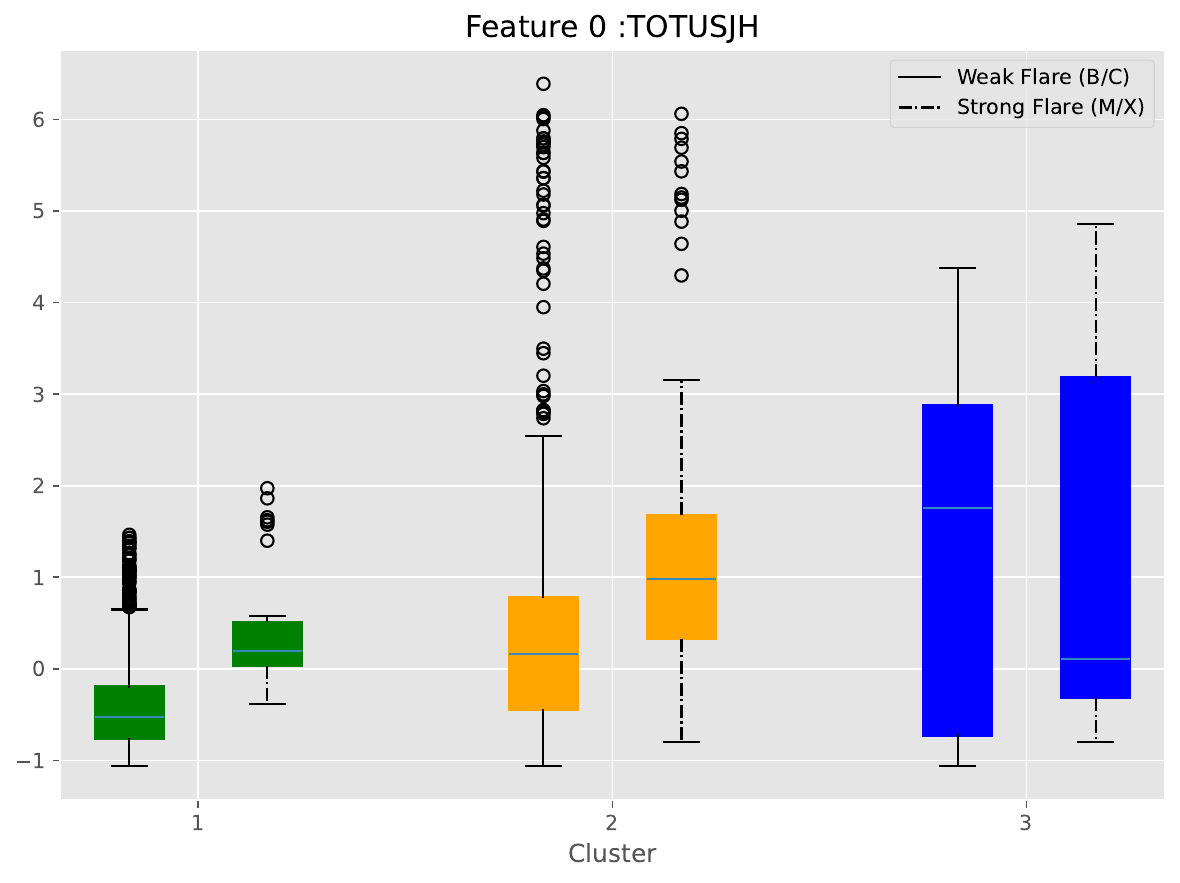}
        \caption{Mixture Model over flare events MM-H (\ref{sec:mm2H}) Total photospheric magnetic energy density in ergs per cubic centimeter.}
        \label{fig:2Hf0}
    \end{subfigure}%
    \begin{subfigure}{0.45\textwidth}
        \centering
        \includegraphics[scale=0.34]{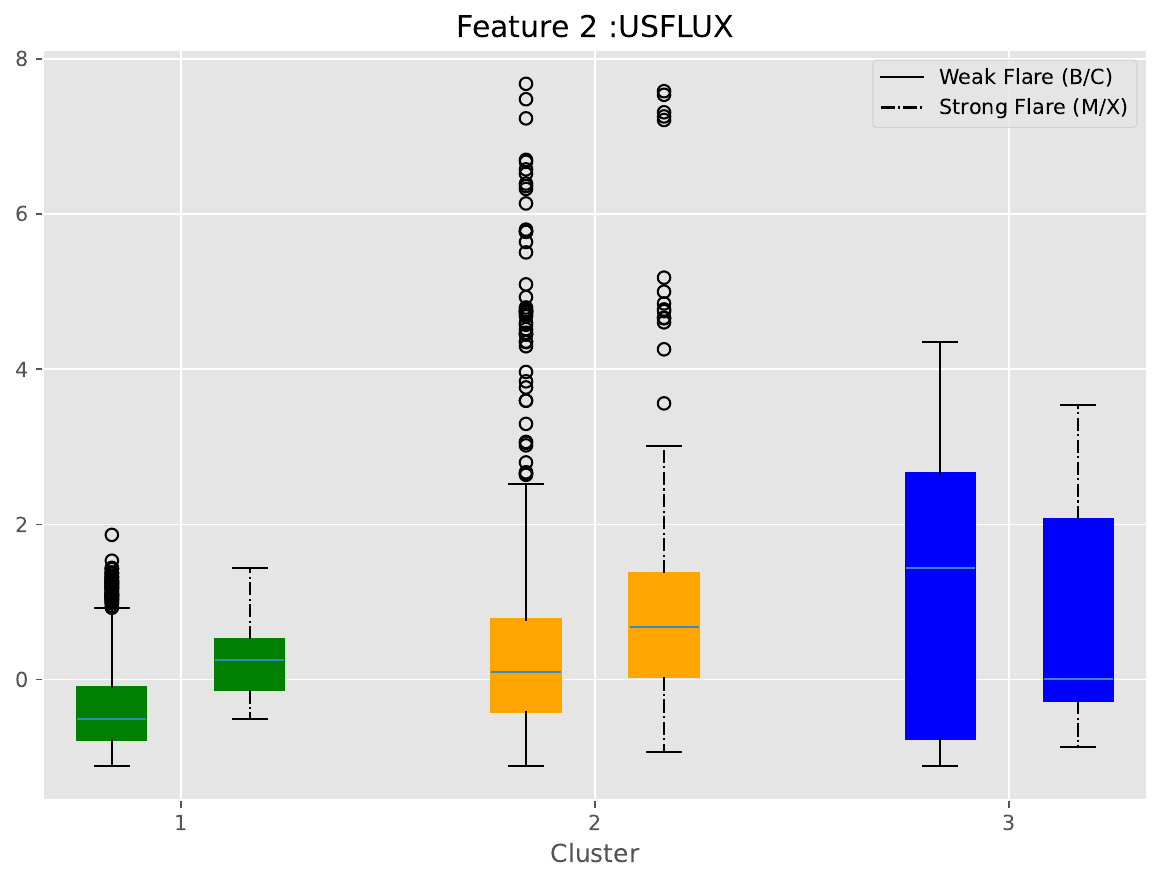}
        \caption{Total unsigned flux in Maxwells.}
        \label{fig:2Hf2}
    \end{subfigure}

    \begin{subfigure}{.45\textwidth}
        \centering
        \includegraphics[scale=0.32]{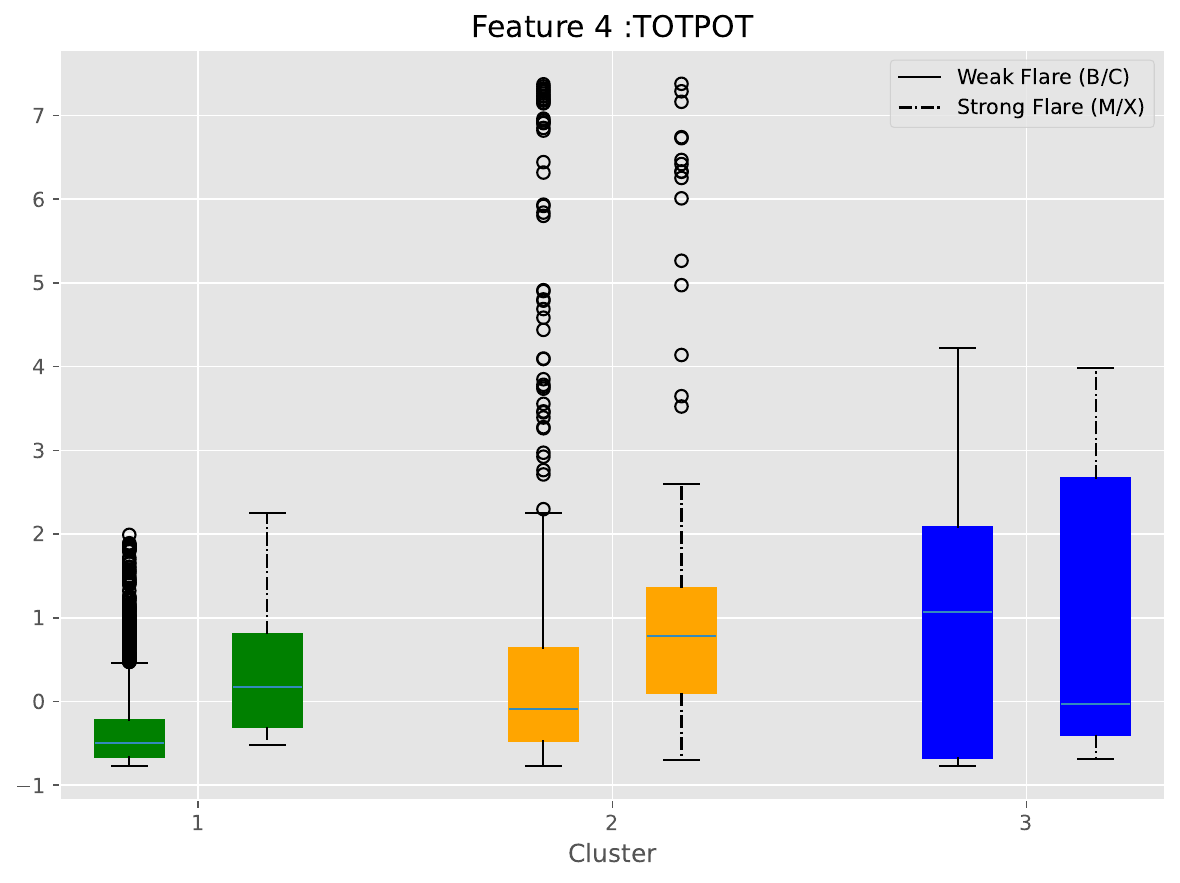}
        \caption{Total photospheric magnetic energy density in ergs per cubic centimeter.}
        \label{fig:2Hf4}
    \end{subfigure}%
    \begin{subfigure}{0.45\textwidth}
        \centering
        \includegraphics[scale=0.32]{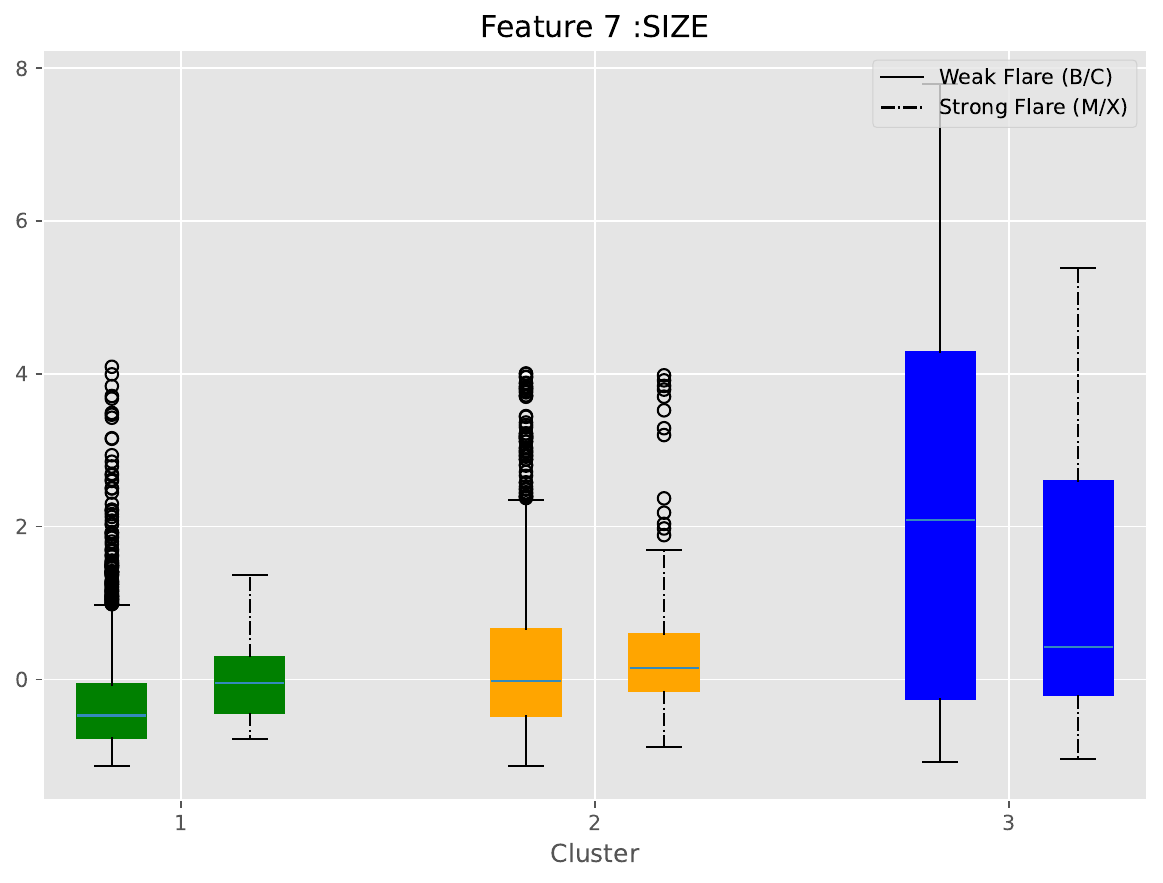}
        \caption{Projected area of patch on image in micro-hemisphere.}
        \label{fig:2Hf7}
    \end{subfigure}

    \begin{subfigure}{.45\textwidth}
        \centering
        \includegraphics[scale=0.32]{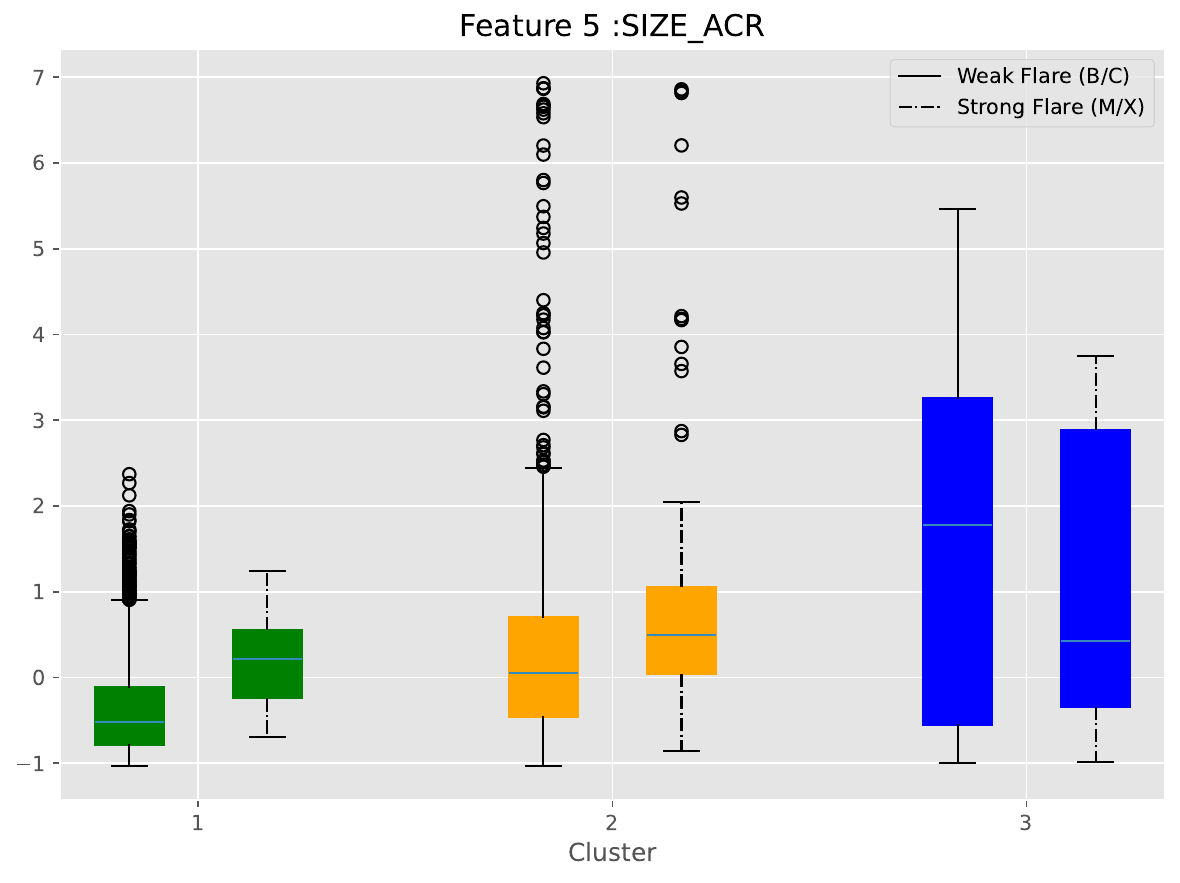}
        \caption{Projected area of active pixels on image in micro-hemisphere.}
        \label{fig:2Hf5}
    \end{subfigure}%
    \begin{subfigure}{0.45\textwidth}
        \centering
        \includegraphics[scale=0.32]{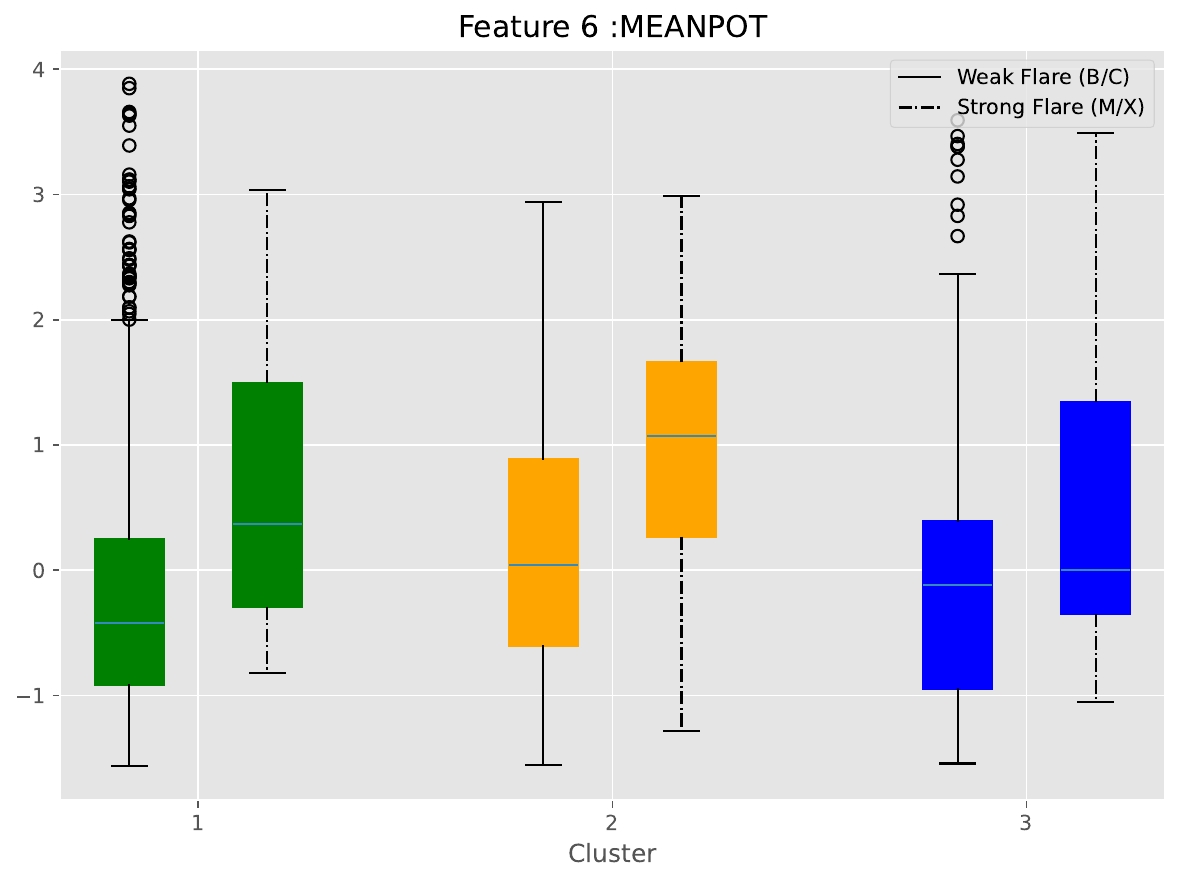}
        \caption{Mean photospheric excess magnetic energy density in ergs per cubic centimeter.}
        \label{fig:2Hf6}
    \end{subfigure}

     \caption{Selected subset of covariate $X$ under each cluster of MM-H.}
\end{figure}

\end{document}